%% file: paper.tex
\documentclass[12pt]{article}

\usepackage[utf8]{inputenc}
\usepackage[margin=1in]{geometry}
\usepackage[T1]{fontenc} 
\usepackage{microtype}
\usepackage[small,compact]{titlesec} 
\usepackage{stix}
\usepackage{graphicx}
\usepackage{amsfonts}
\usepackage{amsmath}
\usepackage{multirow}
\usepackage{setspace}
\usepackage[dvipsnames,table]{xcolor}

\definecolor{redsubsetcol}{RGB}{99,131,175}
\definecolor{subsetcol}{RGB}{191,215,249}
\definecolor{subsetcol}{RGB}{191,215,249}
\definecolor{notsigcolor}{RGB}{119,119,119}

\usepackage{enumitem}
\newlist{Steps}{enumerate}{2}
\setlist[Steps]{label=\textbf{Step \arabic*.},itemindent=*,labelindent=1.5\parindent}

\usepackage{longtable}
\usepackage[linesnumbered,ruled,vlined]{algorithm2e}
\usepackage{hyperref}
\usepackage{natbib}
\usepackage{authblk}
\usepackage{ragged2e}
\usepackage{csvsimple-l3}
\usepackage{booktabs}
\usepackage{siunitx}
\usepackage{xfp}
\usepackage{etoolbox}
\usepackage{ifthen}
\usepackage{threeparttable}
\usepackage{caption}
\usepackage{subcaption}
\usepackage{array}
\usepackage{dsfont}
\usepackage{lscape}

\usepackage[title]{appendix}
\hypersetup{
pdftitle={},
pdfsubject={},
pdfauthor={},
pdfkeywords={mykeywords},
   colorlinks,
    linkcolor={black},
    citecolor={black},
    urlcolor={black},
}
\usepackage{adjustbox}

\input{acronyms}
\input{notation}

\newtoggle{anonymous}
\newtoggle{draft}
\settoggle{anonymous}{false}
\settoggle{draft}{false}

\title{Assessing Utility of Differential Privacy for RCTs}
\iftoggle{anonymous}{
\author[1]{Anonymous}
\affil[1]{University}
}{
\author[1]{Kaitlyn R. Webb}
\author[1]{Soumya Mukherjee}
\author[1]{Aratrika Mustafi}
\author[1]{Aleksandra Slavkovi\'c}
\author[2]{Lars Vilhuber}
\affil[1]{Penn State, Department of Statistics}
\affil[2]{Cornell University, Department of Economics}
}
\date{\today}

\iftoggle{draft}{%
\usepackage{draftwatermark} 
\SetWatermarkScale{0.3}
\SetWatermarkLightness{0.9}
\SetWatermarkText{DRAFT - INCOMPLETE - Do not cite or quote}
}{}

\begin{document}

\maketitle

\begin{abstract}

\input{abstract}
\end{abstract}
\doublespacing

\newpage

\section{Introduction}
\input{section-introduction}


\section{Definitions and Problem setup}
\label{sec:definitions-problemsetup}
\input{section-definitions-problemsetup}

\section{Protecting privacy for RCT data}
\label{sec:synthetic-data}


\input{section-synthetic-data}



\section{Evaluating the Mechanisms}
\label{sec:evaluating}

\input{section-evaluation}



\section{Application to "Reducing Crime and Violence: Experimental Evidence from Cognitive Behavioral Therapy in Liberia" \citep{Reducecrime}}
\label{sec:real-world}
\color{red}

\color{black}
\input{section-real-world}


\section{Discussion}
\input{section-discussion}

\newpage

\bibliographystyle{aea-mod}
\bibliography{references.bib,refs-sesa.bib}

\input{acknowledgements}

\newpage
 \begin{appendices}

\section{Algorithms proposed by \citet{karwa2017finite} in 'Finite Sample Differentially Private Confidence Intervals'}\label{sec:karwa-algos}

\input{section-appendix-syntheticdata}

 \section{Replicating Original Study by \citet{Reducecrime}} \label{sec:replicate-conf}
 
\input{section-appendix-real-world}

\section{Additional Plots and Tables}\label{sec:appendix-plots}
\input{section-appendix-plots-and-tables}

\end{appendices}
\end{document}

%% file: acronyms.tex
\usepackage{acronym}

\acrodef{AEA}{American Economic Association}
\acrodef{AER}{American Economic Review}
\acrodef{API}{application programming interface}
\acrodef{DOI}{Digital Object Identifier}
\acrodef{DP}{differential privacy}
\acrodef{FAIR}{Findable, Accessible, Interoperable, Re-usable}
\acrodef{PSID}{Panel Study of Income Dynamics}
\acrodef{HRS}{Health and Retirement Study}
\acrodef{RCT}{randomized controlled trial}
\acrodef{ICPSR}{Inter-university Consortium for Political and Social Research}
\acrodef{DCAP}{data and code availability policy}
\acrodef{PAP}{pre-analysis plans}
\acrodef{NACJD}{National Archive of Criminal Justice Data}
\acrodef{IAB}{Research Data Center (FDZ) at the Institute for Employment Research}
\acrodef{IRB}{Institutional Review Board}
\acrodef{FSRDC}{Federal Statistical Research Data Centers}
\acrodef{FAQ}{frequently asked questions}
\acrodef{ReStud}{Review of Economic Studies}
\acrodef{EJ}{Economic Journal}
\acrodef{JASA}{Journal of the American Statistical Association}
\acrodef{CJE}{Canadian Journal of Economics}
\acrodef{AJPS}{American Journal of Political Science}
\acrodef{PII}{personally identifiable information}
\acrodef{LMIC}{low- and middle-income countries}
\acrodef{IPA}{\textit{Innovations for Poverty Action}}
\acrodef{SDL}{statistical disclosure limitation}
\acrodef{Algorithm 1}{Hybrid-DP}
\acrodef{Algorithm 2}{Crude Fully-DP}
\acrodef{Algorithm 3}{Advanced Fully-DP}
\acrodef{pureDP}{pure-DP}
\acrodef{appDP}{approximate-DP}
\acrodef{ToT}{Test of Tests}


%% file: notation.tex
\newcommand{\be}{\mathbf{e}}

\newcommand{\bV}{\mathbf{V}}

\newcommand{\bb}{\mathbf{b}}

\newcommand{\bd}{\mathbf{d}}
\newcommand{\bx}{\mathbf{x}}
\newcommand{\bX}{\mathbf{X}}
\newcommand{\bfy}{\mathbf{y}}

\newcommand{\bZ}{\mathbf{Z}}
\newcommand{\bT}{\mathbf{T}}
\newcommand{\bS}{\mathbf{S}}
\newcommand{\bs}{\mathbf{s}}
\newcommand{\mseval}{\xi}
\newcommand{\sanbX}{\widetilde{\mathbf{X}}}
\newcommand{\sanbT}{\widetilde{\mathbf{T}}}
\newcommand{\sanbU}{\widetilde{\mathbf{W}}}
\newcommand{\sanby}{\widetilde{\mathbf{y}}}
\newcommand{\bbeta}{\boldsymbol{\beta}}
\newcommand{\btreat}{\boldsymbol{\tau}}
\newcommand{\bblock}{\boldsymbol{\nu}}
\newcommand{\bcov}{\boldsymbol{\gamma}}
\newcommand{\bxdisc}{\bd^*} 
\newcommand{\uniqr}{\mathbf{r}^*} 
\newcommand{\bdmean}{\upsilon}

\newcommand{\dpout}{\omega}

\newcommand{\data}{D}
\newcommand{\treff}{\tau} 
\newcommand{\covcoef}{\gamma} 
\newcommand{\ncat}{m_{cat}} 
\newcommand{\ncont}{m_{cont}} 
\newcommand{\nuniqr}{q} 
\newcommand{\nbin}{\eta}
\newcommand{\binparam}{\zeta}

\newcommand{\xdisc}{d^*}
\newcommand{\interval}{\mathcal{I}}
\newcommand{\indic}{\mathds{1}}

\newcommand{\mechhist}{M_{hist}}
\newcommand{\mechhybrid}{M_{hybrid}}
\newcommand{\mechdpmod}{M_{GenM}}
\newcommand{\trass}{R}

\newcommand{\absdiff}{|\Delta|}

\newcommand{\cioverlap}{\widetilde{{O}}}

\DeclareMathOperator*{\argmax}{arg\,max} 
\newcommand{\E}{\operatorname{E}} 

\newcommand{\dpmb}[1]{GenM-#1 }
\newcommand{\longdpmb}{GenModel }

%% file: abstract.tex
\Acp{RCT} have become powerful tools for assessing the impact of interventions and policies in many contexts. They are considered the gold standard for causal inference in the biomedical fields and many social sciences. Researchers have published an increasing number of studies that rely on RCTs for at least part of their inference. These studies typically include the response data that has been collected, de-identified, and sometimes protected through traditional disclosure limitation methods. In this paper, we empirically assess the \textcolor{black}{impact of privacy-preserving synthetic data generation methodologies}
on published RCT analyses by leveraging available replication packages (research compendia) in economics and policy analysis. We implement three privacy-preserving algorithms, that use as a base one of the basic differentially private (DP) algorithms, the perturbed histogram, to support the quality of statistical inference. We highlight challenges with the straight use of this algorithm and the stability-based histogram in our setting and described the adjustments needed.  We provide simulation studies and demonstrate that we can replicate the analysis in a published economics article on privacy-protected data under various parameterizations. We find
that relatively straightforward (at a high-level) privacy-preserving methods influenced by DP  techniques allow for inference-valid
protection of published data.
The results have applicability to researchers wishing to share RCT data, especially in the context of \acl{LMIC}, with strong privacy protection. 


%% file: section-introduction.tex

\acresetall

\Acp{RCT}, have become  powerful tools for assessing the impact of interventions and policies in many contexts \citep[e.g., in economics, see the Nobel Prize lecture by ][]{10.1257/aer.110.7.1952}. Today, they are considered the gold standard for inference in the biomedical fields and many social sciences. In economics, their use has expanded rapidly since the 1990s, with studies ranging from small-scale interventions at the individual or village level to national programs. Researchers have published an increasing number of studies that rely on RCTs for at least part of the inference. 
The AEA RCT Registry alone has over 11,500 registrations as of January 2026, with new ones added at a  rate of 1748 per year in 2025. More than 12,700 unique researchers are associated with these registrations \citep{rct-report-2026}.

Alongside the growth of RCTs, the push for research transparency has led journals to require public release of data and code as ``replication packages.'' The \ac{AER}, the flagship journal of the \ac{AEA}, has required analysis data and code since 2004 \citep{bernanke2004}. Some AEA journals launched in 2009 implemented such a policy from creation; see \citet{vilhuber2020} and \citet{vlaeminck2021} for a review of reproducibility practices in economics. 
The broader availability of complete replication packages has enabled re-analyses and meta-analyses that strengthen understanding of both methods and findings. For example, \citet{meager_understanding_2019} re-analyzed multiple RCTs using Bayesian hierarchical analysis (BHA), while \citet{10.1257/aeri.20210236} examined event studies for pre-treatment time trends.
\footnote{It should be noted that \citet{10.1257/aeri.20210236} still had to exclude nearly four times as many papers as they included because data were not readily available.}

A parallel development concerns novel privacy/ confidentiality\footnote{We will use privacy and confidentiality interchangeably to mean data confidentiality.} protection methodologies. While privacy protection has long been a standard part of the toolkit of statistical agencies \citep[e.g.,][]{hundepool2012statistical} and ethical research guidelines, new formal privacy mechanisms --- such as various forms that satisfy \ac{DP}\citep{dwork_calibrating_2006,dwork_calibrating_2016,TheAlgorithmicFoundationsofDPDwork} --- offer transparent and mathematically provable confidentiality.
We briefly describe the basic idea behind DP in the next section.
Nevertheless, the typical (recent) guidance followed by researchers who conduct \acp{RCT} primarily relies on much weaker methods, such as de-identification \citep{department_of_health_and_human_services_methods_2012,kopper_j-pal_2020,dime_-identification_2020}, $k$-anonymity \citep{k-anonymity}, $l$-diversity \citep{MachanavajjhalaGKV06},  and other aggregation-based methods \citep{hundepool2012statistical}. However, most data included within replication packages that allow for exactly reproducing the results in the papers are typically simply de-identified. We are not aware of the application of techniques from \ac{DP} or relaxations of DP in the context of the dissemination of data collected as part of \Acp{RCT}.%
\footnote{Randomized response \citep{RandomizedResponseWarner} is shown to be a (distributed) DP mechanism, used at the collection stage, and has been applied in industry. A recent example explicitly referencing it in data collection within an RCT is \citet{kancharla2021robust}; the earliest (central) DP work is of  \citet{Vu2009}.}
This suggests that much of the current literature on \Acp{RCT} publishes replication packages that contain inadequately protected data. This is particularly concerning in the economic data setting we are exploring, because many of these studies have data from respondents in \acl{LMIC}, where there may be lower legal protections than in Europe or North America.

One of the possible reasons for the absence of strong privacy protection methods in this literature is that no tools are available to non-specialists that would allow for easy but efficient protection \textcolor{black}{consistent with the definition of \acl{DP}}.\footnote{We are aware of the Two Ravens tool \citep{tworavens}, but have not seen usage of it in the space we surveyed. } 
Efficiency here is defined as ``perturbing inference as little as possible compared to the unprotected inference.'' We note that inference even in the ``unprotected'' case is already subject to uncertainty that is often not adequately taken into account \citep{meager_understanding_2019}. This is even more important for the uncertainty and data modifications that are generated through  \ac{SDL}.  \citet{abowd_economic_2015} and \citet{slavkovic_seeman_2022} argue for the need to account for the privacy-preserving noise in analyses. \citet{slavkovic_seeman_2022}, and references therein, discuss a general way to adjust for privacy-preserving noise in addition to other sources of uncertainty.

The present article contributes to the literature on privacy-aware analysis.
Broadly, we aim to contribute on two separate dimensions. 
First, in the context of data collected for \acp{RCT}, we investigate the feasibility of preserving the generic quality of inference obtained using the confidential data even when the same inference is performed using data endowed with privacy protections that are stronger than the simple de-identification usually used.
Second, we do so while maintaining the feasibility of application, here defined as \textcolor{black}{computational feasibility} on commodity hardware used by researchers in these fields (and in particular, by researchers in \acl{LMIC}, where many \acp{RCT} are conducted). We have created an R package\iftoggle{anonymous}{ available on Github (URL redacted)}{, DPrct \citep{DPrct}}, which includes some basic differential privacy algorithms, including the algorithms discussed and proposed in this paper. The package allows researchers to easily implement privacy-preserving methods on their RCT-based data.

Our focus on \acp{RCT} is intentionally narrow.\footnote{A somewhat different approach is taken by \citet{rosenblatt2023b}, who start with frequently-used published datasets and explore the (conceptual) reproducibility of analyses in articles that used such datasets. They, too, focus on the simpler methods and find mixed results.} We believe that exploring the impact of privacy-preserving technologies in the context of \acp{RCT} is useful for several reasons. First, statistical methods are, in general, straightforward: standard linear regression, difference-in-difference methods, and possibly even simple difference in means across treated and untreated populations. These are among the first analysis methods for which adaptations to DP protection have been studied \citep[e.g.,][]{awan2020structure, alabi,slavkovic2021perturbed, barrientos2018, bowen2020synthetic}. If formal privacy-preserving methods cannot be made to work ``out-of-the-box'' and at scale in this context, then it will be much more difficult to argue for broader application. Second, most \acp{RCT} are small-scale, using samples of the overall population, allowing us to avoid computational constraints when algorithms scale with sample size $N$.\footnote{We note that sampling might also allow us to leverage privacy-amplifying methods \citep{balle2018privacy}, though we do not exploit that in this paper. We also note that DP was originally designed with large rather than small samples in mind but the small-scale databases are as equally important in practice.}
Third, \acp{RCT} are often accompanied by pre-analysis plans, with specific hypotheses in mind and with the intent to avoid false discovery. These areas have also been explored within the DP framework    \citep[e.g.,][]{Vu2009, nixon2020, dwork_differentially_2021}). Furthermore, it is already understood in the privacy community that the inherent noisiness of the sampling may affect inference  \citep[e.g., ][]{slavkovic_seeman_2022}. The analogy between adding noise for the purpose of BHA \citep{meager_understanding_2019}, and adding noise for privacy protection may be convenient to improve the acceptance of such methods. 
A similar Bayesian framework can be used to adjust noisy inference due to privacy  \citep[e.g., ][]{seeman2020private}.

Specifically, we explore the impact of privacy-preserving methods in two ways. First, we propose three privacy-preserving mechanisms that use as building blocks some of the basic DP algorithms. Through simulations, we show their appropriateness for the stated goals. We also explain why they are not fully DP,  and identify the underlying data issues that complicate  an easy fully-DP implementation. We then use data and models from a published analysis of an RCT, apply the algorithms to the data prior to reproducing the study's original methods, and verify that inference is not unreasonably affected. In this paper, we use the data from \citet{Reducecrime} [henceforth  ``BJS''] as published in their replication package \citep{ReducecrimePkg}.\footnote{The article was published in the \ac{AER}. One of the authors is the Data Editor of the \ac{AEA} at the time of writing of this article, but the article in question predates his tenure at the \ac{AEA}, and was not subject to his review.} 
%
The protected data is constructed so as to be a drop-in replacement for the data originally published by BJS. Importantly, we explicitly do not change the analysis methods used by the authors, except where necessary to take into account the privacy-preserving mechanism. There are possibly many ways that other researchers might have analyzed the data collected by this (or other) sets of authors. We do not explore those, but note that it is precisely through the availability of replication packages that such differing approaches can be addressed -- through replications.

%% file: section-definitions-problemsetup.tex


Randomized controlled trials are experiments conducted to evaluate the effectiveness of a policy, a drug, or some other intervention. They are particularly popular in biomedical fields and have become increasingly common in policy evaluations in socio-economic contexts. The key feature is that assignment to ``treatment'' is randomized within the eligible population -- whether it means receiving a novel drug or a cash grant. Random assignment can be at the level of individuals, institutions (e.g., schools, hospitals), villages, provinces, etc. Generally, a control group is also chosen at random from the same population, but does not receive the treatment. The popularity and strength of \acp{RCT} comes from the powerful notion of causality, due to the randomization procedure. As long as the randomization procedure actually worked as designed, it is often sufficient to compare means for the two samples (treatment and control) of relevant outcome variables to obtain a valid causal inference as to the effectiveness of the treatment. For the application in this paper, this simplicity also means that methods used to obtain causal inference are (in theory) statistically simple, and thus lend themselves, in theory, more easily to the application of strong privacy-preserving mechanisms.


The key definition of differential privacy, \textbf{DP}, relies on the concept of $\epsilon$-indistinguishability \citep{dwork_calibrating_2006,machanavajjhala2015}. 
An algorithm $M$ satisfies
\textbf{$(\epsilon,\delta)$-differential privacy} (\acl{appDP}) for some $\epsilon,\delta>0$ if for each of its
possible outputs $\dpout$ and for every pair of databases $\data_1 , \data_2$ that differ on the
changing of a single record, \begin{equation}\label{eqn:DPdefn}
P(M(\data_1) = \dpout ) \leq e^\epsilon P(M(\data_2) = \dpout) + \delta.\end{equation} When $\delta=0$, an algorithm $M$ satisfies
\textbf{$\epsilon$-differential privacy} (\acl{pureDP}).
In other words, given the outputs from two databases that differ only in a single record are very ``similar'', it is statistically very difficult (quantified by parameter $\epsilon$, also referred to as a privacy-loss, that is positive but ideally a small value greater than 0) to know if any particular record was included in the database or not. 
Various privacy-preserving methods/mechanisms can satisfy these and related definitions \citep[e.g., see ][]{desfontaines2022sok}. In general, mechanisms inject precisely controlled statistical noise into the data. Randomized response \citep{RandomizedResponseWarner} is an example of \textit{input} noise injection at the collection stage. We will be concerned with \textit{output} noise injection, which involves perturbing data that have already been collected.\footnote{For examples of DP mechanism perturbations, and a much broader taxonomy, see \citet{desfontaines2022sok}.} The noise is typically taken from well-known distributions, such as Laplace and Gaussian (normal).


In what follows, we aim to use terms that are comprehensible to a variety of disciplines. All of our mechanisms will start with \textbf{original} or \textbf{confidential} data $\data$. Parameters estimated using $\data$ are treated as \textbf{confidential}. A mechanism $M(\cdot)$ that transforms $\data$ into \textbf{privacy-preserving} \textcolor{black}{or \textbf{privacy-protected}} output $\dpout$ is said to \textbf{protect} confidential data. In this paper $\dpout$ is schema-equivalent to $\data$, and we will designate the privacy-preserving output as $\widetilde{\data}$. We try to avoid the somewhat ambiguous term ``privatized'' for $\widetilde{\data}$, as many social scientists associate the act of ``making private'' as making something non-public, which is the opposite of the intent of the application of $M(\cdot)$ here. In the context of our analysis of BJS, we will use the term \textbf{original}, because the authors have already transformed the \textbf{confidential} data into the \textcolor{black}{downloadable} and published data in \citet{ReducecrimePkg} using a mechanism $M_{deidentify}$. We will treat those data as if they were ``confidential'' for the purpose of the experiments in this article. Finally, because we are in the context of \acp{RCT}, we also have randomized assignment (possibly stratified) to treatments (including the control group), which we will denote as $\trass$, an (usually public) algorithm that conditions on the data available to the researchers. 

We pursue three goals motivated by a setting where the typical researcher wants to: 

\begin{description}
    \item[Goal 1] publish a sufficiently precise privacy-protected inference of the effect of the treatment on the treated from the model, given the data $\data$; 
    \item[Goal 2] release (publish) the privacy-protected database $\widetilde{\data}$ so that others can scrutinize the analysis, while preserving the confidentiality of the respondents whose data are contained in the database;
    \item[Goal 3] Apply $M(\cdot)$ in a way that is computationally tractable on commodity computer hardware, i.e., a researcher laptop or at most a reasonably dimensioned server. 
\end{description}




We focus on a particular key table in BJS, Table 2. It is the result of several independent regressions, each with a separate dependent (response) variable of interest, measured in the short term (Panel a) and long term (Panel b), controlling for a set of assignment indicators for the various treatments, other covariates, and stratifiers.%
\footnote{\citet{Reducecrime} stratify the treatment across a few subgroups, but for expositional ease, we mostly ignore that aspect. In the empirical analysis, we implement the authors' stratified random assignment as per their methodological description.}
These are ``intent-to-treat'' (ITT) regressions. They are typical of \acp{RCT}, where the experimenter is interested in determining whether a particular treatment has any effect on a response variable when the treatment is applied to an entity, individual, or treatment unit. The experimental treatment is randomly assigned to the treatment units according to some chosen experimental design,  and  response variables are recorded at various points after the treatment. As is typical for this type of analysis, BJS has both discrete and continuous covariates. The response variables are also a mix of continuous and discrete outcomes. 

The various characteristics and attributes of the treatment unit (in this case a person participating in the study),  including membership in different strata and the actual outcomes, may all be sensitive. In BJS, participation in therapy, receipt of cash (two treatments), and self-reported criminal activity (covariates) are all quite sensitive attributes and may pose privacy concerns. While the inclusion of many covariates in the regression analysis can improve the estimate of the treatment effect, these covariates also pose a privacy concern for the treatment units participating in the study. An attacker, who may access the database provided as part of the replication package, may be able to reidentify individuals, and learn new, sensitive information about them.  This would constitute a confidentiality violation for the treatment units.

Consider a confidential dataset $\data = \left [ \bfy, \bT, \bX\right ]$
with $n$ rows and $1+t+p$ columns, where $n$ is the total number of treatment units, a single response variables $\bfy$, $t$  mutually exclusive treatment assignments $\bT$ (the control group being one of them), and $p$ covariates $\bX$\footnote{Additionally, there could be $b$ blocking variables $\bV$, in which case the confidential dataset is $\data = \left [ \bfy, \bT, \bX, \bV\right ]$ with $n$ rows and $1+t+p+b$ columns. The corresponding regression model is $\bfy=\bT\btreat+\bX\bcov+\bV\bblock+\be$ where $\bV$ is  $n\times b$ matrix and $e_i\overset{i.i.d.}{\sim}N(0,\sigma^2)$ for $i=1,\dots,n$. In this case, the randomized treatment assignment may depend on the blocking variables $\bV$.}. $\bT$ is assigned based on some \textcolor{black}{randomized treatment assignment} algorithm $\trass$.
%

The regression model of interest, in the absence of blocking variables
, is given by
\begin{equation}\label{regression model no block}
Y_i = \sum_{k=1}^{t}\treff_{k} T_{k,i} + \sum_{l=1}^{p}\gamma_{l}X_{l,i} +  e_{i}, \quad i=1,\dots,n ,
\end{equation}
or
\begin{equation}\label{regression model vector no block}
\bfy = \bT\btreat  + \bX\bcov +  \mathbf{e},
\end{equation}
where $\bT$ is \textcolor{black}{an} $n \times t$ matrix of exhaustive assignment to treatment arms, $\btreat$ is \textcolor{black}{a} $t$-dimensional vector of \textcolor{black}{(fixed)} treatment effects, $\bX$ is $n \times p$ matrix of covariates, $\gamma$ is a $p$-dimensional vector of regression coefficients \textcolor{black}{(fixed effects) corresponding to the explanatory covariates}, and $\mathbf{e}$ is \textcolor{black}{an} $n$-dimensional error term with $e_{i} \stackrel{i.i.d}{\sim} N(0,\sigma^{2})$ for $i=1,\dots,n$\footnote{In BJS, there are many response variables, \textcolor{black}{but each regression is treated independently, and can be assumed to be based on a different dataset with the first column being the different response variables but with the same $t+p+b$ columns for the treatment assignment, explanatory covariates and blocking variables (if any)}.}
. 




%% file: section-synthetic-data.tex
In the context of \Acp{RCT}, the parameter(s) of interest to the experimenters are the treatment effects $\btreat$. From the experimenter's perspective, statistical utility is preserved if the inference concerning the treatment effects $\btreat$ (such as point estimates, uncertainty quantification of point estimates, interval estimates and $p$-values for significance tests) is minimally affected by the mechanism used to perserve privacy of the data and/or the relevant statistics in order to protect privacy.

We propose three methods to produce privacy-preserving data, $\widetilde{\data}$ designed specifically to preserve the utility of the inference on the treatment effects. The first method samples a perturbed multivariate histogram and is model-agnostic (see, Algorithm \ref{algo:perturbMVhist}). The second two methods are model-informed, but use the sampled perturb histogram to protect the model covariates. Of the two model-informed methods, one provides additional confidentiality protection to the original data by infusing noise to  the estimated model parameters. All three methods provide much stronger privacy protections than de-identification and are built from techniques and ideas from differential privacy (DP). While we denote privacy budgets using $\epsilon$ and $\delta$, the methods themselves do not satisfy strict ($\epsilon,\delta$)-DP definitions. Like in DP, increasing $\epsilon$ or $\delta$ will decrease the privacy protections and increase utility of the privacy-preserving data.

\subsection{Multivariate Histogram Mechanism}\label{sec:mvhistogram}

Sampling from a perturbed multivariate histogram is one of the earliest proposed mechanisms \citep[e.g., see ][]{dwork_calibrating_2006, wasserman_statistical_2010} that satisfies $\epsilon$-DP. The perturbed multivariate histogram mechanism can be extended to  a \textit{stability-based} histogram mechanism to satisfy ($\epsilon,\delta$)-DP  by introducing a probability threshold $c$ \citep{bun16_multilearner,karwa2017finite}. The probability threshold allows to only infuse noise to the observed combinations of column values rather than all possible combinations by reducing the probability that low frequency combinations are sampled. Our multivariate histogram method is based on these traditional mechanisms but does not satisfy strict $\epsilon$-DP or ($\epsilon,\delta$)-DP definitions due to how it discretizes the continuous variables and how it deals with possible combinations of discretized covariates that are not observed in the sample. 
The histogram counts are protected using the Laplace mechanism and 
when necessary, we choose the precision of binning\footnote{The precision of binning prescribes how many bins are created for each continuous variable dimension, as a function of the sample size (for e.g., if $\binparam = \frac{2}{3}$ and $p=2$, with both variables being continuous, the total number of bins in the multivariate histogram is $\left(n^{\frac{2}{3}}\right)^2 = n^{\frac{4}{3}}$).}, denoted by $\binparam$, and discretize continuous variables. It is important to notice the bounds on the discretized bins are informed by the original confidential data. This is the first reason why the Multivariate Histogram Method in Algorithm \ref{algo:perturbMVhist} is not strictly differentially private. Some confidentiality protection comes from adding uniform noise to discretized column values to make them continuous again at the end of sampling from the perturbed histogram.  

The deviation from DP has to do with our threshold $c$ and the unique valued rows $\uniqr_1,\dots, \uniqr_\nuniqr$. Under $\epsilon$-DP, the unique values rows would have to include all possible combinations of the categories and discretized levels. If we had $\ncat=7$ binary categorical variables and $\ncont=7$ continuous variables which were discretized into $\nbin=10$ bins each, then there would be $\nuniqr=2^{\ncat}\times10^{\ncont}=$1.28 billion unique valued rows. However, in Windows the limit for a variable in the 32-bit build of \texttt{R} is 4Gb \citep{Rbase}. Using the \texttt{VGAM} package, generating even $900$ million Laplace random variables to sanitize proportions exceeds the 4.0Gb limit \citep{VGAM}. Since data from \Acp{RCT} studies can have a large number of covariates, we only use combinations of discretized covariates observed in our sample. This would not be a problem for the stability-based histogram which satisfies ($\epsilon,\delta$)-DP, except \Acp{RCT} data often has a small number of observations with its large number of covariates. This data structure can lead to each combination of discretized and categorical covariates observed in the sample to only be observed a once or twice (i.e., $\nuniqr$ is close to $n$). Often, this results in no privacy-preserving proportions being above the threshold $c$.
In the case where no privacy-preserving proportions are initially above the threshold, we set the threshold to $0$.

\SetKwComment{Comment}{/* }{ */}
\begin{algorithm}[hbt!]
\caption{\textit{(MV Histogram Method)} Using a discretized multivariate histogram, we sanitize a $n\times (\ncat+\ncont)$ data matrix, $\data$, where the first $\ncont$ columns are continuous and the next $\ncat$ columns are categorical. Columns are denoted $\bx_{\cdot j}=(x_{1j} \cdots x_{nj})^T$ for $j=1,\ldots,(\ncat+\ncont)$ where $x_{ij}$ is the element in row $i$ and column $j$. Similarly, rows are $\bx_{i\cdot}$. The indicator function is denoted $\indic$ and $\max(\bx_{\cdot j})$ is the maximum element of $\bx_{\cdot j}$.}
\label{algo:perturbMVhist}
\KwData{$\data=[\bd_{\cdot 1}~\bd_{\cdot 2}~\cdots \bd_{\cdot\ncont} \bd_{\cdot (1+\ncont)}\cdots \bd_{\cdot (\ncont+\ncat)}]$ is $n\times (\ncat+\ncont)$ dataset.}
\KwIn{$\epsilon>0$, $\delta\geq 0$, $\binparam\in (0,1)$}
$\nbin\gets n^\binparam$\;
\ForEach{$j=1,\dots,\ncont$}{
\If{count of unique elements of $\bd_{\cdot j}>\nbin$}{
$\interval_k\gets [a_{k-1},a_{k})$ for $k=1,\dots,\nbin-1$ and $\interval_\nbin\gets [a_{k-1},a_k]$ where $a_k=\frac{\max(\bd_{\cdot j})+(\nbin-k)\min(\bd_{\cdot j})}{\nbin}$ for $k=1,\dots,\nbin$ and $a_0=\min(\bd_{\cdot j})$\;
$\bxdisc_{\cdot j}\gets (\xdisc_{1j}\cdots \xdisc_{mj})^T$ where  $\xdisc_{i j}=\sum_{k=1}^\nbin \indic\{x_{ij}\in \interval_k\}\frac{a_{k-1}-a_{k}}{2}$
}\Else{
$\bxdisc_{\cdot j}\gets \bd_{\cdot j}$
}
}
$\data^*\gets[\bxdisc_{\cdot 1} \cdots \bxdisc_{\cdot \ncont}~\bd_{\cdot (\ncont+1)}\cdots \bd_{\cdot(\ncont+\ncat)}]$\;
$\uniqr_1,\ldots,\uniqr_\nuniqr$ are the unique valued rows of $\data^*$\;
$\hat{p}_\ell=\frac{1}{n}\sum_{i=1}^n\indic\{\bd_{i\cdot}=\uniqr_\ell\}$ \Comment*[r]{unique row proportion}
$\widetilde{p_\ell}\gets (\hat{p}_\ell+Z_\ell)$ for $\ell=1,\ldots,\nuniqr$ where $Z_\ell\overset{iid}{\sim}\operatorname{Laplace}\left(0,\frac{1}{n\epsilon}\right)$\;
\If{$\delta>2/\nuniqr$}{
$c=2\frac{\log(2/\delta)}{n\epsilon }+\frac{1}{n}$ \Comment*[r]{approximate DP threshold}
\If{$\sum_{\ell=1}^{\nuniqr}\indic\{\widetilde{p_\ell}>c\}=0$}{
$c=0$ \Comment*[r]{special case adjustment}
}
} \Else{
$c=0$
}
$\widetilde{p_\ell}\gets\widetilde{p_{\ell}}\indic\{\widetilde{p_\ell}>c\}$\;
$\widetilde{p_\ell}^*\gets \widetilde{p_\ell}/(\sum_{k=1}^{\nuniqr}\widetilde{p_k})$ \Comment*[r]{normalized privacy-preserving row proportions}
Draw $n$ rows $\widetilde{\bd_{1\cdot}},\ldots,\widetilde{\bd_{n\cdot}}$ such that $\operatorname{P}(\widetilde{\bd_{i\cdot}}=\uniqr_\ell)=\widetilde{p_\ell}^*$ for $\ell=1,\ldots,\nuniqr$ and $i=1,\ldots,n$.\;
\ForEach{row $i=1,\dots n$}{
\ForEach{$j=1,\dots,\ncont$}{
Find $k$ such that $d_{ij}\in \interval_k$ \Comment*[r]{the bin the value is from}
$d_{ij}\gets U_{k}$ where $U_k\sim\operatorname{Uniform}(a_{k-1},a_k)$\Comment*[r]{undo discretization}
}
}
\KwResult{$\widetilde{\data}\gets \begin{bmatrix}
    \widetilde{\bd_{1\cdot}}\\ \vdots \\ \widetilde{\bd_{n\cdot}}
\end{bmatrix}$}
\end{algorithm}

The perturbed multivariate histogram method does not depend on the model used in the study. It can be generalized to any $\data$ and does not matter which columns are response, treatment variables, covariates, etc. It does depend on what columns are continuous and which are discrete or categorical. To generalize, in Algorithm \ref{algo:perturbMVhist} we allow $\data$ to be a $n\times (\ncat+\ncont)$ database, where the first $\ncont$ columns are continuous and the rest are categorical. We denote the element in row $i$ and column $j$ as $d_{ij}$. The row vectors are $\bd_{i\cdot}$ for $i=1,\ldots,n$ and the column vectors are $\bd_{\cdot j}$ for $j=1,\ldots,(\ncat+\ncont)$. Lines 3 to 7 in Algorithm \ref{algo:perturbMVhist} discretize the continuous variables. The multivariate histogram is created by lines 9 and 10. The histogram is perturbed in line 11 using Laplace noise. The privacy-preserving proportion becomes $0$ if it is below a threshold in lines 12 to 18. Then the unique rows are sampled in line 20 using normalized privacy-preserving proportions. Finally, the continuous covariates are converted back to continuous values using uniform noise in lines 21 to 24. We denote the application of Algorithm \ref{algo:perturbMVhist} to $\data$ with $\epsilon$ and $\binparam$ parameters as $\mechhist(\data; \epsilon, \delta, \binparam)$.



\subsection{Model-informed Methods}

\begin{figure}[!htb]
    \centering
    \includegraphics[width=0.95\linewidth]{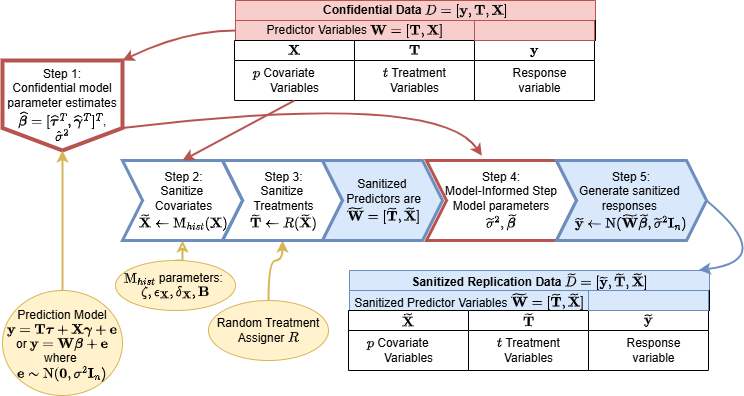}
    \caption{Both model-informed methods of generating synthetic data follow the same five overall steps.}
    \label{fig:modelinformed-flow}
\end{figure}

Within the proposed model-informed methods, the infused-noise can be targeted to prioritize accuracy in  the model summary fitted to the privacy-preserving data. It focuses on maintaining the relationship between the variables that is described by the model.

We propose two model-informed algorithms for sanitizing the data $\data$. Both methods use the MV Histogram Method (Algorithm \ref{algo:perturbMVhist}) and post-processing to sanitize the explanatory variables of a model. They differ on how the response variable is generated, but have the same overall flow (Figure \ref{fig:modelinformed-flow}). The first method is based on approximate DP, but uses the model parameters that are estimated using the confidential data, $\data$. This method, called the \textit{Hybrid Method} (Algorithm \ref{algo:hybrid}), allows flexibility in what models are used but does not satisfy the strict approximate DP definition. The second method, called the \textit{\longdpmb Method} (Algorithm \ref{algo:fullmodelbased}), which preserves the confidentiality of model parameters, but relies heavily on the normality assumptions of multiple linear regression models with Gaussian residuals, such as \eqref{regression model vector no block}.

\subsubsection{Hybrid Method}\label{sec:hybrid}
The Hybrid Method uses the estimated model parameters ($\hat{\btreat},~\hat{\bcov},$ and $\hat{\sigma^2}$) from a prediction model like \eqref{regression model no block} that depends on standard inferential procedures performed on the confidential data, $\data$. 
The parameters of the model ( $\btreat,\bcov,$ and $\sigma^2$) are estimated using a standard inferential procedure, such as ordinary least squares, and the original confidential data \citep{Dunn_Smyth_2018}.  Then the covariate variables are protected using Algorithm \ref{algo:perturbMVhist}, $\sanbX\gets \mechhist(\bX; \epsilon, \delta, \binparam)$. Using the same treatment assignment algorithm as the confidential data ($\trass(\cdot)$) we  generate new treatment indicators based on the $\sanbX$, $\sanbT\gets \trass(\sanbX)$. The generation of $\sanbT$ is a post-processing procedure, and thus does not require privacy budget allocation. Then, $\bfy$ is inputted from the model. 
Finally, the privacy-preserving inference concerning the treatment effects $\btreat$ is obtained using a post-processing procedure, which involves performing standard  inferential procedures (the same procedures that one may use when privacy preservation is not a concern) on the privacy-preserving replication data [$\widetilde{\bfy}$,  $\sanbT$, $\sanbX$]. By using the parameter estimates obtained from the confidential data $\data$ in this algorithm, it is expected that inference validity is preserved by the sanitization procedure. The released $\widetilde{\btreat}$ is nevertheless perturbed and therefore conventionally protected because it is based on $\epsilon$-DP-protected $\sanbX$ and model-informed imputation (as outlined below), but is not formally DP-protected. Since the $\widetilde{\bfy}$ is imputed using the estimated model coefficients which are estimated with the confidential data, these values do not satisfy approximate DP. 

Algorithm \ref{algo:hybrid} uses the ($\epsilon,\delta$)-DP protected treatment indicators $\sanbT$ and covariates $\sanbX$, the confidential parameter estimates $\hat{\btreat}$, and a suitable model for the privacy-protected response variable $\widetilde{\bfy}$. The confidential $\hat{\btreat}$ are not published; rather privacy-protected $\widetilde{\data}(\epsilon,\binparam) = \left [ \widetilde{\bfy}, \sanbT, \sanbX \right ]$ is published as part of the replication package, and $\widetilde{\btreat}$, its associated standard errors and inferential statistics based on \eqref{regression model no block}, are estimated using $\widetilde{\data}$, and published as part of the relevant article. It is possible that information about the confidential estimated effects and thus about the confidential data could leak through the release of $\widetilde{\data}$. However, this risk is minimal compared to the risk from publishing the estimated effects derived from the confidential data. 

\begin{algorithm}
    \caption{\textit{(Hybrid Method)} The Hybrid Method does not satisfy $\epsilon$-DP, but it based on the privacy-preserving covariates and treatment matrices, $[\sanbT~\sanbX]$ that are generated with Algorithm \ref{algo:perturbMVhist} which satisfies $\epsilon$-DP. The confidential estimated model parameters ($\hat{\btreat},\hat{\bcov},$ and $\hat{\sigma^2}$) are not released. Instead, we can release $\widetilde{\btreat}$,$\widetilde{\bcov}$, and $\widetilde{\sigma^2}$, which are the estimated from fitting the model $\widetilde{\bfy}=\btreat\sanbT+\bcov\sanbX+N(0, \sigma^2)$.}\label{algo:hybrid}
    \KwData{$\data=[\bfy ~\bT~\bX]$}
\KwIn{$\epsilon>0$, $\binparam\in (0,1)$, $\trass(\cdot)$}
Estimate $\hat{\bcov}$, $\hat{\btreat}$, $\hat{\sigma^2}$ from $\bfy=\bT\btreat+\bX\bcov+\mathbf{e}$ where $e_i\overset{iid}{\sim}N(0,\sigma^2)$\;
$\sanbX\gets M_{hist}(\bX;\epsilon,\binparam)$\;
$\sanbT\gets \trass(\sanbX)$\;
$\widetilde{\bfy}\gets \hat{\btreat}\sanbT+\hat{\bcov}\sanbX+N(0, \hat{\sigma}^2)$\;
\KwResult{$\widetilde{\data}=[\widetilde{\bfy}~\sanbT~\sanbX]$}
\end{algorithm}

We also consider a model-informed method that satisfies approximate DP to sanitize $\data$ when the regression model is of the form \eqref{regression model no block} or similar models with a normally distributed error term. We take advantage of the fact that the estimated coefficients and residuals are generated from a Gaussian distribution, and use two algorithms proposed by \cite{karwa2017finite}. We refer to these as the DP Variance Algorithm and DP Range Algorithm which are detailed in Appendix \ref{sec:karwa-algos}. 

The privacy-preserving dataset outputted by the Hybrid method is denoted $\widetilde{\data}\gets \mechhybrid(\data; \epsilon, \delta, \binparam, \trass).$

\subsubsection{\longdpmb Method}\label{sec:dpmodelbased}
\begin{figure}[!htb]
    \centering
    \includegraphics[width=\linewidth]{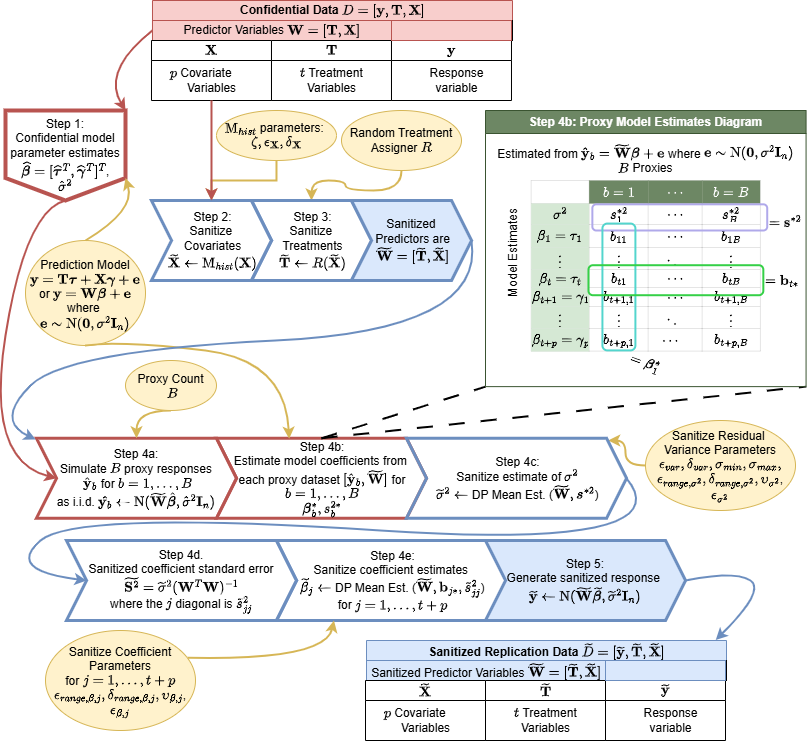}
    \caption{\longdpmb from Algorithm \ref{algo:fullmodelbased} is visualized above. Yellow represents parameters and model inputs. Red indicates the output of a step contains confidential information. Blue indicates the output satisfies approximate DP if the multivariate histogram satisfies approximate DP.}
    \label{fig:dpmb-flow}
\end{figure}

We also propose a model-informed method, \textit{\longdpmb}, which infuses noise to the model parameters. The algorithm is detailed in Algorithm \ref{algo:fullmodelbased} and visualized in Figure \ref{fig:dpmb-flow}. The process of infusing noise to the model parameters in this method would satisfy approximate DP if the privacy-preserving covariates where generated in a way that satisfied approximate that is ($\epsilon,\delta$)-DP, unlike our adaptation of the perturbed multivariate histogram in Algorithm \ref{algo:perturbMVhist}. The \longdpmb method requires the generating model to have Gaussian distributed error terms. Additionally, the number of rows in the data, $n$, needs to be large enough to use the central limit theorem such that the sample mean of $n$ random variables from a $\chi^2$ distribution with 1 degree of freedom is approximately Gaussian. When comparing results of datasets protected by the \longdpmb algorithm with different privacy budgets in Sections \ref{sec:evaluating} and \ref{sec:real-world}, we abbreviate the algorithm as \dpmb{} followed by the privacy budget (e.g., \dpmb{1}). 

The first step of the \longdpmb algorithm is to sanitize the covariate dataset using $\epsilon_{\bX},\delta_{\bX}$ privacy parameters with the MV Histogram algorithm (Algorithm \ref{algo:perturbMVhist}). Then privacy-preserving treatment levels are assigned using the randomized treatment assignment algorithm, $\trass$. Then we sanitize the estimates of the model parameters $\widetilde{\btreat},~\widetilde{\bcov},$ and $\widetilde{\sigma^2}$. With the privacy-preserving parameter estimates, covariate data, and treatment assignments, we impute $\bfy$.
The key idea behind sanitizing the model parameters utilizes two important statistical properties of inference in a regression model. We first construct $B$ independent proxies of the response variable $\bfy$ (denoted by $\hat{\bfy}_{b}$, $b=1,\dots,B$) using the prediction model with the privacy-preserving predictor variables
\begin{equation}\label{Prediction model Algorithm 3 no block with san predictors}
\hat{\bfy}_{b}= \sanbT\hat{\btreat} + \sanbX\hat{\bcov} + \mathbf{e}_{b},\end{equation}
where $\mathbf{e}_{b} \overset{i.i.d}{\sim} N(0,\hat{\sigma}^2 I_n )$ for $b=1,\dots,B$. We then perform the regression of these $B$ proxy responses on the privacy-preserving data $\sanbU = [\sanbT,\sanbX]$ separately for each $b=1,\dots,B$ to obtain $B$ independently and identically distributed sets of estimates. Denote the coefficient estimates as $\bbeta^{*}_{b}=[\btreat^{*}_{b},\bcov^{*}_{b}]$ for $b=1,\dots,B$ where $\bbeta^{*}_b$ is a $t+p$-dimensional vector stacking the $t$-dimensional $\btreat^{*}_b$ and $p$-dimensional $\bcov^{*}_b$. Similarly, denote the $\hat{\sigma^2}$ estimates as $\bs^{*2}=[\mseval_1,\dots, \mseval_B]$.

\begin{algorithm}[!hbt]
\caption{Using Iterations, we generate privacy-preserving $\widetilde{\mathbf{y}}$ from the regression model $\mathbf{y}=\bX \bbeta+N(0,\sigma^2I)$. The using the properties of linear regression coefficients and the central limit theorem on the mean squared error (MSE) we get privacy-preserving estimates of all model parameters to generate a synthetic response variable from.}
\KwData{$\data=[\bfy,\bT,\bX]$ the original dataset}\label{algo:fullmodelbased}
\KwIn{
$B$ \Comment*[r]{number of iterations}\\
$\sigma_{min},\sigma_{max}$ \Comment*[r]{bounds on $\hat{\sigma^2}$}\\
$\bdmean_{\sigma^2}$,$\bdmean_{\beta,j}$ for $j=1,\dots,p+t$ \Comment*[r]{bounds s.t. $|\E(\beta_j)|<\bdmean_{\beta,j}$}  \\
$\alpha_{range,\sigma^2}$, $\alpha_{range,\beta,j}$ for $j=1,\dots,p+t$ \Comment*[r]{for DP range algorithm}  \\
$\epsilon_{range,\beta,j},\delta_{range,\beta,j}$ and $\epsilon_{\beta,j}$ for $j=1,\dots,t+p$\Comment*[r]{privacy parameters}  \\
$\epsilon_{range,\sigma^2},\delta_{range,\sigma^2}$, $\epsilon_{var},\delta_{var}$, and $\epsilon_{\sigma^2}$ \Comment*[r]{privacy parameters}\\
$\binparam$, $\epsilon_{\bX}$, $\delta_{\bX}$ \Comment*[r]{parameters for $\mechhist$}
$\trass$ \Comment*[r]{treatment assignment algorithm}
}
Estimate $\hat{\bbeta}=(\hat{\btreat}^T,\hat{\bcov}^T)^T$ and $\hat{\sigma^2}$ from $\bfy=\bT\btreat+\bX\bcov+\mathbf{e}$ where $\mathbf{e}\sim N(\mathbf{0},\sigma^2I_n)$\;
$\sanbX\gets \mechhist(\bX; \epsilon_{\bX},\delta_{\bX}, \binparam)$\;
$\sanbT\gets \trass(\sanbX)$\;
$\sanbU=[\sanbT ~\sanbX]$\;
\ForEach{$b=1,\dots,B$}{
$\bfy_b\gets \sanbX\hat{\bbeta}+\hat{\sigma}\bZ_b$ where $\bZ_b\overset{iid}{\sim}N(0,I_n)$\Comment*[r]{generate proxy responses};
$\bbeta^*_b\gets (\sanbU^T\sanbU)^{-1}\sanbU^T\bfy_b$
\Comment*[r]{fit regression model};
$\mseval_b\gets \frac{1}{n}\operatorname{sum}(\bfy_b-\sanbU\bbeta^*_b$)\Comment*[r]{MSE of model}
}

$\bs^{*2}=(\mseval_{1},\dots \mseval_{B})^T$ is the vector of $B$ estimates of $\sigma^2$\;
$v_{0}\gets $DP Variance Algorithm($\bs^{*2}$
 $\epsilon_{var}$, $\delta_{var}$, $\sigma_{min}^2$, $\sigma_{max}^2$)\;
 $d_{0}\gets $DP Range Algorithm($\bs^{*2}$,
$\epsilon_{range,\sigma^2}$, $\delta_{range,\sigma^2}$, $\bdmean_{\sigma^2}$, $\sqrt{v_0}$)\;
$\widetilde{\sigma^2}\gets \left(\frac{1}{B}\sum_{\ell=1}^B \mseval_{\ell}\right)+Z$ where $Z\sim$Laplace($0,d_{0}/\epsilon_{\sigma^2}$)\;
$\widetilde{\bS}\gets \frac{n\widetilde{\sigma}^2}{n-p}(\sanbU^T\sanbU)^{-1}$ with elements denoted as $\widetilde{s_{ij}}$\;
\ForEach{$j=1,\dots,p+t$}{
$\bb_{j,*}=(b_{j,1},\dots b_{j,B})^T$ is the vector of $B$ estimates of $\beta_j$ coefficient\;
$d_{j}\gets $DP Range Algorithm($\bb_{*,j}$, $\epsilon_{range,\beta,j}$, $\delta_{range,\beta,j}$, $\bdmean_{\beta,j}$, $\widetilde{s_{j,j}}$)\;
$\widetilde{\beta_j}\gets \left(\frac{1}{B}\sum_{\ell=1}^Bb_{j,\ell}\right)+Z_j$ where $Z_j\overset{iid}{\sim}$Laplace($0,d_{j}/\epsilon_{\beta,j}$)\;
}
$\widetilde{\bbeta}\gets (\widetilde{\beta_1},\dots,\widetilde{\beta_{p+t}})^T$\;
$\widetilde{\bfy}\gets \sanbU\widetilde{\bbeta} + N(0,\widetilde{\sigma^2}I_n)$\;
\KwResult{$\widetilde{\data}=[\widetilde{\bfy}~\sanbT~\sanbX]$}
\end{algorithm}

The first key  statistical property that we leverage is that $\beta^{*}_{b}$ follows a multivariate Gaussian distribution with mean $\hat{\bbeta} = [\hat{\btreat},\hat{\bcov}]$  and covariance matrix $\hat{\sigma}^2 (\sanbU^{T}\sanbU)^{-1}$. Thus, $\bbeta^{*}_{1},\dots,\bbeta^{*}_{B}$ can be considered as an i.i.d sample from a multivariate Gaussian distribution\footnote{If the model includes blocking variables, $\bV$, and is of the form $\bfy=\bT\btreat+\bX\bcov+\bV\bblock+\boldsymbol{\epsilon}$, then the data consisting of $[\bX~\bV]$ is protected with the MV Histogram algorithm. The treatments are randomly assigned. The privacy-preserving predictors are $\sanbU=[\sanbT,~\sanbX,~\widetilde{\bV}]$. Additionally, $\bbeta^{*}_{b}=[\btreat^{*}_{b},\bcov^{*}_{b},\bblock^{*}_{b}]$ are used to sanitize the model coefficients.}.
In particular, 
the $\btreat^{*}_{b,j}$'s constitute an i.i.d sample of size $B$ from an univariate Gaussian distribution with mean $\hat{\btreat}_{j}$ and variance $\hat{\sigma}^2 ((\sanbU^{T}\sanbU)^{-1})_{jj}$ for $j=1,\dots,t+p$. For $j=1,\dots,t$, these are the point estimate and the estimated variance of the treatment effect $\btreat_{j}$. For $j=t+1,\dots,t+p$, these are the estimated covariate coefficient, $\widehat{\covcoef_{j-t}}$, and its estimated variance. We use two \acl{appDP} algorithms by \citet{karwa2017finite} 
to sanitize the point estimates of the estimated regression coefficients and the residual standard errors. Additionally, the residual mean squared error is $\mseval_b\sim\frac{\sigma^2}{n}\sum_{i=1}^nZ_i$ where $Z_i\overset{i.i.d.}{\sim} \chi^2(1)$ conditional on $\sanbU$. For $n$ large enough, $\mseval_b$ is approximately Gaussian with mean $\sigma^2$ and variance $2\sigma^2/n$. The residual mean squared errors are protected in a similar fashion to the coefficients. First we use the DP Variance algorithm proposed by \citet{karwa2017finite} (Algorithm \ref{algo:dpvar} in Appendix) with privacy parameters $\epsilon_{var},\delta_{var}$ and standard deviation bound parameters $\sigma_{min}$ and $\sigma_{max}$. Then we use the DP Range algorithm (Algorithm \ref{algo:dprange} in Appendix) with privacy parameters $\epsilon_{var},\delta_{var}$ and mean bound parameter $\bdmean_{\sigma^2}$. Then Laplace noise is added to the mean of the $B$ MSE values with privacy parameter $\epsilon_{\sigma^2}$. The privacy-preserving estimate of the residual variance is denoted $\widetilde{\sigma^2}$. This can be used to estimate the coefficient covariance matrix as $\widetilde{\sigma}^2 (\sanbU^{T}\sanbU)^{-1}$ \citep{Dunn_Smyth_2018}. A privacy-preserving mean of of the sample $\btreat^{*}_{b,j}$ for $b=1,\dots, B$ would make an appropriate privacy-preserving estimate of $\beta_{j}$. For the $k$ treatment effect, we get a privacy-preserving range of the mean based on the sample using the $k$th diagonal value of $\widetilde{\sigma}^2 (\sanbU^{T}\sanbU)^{-1}$ as the estimated standard deviation in the DP Range algorithm proposed by \citet{karwa2017finite} (Algorithm \ref{algo:dprange} in Appendix) with privacy parameters $\epsilon_{range,\treff, k},\delta_{range,\treff, k}$, range confidence parameter $\alpha_{range,\treff, k}$, and mean bound parameter $\bdmean_{\treff,k}$. Then we generate Laplace noise with scale parameter equal to the output of DP Range over $\epsilon_{\treff,k}$, and add it to the sample mean.  We repeat this with all the covariate coefficients $\ell=1,\dots,p$ (reindexing $\ell=j-t$) with privacy parameters $\epsilon_{range,\covcoef, \ell},\delta_{range,\covcoef, \ell}, \epsilon_{\covcoef, \ell}$ and range algorithm parameters $\alpha_{range,\covcoef,\ell}$ and $\bdmean_{\covcoef,\ell}$. More generally, several parameters can be expressed in terms of $\beta$ such as $\epsilon_{range,\beta,j}=\epsilon_{range,\treff,j}$ when $j\leq t$ and $\epsilon_{range,\beta,j}=\epsilon_{range,\covcoef,j-t}$ otherwise. In the end, the \longdpmb method has an overall privacy budget of $(\epsilon,\delta)$ where $\epsilon=\epsilon_{\bX}+\epsilon_{var}+\sum_{j=1}^{t+p}\epsilon_{range,\beta,j}+\epsilon_{\beta,j}$ and $\delta=\delta_{\bX}+\delta_{var}+\sum_{j=1}^{t+p}\delta_{range,\beta,j}$.  
We find it useful to denote the privacy budget for generating the response separate from the budget for sanitizing the covariates, $\epsilon_{\bfy}=\epsilon-\epsilon_{\bX}$ and $\delta_{\bfy}=\delta-\delta_{\bX}$. The application of the \longdpmb method on a dataset $\data$ is denoted $\mechdpmod(\data;~ \boldsymbol{\epsilon},~\boldsymbol{\delta},~B,~ \sigma_{min},~\sigma_{max},~\boldsymbol{\bdmean},~\boldsymbol{\alpha}_{range},~\binparam,~\trass)$. Note that $B$, the number of generated proxy response variables, is a user choice and can be made as large as required.\footnote{$B\geq 5000$ has been suggested, in personal communication with one of the authors of \cite{karwa2017finite}, for the validity of the algorithms to be subsequently applied.}

The $\widetilde{\sigma^2}$ produced in line 17 of Algorithm \ref{algo:fullmodelbased} is accurate to $\sigma^2$ up to a factor according to \citet{karwa2017finite}. It may add too much noise to generate $\widetilde{\bfy}$ and utility will be lost. We discuss several potential variants. All the variants produce a $\widetilde{\bfy}$ that is based on some elements that are generated with methods that satisfy approximate DP. However, the variant methods use $\hat{\sigma^2}$ from the analysis with the confidential data. These would be a relaxation of the confidentiality protection. 
One variant uses the estimated $\hat{\sigma^2}$ that was fitted on the confidential data  rather than $\widetilde{\sigma^2}$ in lines 24 to generate $\widetilde{\bfy}$. Similarly, $\hat{\sigma^2}$ could be used in line 17 and line 18. This would eliminate the need for line 15 and the inputs, $\epsilon_{var},\delta_{var},\sigma_{min},$ and $\sigma_{max}$.

The \longdpmb method could be extended to a generalized linear model using asymptotic properties of coefficient estimates. Given a generalized linear model, $\hat{\btreat}$ are asymptotically (as sample size $n$ increases) normally distributed. In the current work, we only consider linear regression model.  




%% file: section-evaluation.tex
To compare the three proposed methods (Algorithms \ref{algo:perturbMVhist}, \ref{algo:hybrid}, and \ref{algo:fullmodelbased}), we will use a simulated dataset, $\data_{sim}$. In section \ref{sec:makesimdata}, we discuss how the simulated dataset is generated. We will make comparisons across seven privacy budgets in Section \ref{sec:sim-privbudget-compare}. Then in Section \ref{sec:sim-model-compare}, we compare across 11 models that differ by number of continuous and binary covariates and the type of continuous variables (finite-valued or unbounded). We focus on metrics for comparing the estimated treatment effects from the analysis on the sanitized and confidential datasets. These metrics are called \textit{utility measures} and are described in Section \ref{sec:utility}. The analysis methods remain the same across the confidential and sanitized data, but further improvements could be made by developing new analysis methods to account for the noise added into the sanitized datasets.

\subsection{Utility measures}\label{sec:utility}
An existing utility metric relies on the overlap between the 95\% confidence intervals (CI's) of the treatment effects from the sanitized and confidential estimated treatment effects for each treatment level \citep{doi:10.1198/000313006X124640}.  The sanitized treatment effects, $\widetilde{\treff}$, are estimated using the synthetic dataset produced by one of our three proposed algorithms. The confidential treatment effects, $\widehat{\treff}$, are estimated from the confidential dataset. 
For a treatment effect, $\treff$, let $(L,U)$ be the CI for the treatment effect computed based on the confidential dataset and $(\widetilde{L},\widetilde{U})$ be the CI computed based on the sanitized dataset. If the confidence intervals overlap, then their intersection is $(L^{over},U^{over})$, where $L^{over} = \max (L,\widetilde{L} )$ and $U^{over} = \min (U,\widetilde{U})$. In this case, the CI overlap metric is the ratio of the width of the intersection and the average CI width. When the CI's do not overlap, the overlap metric takes the value $0$. This overlap metric, denoted as $\cioverlap$, is
    \begin{align}\label{eqn:cioverlap}
    \cioverlap = \begin{cases}\frac{1}{2} \left[ \frac{U^{over} - L^{over}}{U-L} + \frac{U^{over} - L^{over}}{\widetilde{U}-\widetilde{L}}\right] \textrm{ if } U^{over} - L^{over}>0 ,\\
    0 \textrm{ otherwise .}
    \end{cases}
    \end{align}
    The confidence interval overlap can vary between $0$ and $1$, with values near $1$ indicating that there is a large degree of overlap. Thus, higher values (near 1) indicate that the deviation of the inference regarding the treatment effect based on the confidential dataset from the same inference based on the sanitized dataset is small. Lower values (near 0) can be a result of the centers of the intervals being far apart or unequal size of the intervals. If the confidence intervals are similarly sized, the further apart the centers are the smaller the intersection of the CI's is.  Since the metric is scaled by the average interval width, if one CI is much wider than the other, then the overlap metric will be small even if the smaller interval is completely contained within the larger one. 


Another utility metric we use is absolute difference. For our experiments, we can use the true treatment effect value, since it is known. The absolute difference is $\absdiff=|\widetilde{\treff}-\treff|$. For real-world examples, we can approximate this by using the estimated treatment effect from analysis on confidential data, $\hat{\treff}$. The estimated absolute difference is $\widehat{\absdiff}=|\widetilde{\treff}-\widehat{\treff}|$. This metric allows us to evaluate the sanitized point estimate, but does not take into account the standard error of the sanitized point estimate. Further research could be done on these algorithms to improve standard error estimates to be used in inference based on the sanitized data. Thus, absolute error can help show potential for a mechanism which needs the inference on the sanitized data to be adjusted. 

\subsection{Comparing Treatment Effect Utility Experiments}\label{sec:makesimdata}

Using the utility metrics discussed in Section \ref{sec:utility}, we compare our three proposed sanitizing algorithms (Algorithms \ref{algo:perturbMVhist}, \ref{algo:hybrid}, and \ref{algo:fullmodelbased})  using simulated data sets in two experiments. The first experiment compares across privacy budgets ($\epsilon$ and $\delta$ values). The second experiment focuses on comparing the mechanism applications  across models with different counts of covariate variables  and different types of covariate variables. Each experiment uses a different simulated dataset with 1,000 observations. There is a control and treatment group assigned using complete random assignment. The data generating models are multiple regression models with the true treatment effect $\treff=5$ and true residual standard deviation $\sigma=2$. The model intercept is $1$ for both experiments. For each simulated dataset, the coefficients are generated from a geometric series so that there is a diminishing influence of each additional covariate in the model.

The coefficients for categorical variables are from a geometric series: $3(7/11)^{r-1}$ for $r=1,\dots, \ncat$ where $\ncat$ is the number of categorical covariates. Each categorical covariate is binary. To simulate each column, we randomly generate $c_{bin}$ from $\operatorname{Bin}(n=1000,~p=1/2)$. Then $c_{bin}$ observations are randomly sampled and given value $1$ in the column. Otherwise, the observations have value $0$. 

The coefficients for continuous variables are from a geometric series: $0.99(2/3)^{r-1}$ for $r=1,\dots, \ncont$ where $\ncont$ is the number of continuous covariates. 
 We use two alternating methods to generate the continuous covariate columns. The alternating types are unbounded continuous covariates and continuous columns that have a finite set of possible values ($0,0.01,0.02,\dots,0.2$). Starting with the first continuous column, all the unbounded covariate observations are generated from a Gaussian distribution with mean $0$ and standard deviation $2$. Finite-valued covariate observations are generated from a $\operatorname{Uniform}(0,0.2)$ and then rounded to two decimal places.  Due to how the continuous variables are generated the scale of the unbounded values is likely larger than the finite-valued variables. Thus, the Gaussian generated variables will have a larger impact on the generated response variables than the finite-valued variables. 

For our two experiments, we refer to the simulated data as the confidential data since that is what it represents in practice.
 Many algorithm parameters are fixed across these two experiments.
When comparing the methods, we use the bin parameter $\binparam=2/3$ for the MV Histogram algorithm and the internal MV Histogram steps of the model-informed algorithms. In the \longdpmb method, 50\% of the privacy budget is allocated to sanitizing the covariates.  To sanitize the residual variance  5\% of the budget is allocated to sanitizing the residual variance and split evenly among the steps. The standard error bound parameters are set as $\sigma_{min}=2^{-15}$ and $\sigma_{max}=2^{15}$. The treatment effect is sanitized using 15\% of the privacy budget. 
The remaining privacy budget is split evenly to sanitize the remaining model coefficients. For each covariate coefficient,  treatment effect, and variance parameter, the budget is split evenly between the DP Range algorithm of  \citet{karwa2017finite} (Algorithm \ref{algo:dprange} in Appendix) and the infusion of Laplace noise. The range algorithm uses parameters $\bdmean=50$ and $\alpha_{range}=0.05$.  There are $B=5000$ proxies generated in the \longdpmb Algorithm\footnote{The parameters selected are consistent with the real data example in Section \ref{sec:real-world}. A summary of the parameters for \longdpmb across the experiments and real data is in Appendix Table \ref{tab:algoparams}.}.

\subsection{Comparison across Privacy Budget}\label{sec:sim-privbudget-compare}

To compare utility across privacy budgets, we simulate a dataset with 4 binary covariates and 4 continuous covariates. Let $T_i$ be an indicator that observation $i$ is assigned to the treatment group. We simulate one dataset using \begin{align*}y_i\gets& 1+5T_i+\sum_{k=1}^{4}0.99\left(\frac{2}{3}\right)^{k-1}X_{k,i}+\sum_{k=1}^{4}3\left(\frac{7}{11}\right)^{k-1}X_{4+k,i}+2Z_i,\quad i=1,\dots,1000
\end{align*},
where $Z_i,X_{1,i},X_{3,i}\overset{iid}{\sim}N(0,1)$, $X_{4+k,i}$ for $k=1,\dots,4$ are binary variables, and $X_{2,i}$, $X_{4,i}$ are the rounded uniform variables.

\input{tables-main/sim1_budget_wide_v0224_changecatnum}

We compare confidence interval overlap and absolute difference metrics across several privacy budgets. We consider $\epsilon\in \{0.5,~1,~2,~4,~5000\}$ under a pure DP framework $(\delta=0)$. With an extremely large privacy budget, such as $\epsilon=5000$, there are essentially no privacy guarantees. The algorithms simply become resampling and regenerating algorithms rather than sanitizing algorithms. When $\epsilon=4$, the \longdpmb privacy budget for the multivariate histogram step is $\epsilon_{\bX}=2$. Similarly for $\epsilon=2$ the multivariate histogram step has a budget $\epsilon_{\bX}=1$. By comparing the \longdpmb results with twice the privacy budget to the Hybrid results, we can see how much utility is lost when using sanitized model parameter estimates in the \longdpmb method rather than the confidential estimates in the Hybrid method.  
We perform 20 repetitions. Within each repetition, a sanitized dataset is generated using each privacy method, and metrics about the sanitized treatment effect are recorded. The original confidential data remains the same throughout all repetitions and privacy budgets.  The CI overlaps and the treatment effect point estimates are then averaged across the 20 repetitions. The median absolute difference measure is reported; this gives one metric which is robust to outlier repetitions. From our resulting plots, we determine\longdpmb is prone to low-valued outliers for the sanitized treatment effect and thus the absolute difference metric.

The confidential estimated treatment effect is $\widehat{\treff}=5.2$ with estimated standard error $0.124$.  The MV Histogram and Hybrid methods result in estimated standard errors between $0.112$ and $0.136$. However, the \longdpmb method results in estimated standard errors in a much wider range between $0.002$ and $0.484$.  In Table \ref{tab:sim-privbudget}, we see that the Hybrid sanitized datasets generally have the highest average CI overlaps and the lowest median absolute difference for low privacy budgets. The MV Histogram sanitized datasets result in higher mean CI overlaps than the \longdpmb datasets, especially for larger privacy budgets. This is likely because some \longdpmb sanitized datasets have extremely small estimated standard errors produced by the original inference applied to the \longdpmb sanitized dataset. There are variants of \longdpmb we discussed in Section \ref{sec:dpmodelbased} which could address the wide range of estimated standard errors but do not satisfy the DP framework. Future work could focus on adjusting the post-sanitization analysis to account for the sanitizing noise, or improving the sanitized estimate of residual variance. Since the widths of the confidence intervals make it difficult to compare the \longdpmb method to other methods, we can also consider the median absolute difference metric ($\absdiff$).

\begin{figure}[!htb]
    \centering
    \includegraphics[width=0.9\linewidth]{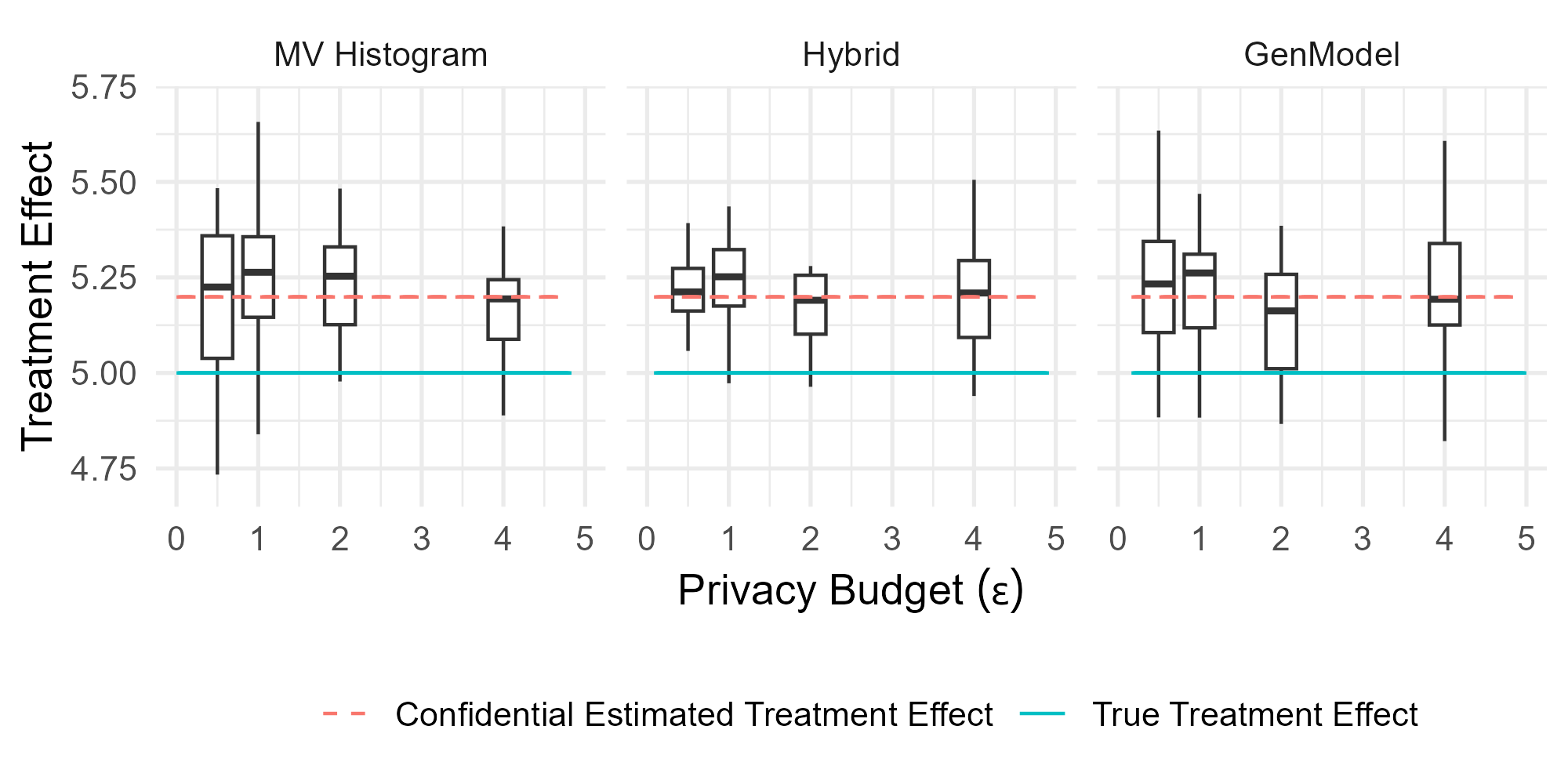}
    \caption{The \longdpmb and MV Histogram sanitized treatment effects take a wider range of values across the 20 repetitions while the Hybrid sanitized treatment effects are more concentrated near the confidential estimated treatment effect (red dashed line). The outliers have been removed here, but plots including outlier can be found in Figure \ref{fig:apdx-sim-privbudget} in Appendix \ref{sec:appendix-plots}}
    \label{fig:sim-privbudget-boxplots}
\end{figure}

 The median absolute difference decreases as the privacy budget increases for the MV Histogram and \longdpmb methods. The mean absolute difference from the \longdpmb method is less than 0.25.  The \longdpmb method with twice the privacy budget of the Hybrid method results in median $\absdiff$ within $0.01$ of the the Hybrid method, with the exception of the $\epsilon_\bX=0.5$ case. For $\epsilon=2$ and $\epsilon=4$, the \longdpmb method has a slightly lower median $\absdiff$ than the MV Histogram. As shown in Figure \ref{fig:sim-privbudget-boxplots}, the spread of the sanitized treatment effects is slightly larger for the \longdpmb method. 

 Overall, the three algorithms have mean estimated treatment effects near the confidential estimated treatment effect $\widehat{\treff}=5.20$. The \longdpmb method results in an average sanitized treatment effect closer to the true value of $\treff=5$ for $\epsilon\in \{5000,4,2\}$. Figure \ref{fig:sim-privbudget-boxplots} shows the distribution over the 20 repetitions of the sanitized treatment effects, $\widetilde{\treff}$, from the sanitized data from three methods for four $\epsilon$ values.  The spread across the 20 repetitions of $\widetilde{\treff}$ from the \longdpmb data is much wider than the spread of $\widetilde{\treff}$ from the other two methods. The centers of the \longdpmb $\widetilde{\treff}$  for each privacy budget are closer to or equidistance from the true treatment effect, $\treff=5$ than the other two methods. The  MV Histogram and Hybrid methods have sanitized treatment effects that are tightly focused around the confidential data's estimated treatment effect.

\subsection{Comparison across Regression Models} \label{sec:sim-model-compare}

To compare utility across models with various number and types of covariate variables, we simulate a dataset with 10 binary covariates and 10 continuous covariates. We simulate one dataset using the following generating model: \begin{gather}\label{eq:sim-compare-models-generate}y_i\gets 1+5T_i+\sum_{k=1}^{10}0.99\left(\frac{2}{3}\right)^{k-1}X_{k,i}+\sum_{k=1}^{10}3\left(\frac{7}{11}\right)^{k-1}X_{10+k,i}+2Z_i, \quad i=1,\dots,1000,
\end{gather}
where $Z_i,X_{2j-1,i}\overset{iid}{\sim}N(0,1)$ for $j=1,\ldots,5$; $X_{2j,i}$ for $j=1,\dots,5$ are the rounded uniform observations; and $X_{j,i}$ for $j=11,\dots,20$ are binary.

The models for analysis that are considered use the first $2$, $5$, or $10$ binary variables and continuous variables. From these, we consider nine models: one for each combination of possible number and types of variables. Specifically, the regression models are of the following form:
\begin{align} \label{eq:mod-sim-compare-models}
y_i=&1+\treff T_i+\sum_{k=1}^{\ncont^{(j)}}\covcoef_{k}X_{k,i}+\sum_{k=1}^{\ncat^{(j)}}\covcoef_{k+\ncont^{(j)}}X_{10+k,i}+e_i &\text{ where }e_i\overset{iid}{\sim}N(0,\sigma^2)\text{ for }i=1,\dots,n
\end{align}

The model $j$ values for the number of binary covariates, $\ncat^{(j)}$,  and the number of continuous covariates, $\ncont^{(j)}$,  for $j=1,\dots,9$ can be found in Table \ref{tab:models-sim-compare}. We also consider two special case models. Both use five continuous covariates and the first two binary covariates. The Unbounded Model uses the five unbounded continuous covariates ($X_{1},X_{3},\dots,X_{9}$) and the Finite Set Model uses the five continuous covariates that take a finite set of values ($X_{2},X_{4},\dots,X_{10}$). 
\begin{align}
    &\text{Unbounded Model:}& y_i=&\covcoef_0+\treff T_i+\sum_{k=1}^{5}\covcoef_{k}X_{2(k-1),i}+\sum_{k=1}^{2}\covcoef_{k+5}X_{10+k,i}+e_i \nonumber\\
    &\text{Finite Set Model:}& y_i=&\covcoef_0+\treff T_i+\sum_{k=1}^{5}\covcoef_{k}X_{2k,i}+\sum_{k=1}^{2}\covcoef_{k+5}X_{10+k,i}+e_i \nonumber\\
   &&  &\text{ where }e_i\overset{iid}{\sim}N(0,\sigma^2)\text{ for }i=1,\dots,n  \label{eq:mod-sim-compare-models-spCs}
\end{align}
The simulated data is generated using \eqref{eq:sim-compare-models-generate}, but most of the regression models for the analysis omit variables. Thus, we risk an omitted variable bias in the models \citep{Dunn_Smyth_2018}. However, we check the regression assumptions for the models fitted with the confidential data (Figure \ref{fig:sim-model-fits} in Appendix \ref{sec:appendix-plots}). Based on the plots, the regression assumptions are met and there is not evidence that the models have omitted variable bias. 


\input{tables-main/table-models-for-simulation-experiment}

Using a pure DP framework ($\delta=0$), we fix the privacy budget to be $\epsilon=1$ for MV Histogram (Algorithm \ref{algo:perturbMVhist}) and Hybrid (Algorithm \ref{algo:hybrid}). We also compare sanitized datasets from the \longdpmb algorithm (Algorithm \ref{algo:fullmodelbased}) at two privacy levels. One method uses the same overall budget $\epsilon=1$ and is denoted \dpmb{1}. The other, denoted \dpmb{2}, uses $\epsilon=2$. For \dpmb{2} the privacy allocation $\epsilon_{\bX}=1$ is equal to privacy budget for the Hybrid and MV Histogram methods.

Similar, to the comparisons made for the privacy budgets, we average CI overlap, and estimated/sanitized treatment effect across 20 repetitions. We take the median of the absolute difference measures.  These results for models with two binary covariates are shown in Table \ref{tab:sim-numcovariates-cat4}. The other models are in Table \ref{tab:sim-numcovariates-othercats}. Again, the estimated standard errors from the \longdpmb sanitized data are from a much larger range ($0.002$ to $0.472$) than other algorithms. The estimated standard error of the treatment effects for MV Histogram and Hybrid method range from $0.123$ to $0.219$. For the confidential data, these range from $0.136$ to $0.203$. 

\begin{figure}[!htb]
     \centering
     \includegraphics[width=0.8\linewidth]{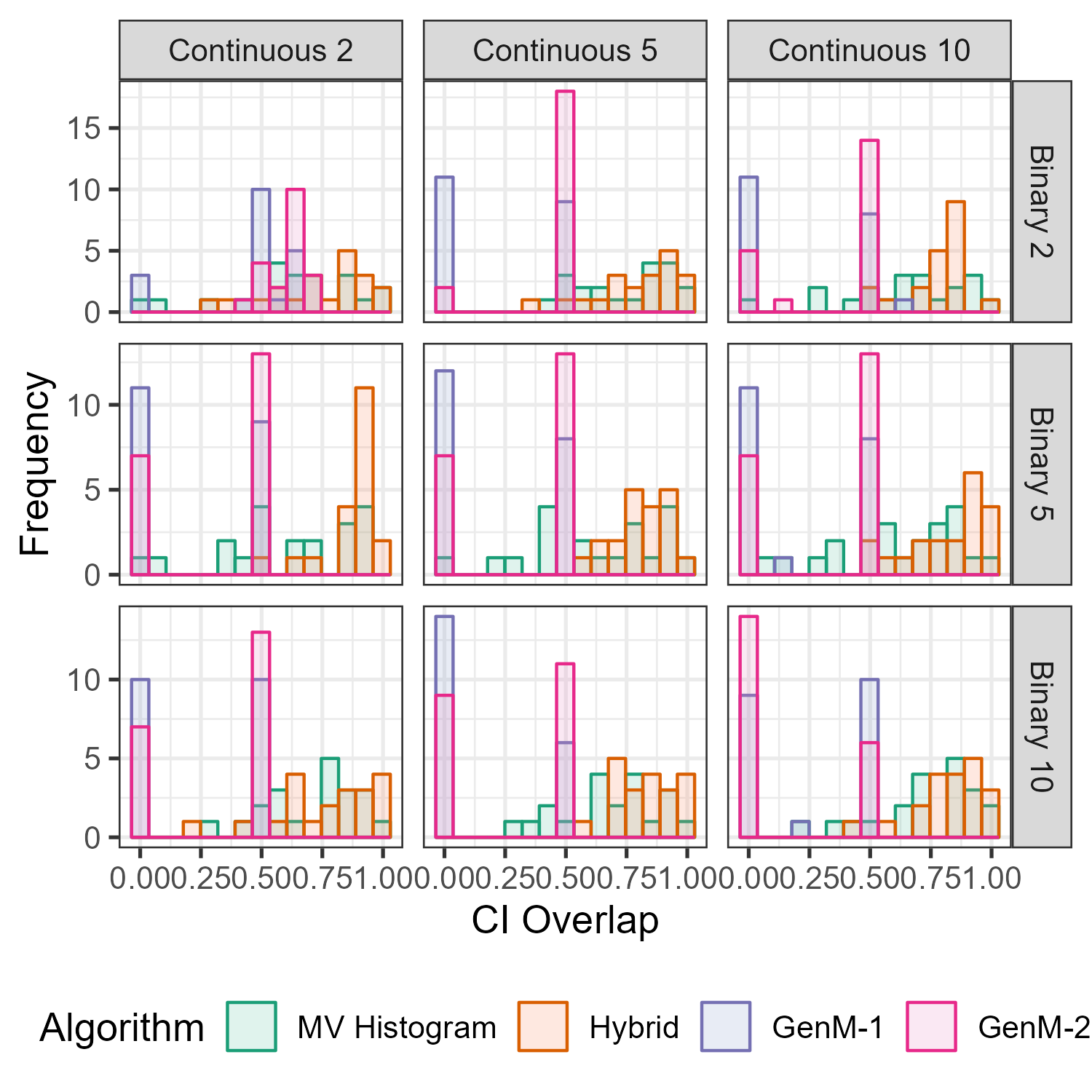}
     \caption{The distribution of CI overlap values for Models 1 to 9 based on the number and type of covariates for various sanitizing methods.}
     \label{fig:sim2-cioverlap-hist}
 \end{figure}

Other than Model 1, the average CI overlaps for the \longdpmb data are quite small. In Model 1, which has only four covariates, the \dpmb{2} average CI overlap is only $0.1$ less than that of the MV Histogram. The CI overlaps for \longdpmb are often near $0$ or near $0.5$ as shown in Figure \ref{fig:sim2-cioverlap-hist}. The MV Histogram has a large spread of CI overlap values, especially for models with 5 or less binary variables (Figure \ref{fig:sim2-cioverlap-hist}.  Over all the models, the Hybrid method has the highest average CI overlap compared to the other sanitization method and often has the lowest median absolute difference. In Table \ref{tab:sim-numcovariates-cat4}\footnote{Since each model is only fit once with the confidential data, the average $\hat{\treff}$, average $\cioverlap$, and median $\absdiff$ reported in Tables \ref{tab:sim-numcovariates-cat4} and \ref{tab:sim-numcovariates-othercats} are the fitted model results rather than summarized results across repetitions.}, \dpmb{2} results in a median absolute difference better than or within $0.1$ of the MV Histogram.  This is also true of Model 4 in Table \ref{tab:sim-numcovariates-othercats}. As expected from higher privacy budgets, \dpmb{2} has better utility than \dpmb{1} when using the median $\absdiff$ metric. It is better when using the average $\cioverlap$ metric for all models but Model 9. For  Model 1, Unbounded Model, and Model 4, the median absolute difference of \dpmb{1} is better than the MV Histogram which has the same overall budget. 

\input{tables-main/sim2_models_split_cat4}


The \longdpmb performs particularly poorly in our utility metrics for models with ten binary covariates (Table \ref{tab:sim-numcovariates-othercats}).
 The privacy budget has to be split to sanitize each covariate coefficient, thus increasing the number of covariates increases the amount of noise infused to the sanitized coefficients. \dpmb{1} has its worst utility for Model 8 (average $\cioverlap=0.15$ and median $\absdiff=0.74$). \dpmb{2} has its worst utility in Model 9 (average $\cioverlap=0.15$ and median $\absdiff=0.44$) For a fixed number of total covariates, we find \longdpmb has lower median absolute difference and higher average CI overlap when the model has more continuous covariates than binary covariates. For example, \dpmb{1} has average $\cioverlap=0.21$ and median $\absdiff=0.33$ and \dpmb{2} has average $\cioverlap=0.33$ and median $\absdiff=0.18$ for Model 6 (10 continuous and 5 binary). For Model 8 (5 continuous and 10 binary), \dpmb{1} has average $\cioverlap=0.15$ and median $\absdiff=0.74$ and \dpmb{2} has average $\cioverlap=0.28$ and median $\absdiff=0.25$.  The \longdpmb methods have better results in Model 3 than Model 7 and better results in Model 2 than Model 4. There is not a clear pattern when comparing models with the same number of covariates for the MV Histogram. This could be due to how the response variables were generated or something systematic in the algorithm. 
 All methods did better on the Unbounded model than the Finite Set Model; this might be due to the Finite Set covariate having a smaller scale of impact of the response variable than the unbounded covariates in the generation of the response variables.

\input{tables-main/sim2_models_split_othercats}

\begin{figure}[!htb]
    \centering\includegraphics[width=0.98\linewidth]{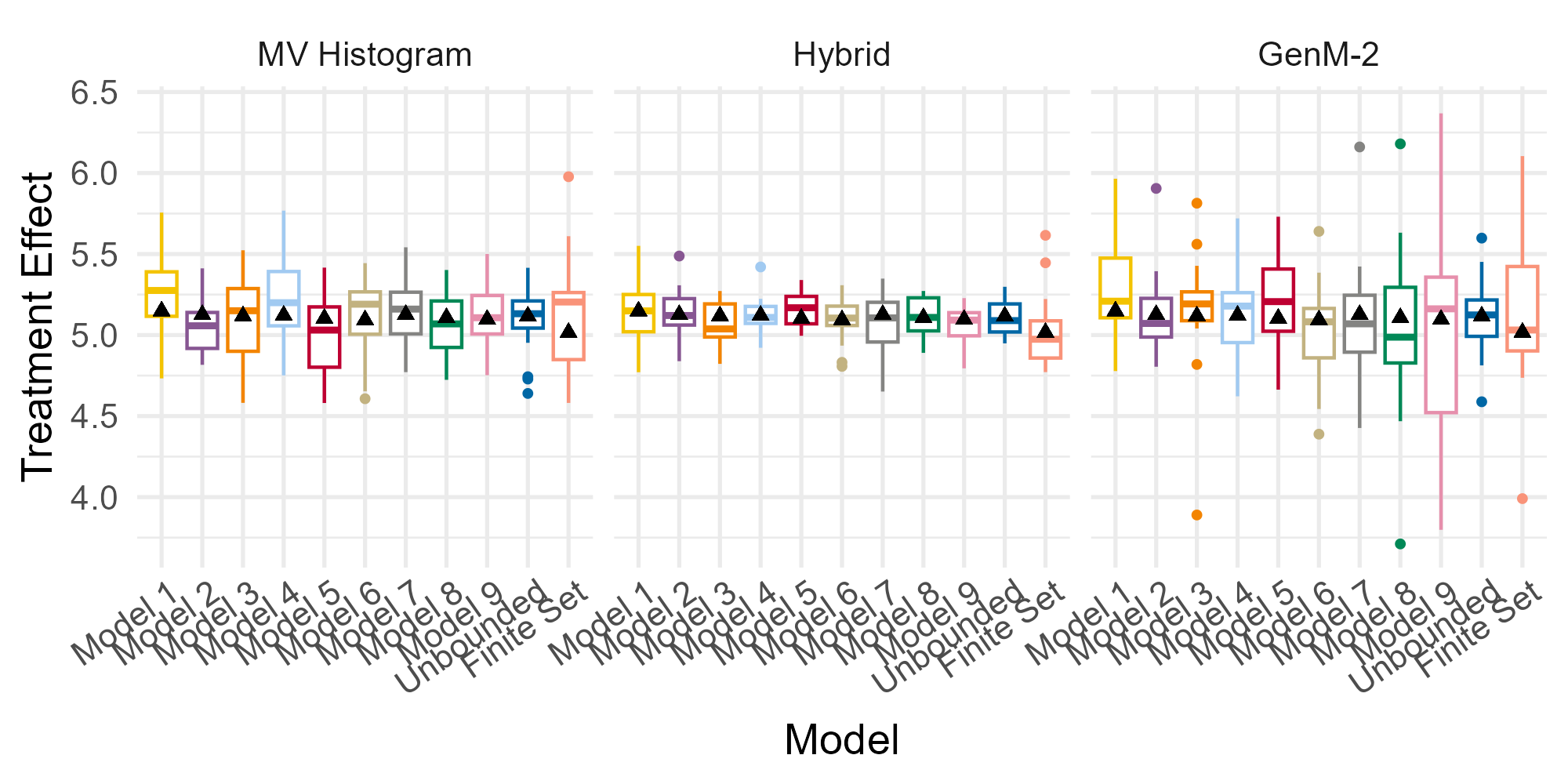}
    \caption{Estimated treatment effects by sanitization method with the confidential estimator (black triangle) marked for each model. The MV Histogram and Hybrid results use $\epsilon=1$. The \dpmb{2} method uses $\epsilon=2$, which fixes the $\epsilon_{\bX}=1$ equal to the other methods. }
    \label{fig:simulations-models-treff}
\end{figure}

\begin{figure}[!htb]
    \centering
    \includegraphics[width=0.55\linewidth]{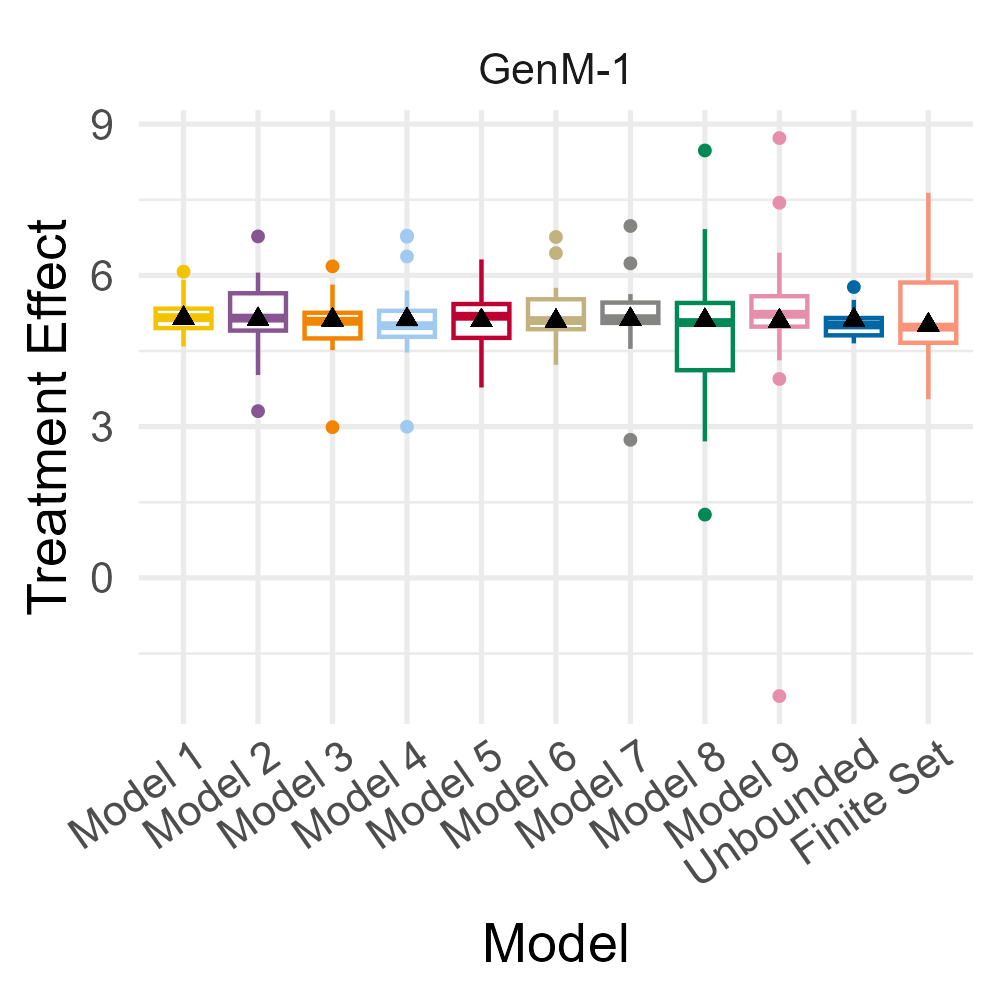}
    \caption{For the \longdpmb method with $\epsilon=1$, the spread of the sanitized treatment effects is large with several outliers.}
    \label{fig:sim2-modelX-genm1}
\end{figure}

When we look at the distribution of the estimated treatment effects (Figure \ref{fig:simulations-models-treff}), the Hybrid method has the smallest spread of sanitized treatment effects. The spread of the $\widetilde{\treff}$ from the \longdpmb methods is much larger than the other sanitization methods. For \dpmb{1} the spread is so great, the plot should not share an axis with the other three methods (Figure \ref{fig:sim2-modelX-genm1}. 

There are many outliers for the \longdpmb methods, \dpmb{1} even reports a negative sanitized treatment effect for one of its repetitions. Additionally, we see the increase in spread from Model 7 to 9 with the \longdpmb methods. The Hybrid method is the most concentrated around the confidential estimated treatment effects. For the first five models, the MV Histogram has a similar spread on the center 50\% of repetitions of sanitized treatment effects as the \dpmb{2} method, but the MV Histogram doesn't have outliers.  

\subsection{Summarizing Experimental Results}

Our two experiments described in Sections \ref{sec:sim-privbudget-compare} and \ref{sec:sim-model-compare} compare the three proposed sanitization methods across various privacy budgets and regression models. The Hybrid method consistently out preforms the MV Histogram and the \longdpmb. However, if a researcher wants to provide confidentiality protection to the estimated model parameters, then the MV Histogram or \longdpmb methods are also reasonable options. The MV Histogram does well when there is a large number of covariates. 

The \longdpmb method performs well with a slightly higher privacy budget and a low number of covariates. However, the estimated standard errors of treatment effects from inference on \longdpmb sanitized data can vary wildly. This makes \longdpmb results hard to evaluate with the CI overlap utility metric, but the utility can be seen in the absolute difference metric. There are potential variants discussed in \ref{sec:dpmodelbased} which could address the standard error estimate issues. However, those variants use the residual standard error from the model fit on the confidential data and reduce the confidentiality protections. We did not implement these variants since the Hybrid method is a much simplified model-informed sanitization method than these variants. It could be argued these variants provide more privacy protection since they only depend on the confidential residual error rather than all of the fitted model coefficients and residual standard error from the confidential data. However, it is unclear how to measure the difference in privacy protections between the \longdpmb variants and the Hybrid method. Further work should be done to improve the implementation and analysis of the \longdpmb method. Even with these improvements, the \longdpmb method should be used for datasets with a limited number of covariates.

%% file: tables-main/sim1_budget_wide_v0224_changecatnum.tex
\begin{table}[!htb]
 \centering
    \small
\begin{tabular}{llrrrrrrrrr}

\toprule
\multicolumn{1}{c}{} & \multicolumn{1}{c}{} & \multicolumn{3}{c}{MV Histogram} & \multicolumn{3}{c}{Hybrid} & \multicolumn{3}{c}{\longdpmb} \\
\cmidrule(l{3pt}r{3pt}){3-5} \cmidrule(l{3pt}r{3pt}){6-8} \cmidrule(l{3pt}r{3pt}){9-11}
$\epsilon$ & $\delta$ & Avg. $\widetilde{\treff}$ & Avg. $\cioverlap$ & Med. $\absdiff$ & Avg. $\widetilde{\treff}$ & Avg. $\cioverlap$ & Med. $\absdiff$ & Avg. $\widetilde{\treff}$ & Avg. $\cioverlap$ & Med. $\absdiff$\\
\midrule
5000 & 0 & 5.21 & 0.84 & 0.21 & 5.23 & 0.84 & 0.22 & 5.20 & 0.51 & 0.20\\
4 & 0 & 5.16 & 0.80 & 0.19 & 5.21 & 0.75 & 0.21 & 5.22 & 0.47 & 0.20\\
2 & 0 & 5.22 & 0.68 & 0.26 & 5.16 & 0.83 & 0.19 & 5.11 & 0.38 & 0.24\\
1 & 0 & 5.23 & 0.67 & 0.26 & 5.23 & 0.77 & 0.25 & 5.10 & 0.39 & 0.29\\
0.5 & 0 & 5.19 & 0.64 & 0.25 & 5.22 & 0.82 & 0.21 & 5.29 & 0.44 & 0.31\\
\bottomrule
\end{tabular}
\caption{We sanitized a simulated dataset with 1,000 observations using 3 methods. The true treatment effect
is $\treff$ = 5. The analysis on the confidential data estimates the treatment effect to be $\widehat{\treff}=5.20$ with estimated standard error $0.12$. The confidence interval overlaps ($\cioverlap$) compare the intervals from the synthetic data and
the confidential data. The sanitized treatment effects $\widetilde{\treff}$ and the confidence interval overlaps and averaged over 20 repetitions. The absolute difference ($\absdiff$) measurements compare the sanitized treatment effect and the true treatment effect. The median $\absdiff$ over 20 repetitions is reported.}
\label{tab:sim-privbudget}
\end{table}

%% file: tables-main/table-models-for-simulation-experiment.tex
\begin{table}[!htb]
    \centering
    \begin{tabular}{c c|c c c}
    \toprule
    \multicolumn{5}{c}{$\ncat^{(j)}$ and $\ncont^{(j)}$ for Model $j$}\\ \midrule
   & & \multicolumn{3}{c}{$\ncat^{(j)}$}\\ 
     & & 2 & 5 & 10\\
       \multirow{3}{*}{ \rotatebox{90}{$\ncont^{(j)}$}}  & 2 &  Model 1 & Model 2 & Model 3\\
         & 5 & Model 4 & Model 5 & Model 6\\ 
         & 10 & Model 7 & Model 8 & Model 9\\ \bottomrule
    \end{tabular}\hspace{1em}
    \begin{tabular}{c |c c c}
    \toprule
    \multicolumn{4}{c}{$\ncat$ and $\ncont$ for Special Case Models}\\ \midrule
   Model & \multicolumn{1}{c}{$\ncont$} & \multicolumn{1}{c}{$\ncat$} & Cont. Variables\\ \midrule
    Unbounded & 5 & 2 &  $X_{1},X_{3},X_{5},X_{7},X_{9}$\\
       Finite Set & 5 & 2 & $X_{2},X_{4},X_{6},X_{8},X_{10}$\\ \bottomrule
    \end{tabular}
    \caption{There are nine standard models using  regression models of the form \eqref{eq:mod-sim-compare-models} and two special case models with regression models of the form \eqref{eq:mod-sim-compare-models-spCs} to compare across in our experiment. The simulated data is generated with \eqref{eq:sim-compare-models-generate}.}
    \label{tab:models-sim-compare}
\end{table}

%% file: tables-main/sim2_models_split_cat4.tex
\begin{table}[!htb]
    \centering
    \caption{We compared metrics across the number and type of covariate variables by sanitizing algorithm summarized over 20 repetitions. When focusing on the model with 2 binary variables, the Hybrid algorithm consistently produced datasets that had the highest mean CI overlap and lowest mean absolute difference of the sanitization methods.}
\label{tab:sim-numcovariates-cat4}
\begin{tabular}{ >{\centering\arraybackslash}p{0.01\textwidth} >{\centering\arraybackslash}p{0.06\textwidth} >{\centering\arraybackslash}p{0.06\textwidth} >{\raggedright\arraybackslash}p{0.17\textwidth} >{\centering\arraybackslash}p{0.09\textwidth} 
>{\centering\arraybackslash}p{0.09\textwidth}   >{\centering\arraybackslash}p{0.1\textwidth} 
}
\toprule
&  \multicolumn{2}{ >{\centering\arraybackslash}p{0.13\textwidth} }{\# Covariates}  &\multicolumn{1}{c }{ } &  \multicolumn{1}{ >{\centering\arraybackslash}p{0.1\textwidth} }{Avg. $\treff$} &  \multicolumn{2}{c }{ }\\
& Cont. & Bin. & Algorithm & Estimate & Avg. $\cioverlap$ &  Median  $\absdiff$\\
\midrule
\multirow{5}{*}{\rotatebox{90}{ {Model 1} }}& 2 & 2 & Confidential & 5.15 & - & 0.15\\
& 2 & 2 & MV Histogram & 5.26 & 0.63 & 0.29\\
& 2 & 2 & Hybrid & 5.13 & 0.74 & 0.19\\
& 2 & 2 & \dpmb{1} & 5.20 & 0.47 & 0.24\\
& 2 & 2 & \dpmb{2} & 5.31 & 0.62 & 0.22\\
\addlinespace
\multirow{5}{*}{\rotatebox{90}{ {Model 2} }}
& 5 & 2 & Confidential & 5.13 & - & 0.13\\
& 5 & 2 & MV Histogram & 5.06 & 0.74 & 0.10\\
& 5 & 2 & Hybrid & 5.13 & 0.79 & 0.14\\
& 5 & 2 & \dpmb{1} & 5.18 & 0.23 & 0.32\\
& 5 & 2 & \dpmb{2} & 5.14 & 0.46 & 0.10\\
\addlinespace
\multirow{5}{*}{\rotatebox{90}{ {Model 3} }}
& 10 & 2 & Confidential & 5.12 & - & 0.12\\
& 10 & 2 & MV Histogram & 5.11 & 0.64 & 0.19\\
& 10 & 2 & Hybrid & 5.07 & 0.78 & 0.13\\
& 10 & 2 & \dpmb{1} & 5.00 & 0.24 & 0.27\\
& 10 & 2 & \dpmb{2} & 5.16 & 0.37 & 0.20\\
\addlinespace
\multirow{5}{*}{\rotatebox{90}{ {Unbounded} }}
& 5 & 2 & Confidential & 5.12 & - & 0.12\\
& 5 & 2 & MV Histogram & 5.10 & 0.73 & 0.18\\
& 5 & 2 & Hybrid & 5.11 & 0.84 & 0.09\\
& 5 & 2 & \dpmb{1} & 5.05 & 0.28 & 0.17\\
& 5 & 2 & \dpmb{2} & 5.12 & 0.36 & 0.15\\
\addlinespace
\multirow{5}{*}{\rotatebox{90}{ {Finite Set} }}
& 5 & 2 & Confidential & 5.02 & - & 0.02\\
& 5 & 2 & MV Histogram & 5.12 & 0.65 & 0.24\\
& 5 & 2 & Hybrid & 5.01 & 0.78 & 0.14\\
& 5 & 2 & \dpmb{1} & 5.24 & 0.23 & 0.47\\
& 5 & 2 & \dpmb{2} & 5.13 & 0.33 & 0.23\\
\bottomrule
\end{tabular}
\end{table}

%% file: tables-main/sim2_models_split_othercats.tex
\begin{table}[!htb]
    \centering
    \caption{We compared metrics across the number and type of covariate variables by sanitizing algorithm summarized over 20 repetitions. For models with more than 2 binary covariates, the Hybrid method has the highest average confidence interval overlap. With more predictors, the \longdpmb utility is substantially decreased.}
\label{tab:sim-numcovariates-othercats}
\begin{tabular}{ >{\centering\arraybackslash}p{0.01\textwidth} >{\centering\arraybackslash}p{0.06\textwidth} >{\centering\arraybackslash}p{0.06\textwidth} >{\raggedright\arraybackslash}p{0.17\textwidth} >{\centering\arraybackslash}p{0.09\textwidth} 
>{\centering\arraybackslash}p{0.09\textwidth}   >{\centering\arraybackslash}p{0.1\textwidth} 
}
\toprule
&  \multicolumn{2}{ >{\centering\arraybackslash}p{0.13\textwidth} }{\# Covariates}  &\multicolumn{1}{c }{ } &  \multicolumn{1}{ >{\centering\arraybackslash}p{0.1\textwidth} }{Avg. $\treff$} &  \multicolumn{2}{c }{ }\\
& Cont. & Bin. & Algorithm & Estimate & Avg. $\cioverlap$ &  Median  $\absdiff$\\
\midrule
\multirow{5}{*}{\rotatebox{90}{ {Model 4} }}
& 2 & 5 & Confidential & 5.12 & - & 0.12\\
& 2 & 5 & MV Histogram & 5.21 & 0.62 & 0.24\\
& 2 & 5 & Hybrid & 5.13 & 0.87 & 0.11\\
& 2 & 5 & \dpmb{1} & 5.12 & 0.23 & 0.25\\
& 2 & 5 & \dpmb{2} & 5.15 & 0.33 & 0.23\\
\addlinespace
\multirow{5}{*}{\rotatebox{90}{ {Model 5} }}
& 5 & 5 & Confidential & 5.10 & - & 0.10\\
& 5 & 5 & MV Histogram & 5.01 & 0.62 & 0.20\\
& 5 & 5 & Hybrid & 5.17 & 0.81 & 0.17\\
& 5 & 5 & \dpmb{1} & 5.12 & 0.20 & 0.36\\
& 5 & 5 & \dpmb{2} & 5.22 & 0.33 & 0.27\\
\addlinespace
\multirow{5}{*}{\rotatebox{90}{ {Model 6} }}
& 10 & 5 & Confidential & 5.09 & - & 0.09\\
& 10 & 5 & MV Histogram & 5.12 & 0.64 & 0.23\\
& 10 & 5 & Hybrid & 5.10 & 0.82 & 0.13\\
& 10 & 5 & \dpmb{1} & 5.23 & 0.21 & 0.33\\
& 10 & 5 & \dpmb{2} & 5.04 & 0.33 & 0.18\\
\addlinespace
\multirow{5}{*}{\rotatebox{90}{ {Model 7} }}
& 2 & 10 & Confidential & 5.13 & - & 0.13\\
& 2 & 10 & MV Histogram & 5.14 & 0.72 & 0.19\\
& 2 & 10 & Hybrid & 5.08 & 0.76 & 0.13\\
& 2 & 10 & \dpmb{1} & 5.19 & 0.26 & 0.39\\
& 2 & 10 & \dpmb{2} & 5.08 & 0.33 & 0.25\\
\addlinespace
\multirow{5}{*}{\rotatebox{90}{ {Model 8} }}
& 5 & 10 & Confidential & 5.11 & - & 0.11\\
& 5 & 10 & MV Histogram & 5.06 & 0.69 & 0.17\\
& 5 & 10 & Hybrid & 5.11 & 0.82 & 0.11\\
& 5 & 10 & \dpmb{1} & 4.85 & 0.15 & 0.74\\
& 5 & 10 & \dpmb{2} & 5.02 & 0.28 & 0.25\\
\addlinespace
\multirow{5}{*}{\rotatebox{90}{ {Model 9} }}
& 10 & 10 & Confidential & 5.10 & - & 0.10\\
& 10 & 10 & MV Histogram & 5.12 & 0.74 & 0.14\\
& 10 & 10 & Hybrid & 5.06 & 0.82 & 0.11\\
& 10 & 10 & \dpmb{1} & 5.11 & 0.26 & 0.32\\
& 10 & 10 & \dpmb{2} & 5.01 & 0.15 & 0.44\\
\bottomrule
\end{tabular}
\end{table}

%% file: section-real-world.tex
In this section, we apply our proposed methodos (Algorithms \ref{algo:perturbMVhist}, \ref{algo:hybrid}, and \ref{algo:fullmodelbased}) on a real-world randomized control trial and evaluate their performance, focusing on the analyses as reported in [BJS]~\cite{Reducecrime}. The associated replication files, including the de-identified data, are available in \citet{ReducecrimePkg}.

\subsection{Data and Original Analysis}
\label{subsec:Liberiasetup}
 For our evaluation, we focus on the results reported in Table 2, Panel B of BJS. 
 Specifically, BJS consider the long-term (12-13 months after the program) \footnote{Formally, we use Round 5 data, as the original code \citep{ReducecrimePkg} specifies.}
effect of therapy and cash grant on eight outcome variables related to antisocial behavior, individually and through one a summary index called  {\it Antisocial behaviors z-score} (referred to as \texttt{fam\_asb\_lt}).  The seven individual outcome variables shown in Table \ref{tab:response-vars} include indicators for anti-social behaviors like selling drugs, standardized specific behavior index values like an index of domestic verbal abuse, and the count of recent stealing activities. The sample is composed of $947$ high-risk youths in Monrovia, Liberia\footnote{One observation of the original data reported $0.1$ for the  value for the number of women the respondent was supporting. This entry is changed to be $1$.}. A $2 \times 2$ factorial design is used with two stratification variables based on the groups the youths were in when they were randomly assigned treatments, once at the time of being assigned to therapy (there were 55 such groups), and once at the time of being assigned to receive a cash grant of $200$ USD (there were 20 such groups). BJS find that neither cash nor therapy alone has a lasting (12-13 month) effect, but that the combination of both treatments does reduce ``antisocial behavior''. 

\input{tables-main/table-response-variables}

The analysis data are obtained from the file named \texttt{STYL\_Final.dta} as provided in the replication package \citep{ReducecrimePkg}. 
The treatment assignments are encoded using three binary treatment variables 
\texttt{tpassonly} (indicating that only therapy was received),
\texttt{cashassonly} (indicating that a cash-only grant was received), 
and \texttt{tpcashass} (indicating that both therapy and cash grant were received). The blocking variable based on therapy assignment is \texttt{tp\_strata\_alt} while the blocking variable based on cash grant is \texttt{cg\_strata}. Fifty-five baseline covariates measured at the beginning of the study period are included in the analysis. The binary baseline covariates and the continuous ones are detailed in Tables \ref{tab:bincovariates} and Table \ref{tab:covariatevars-nonbinary} in the Appendix \ref{sec:replicate-conf}. 

Even though, fifteen baseline covariates and three response variables\footnote{The response variables are averaged over two months per respondent.} are indicator variables, the analysis preformed by BJS treat all variables as continuous. We mirror this choice so our results reflect what BJS could have done to release a privacy-preserving replication package. However, it is important to note that the original analysis reports results for fitted models which do not meet the necessary model assumptions. Without the model assumptions the validity of the model to capture and explain the sample and population data is questionable. Specifically, multiple regression models assume the residuals are normally distributed with equal variance (i.e., homoscedastic) and mean $0$. This is often checked with a QQ-Plot for normality and by plotting the residuals against the fitted values. We recreate the analysis as described in \citet{ReducecrimePkg}. Using a QQ-Plot for normality and a plot of the residuals over the fitted values as diagnostic tools,  the regressions assumptions are not met in seven out of eight models (Figure \ref{fig:2bdiagnostics} in Appendix). All response variables, except the standardized aggressive behavior index (\texttt{abshostilstd\_ltav}), result in residuals that deviate from the normal distribution on the upper tail. The spread of the residuals increase as the fitted value increase for the models with \texttt{fam\_asb\_it}, \texttt{stealnb\_ltav, disputes\_all\_z\_ltav,} and \texttt{domabus\_z\_ltav} response variables. There are clear banded trends in the residuals over the fitted values for the models with \texttt{drugssellever\_ltav, carryweapon\_ltav, arrested\_ltav}, and \texttt{domabuse\_z\_ltav}.The residuals for the model with the standardized index of domestic abuse response (\texttt{domabuse\_z\_ltav}) do follow the normal distribution, and the the residual variance appears reasonably constant over the fitted values. These concerns with the model fit for seven out of eight models may have adverse effects on the utility of our model-informed algorithms: the Hybrid (Algorithm \ref{algo:hybrid}) and \longdpmb (Algorithm \ref{algo:fullmodelbased}) algorithms. The model-informed algorithms assume the model accurately represents the data generating model. The \longdpmb method also heavily relies on the residual distribution being Gaussian.

In addition to the analysis with the full 55 baseline covariates predictors, we considered analysis with a subset of ten covariates, and a reduced subset of seven variables to better assess computational feasibility. 
The reduced subset includes: \texttt{age\_b}, \texttt{asbhostil\_b}, \texttt{drugssellever\_b}, \texttt{drinkboozeself\_b}, \texttt{druggrassself\_b}, \texttt{harddrugsever\_b}, and \texttt{steals\_b}. 
The first two covariates are age and baseline standardized antisocial behavior index  for the individuals participating in the study. The remaining covariates record the baseline antisocial behavior of the youths in terms of whether they: have ever sold drugs; drink alcohol; smoke grass/opium; have ever consumed hard drugs; and have exhibited stealing behavior in the two weeks prior to their baseline interview, respectively. The covariate variables in the reduced subset are shown in the dark blue cells of Appendix Tables~\ref{tab:bincovariates} and \ref{tab:covariatevars-nonbinary}. In addition to the reduced subset variables, the subset of the covariates includes three other variables: $\texttt{school\_b}$, $\texttt{wealth\_indexstd\_b}$, and $\texttt{cognitive\_score\_b}$. These are an individual's years of schooling, baseline standardized index of wealth, and baseline standardized index of cognitive function, respectively (light blue cells in Appendix Table \ref{tab:covariatevars-nonbinary}). In terms of the number of predictor variables, the full set has 60 since there are three treatment variables and two blocking variables. The subset has 15 and the reduced subset has 12. The models have 61, 16, and 13 coefficients including the intercept that the \longdpmb algorithm will have to sanitize.

The original analysis by BJS reports the estimated treatment effects, which they denote as ITT, robust standard errors, and adjusted p-values for the cash only, therapy only, and both cash and therapy treatment levels. 
With the original data, we find similar results with the full baseline covariate set, the 15 predictor subset, and the 12 predictor reduced subset as shown in Appendix Table \ref{tab:apdx-itt-orig}. At the significance level of $\alpha=0.05$, the therapy only and cash only treatments do not have any significant treatment effects for any response variables across the three covariate sets. The treatment with both therapy and cash is significant for the summary index (\texttt{fam\_asb\_lt}) and the aggressive behavior index (\texttt{asbhostilstd\_ltav}). Additionally the estimated treatment effects for the subset and reduced subset are within $0.02$ of the treatment effect from the full baseline covariate set for both the  \texttt{fam\_asb\_lt} ($\widehat{\treff}_{both}=-0.26$) and the \texttt{asbhostilstd\_ltav} ($\widehat{\treff}_{both}=-0.31$) responses.

\subsection{Comparing Treatment Effects Estimated from Privacy-Preserving Data}\label{sec:liberia-compare-san}

To compare the analyses from the original dataset and the privacy-preserving datasets, we use the confidence interval (CI) overlap  (Equation \ref{eqn:cioverlap}) and absolute difference from Section \ref{sec:evaluating}. We also look at the adjusted p-values from the privacy-preserving datasets to see if similar conclusions would be drawn compared to the original study by BJS\footnote{BJS use a Westfall-Young correction to adjust the p-values. For the anti-social summary index, the p-values are adjusted for nine comparisons to correspond to Figure 1 in their paper. For the individual responses, the p-values are adjusted for 21  comparisons (three treatment levels and seven individual response variables). Our analysis uses a Bonferroni correction and adjusts the summary index p-values for three comparisons (three treatment levels) and the individual p-values for 21 comparisons. We will use a significance level of $\alpha=0.05$ throughout our analysis.}.

For all three covariate sets, we compare the MV Histogram algorithm (Algorithm \ref{algo:perturbMVhist}), the Hybrid algorithm (Algorithm \ref{algo:hybrid}), and \longdpmb algorithm (\ref{algo:fullmodelbased}). We generate five privacy-preserving datasets: one from MV Histogram, one from Hybrid, and three from \longdpmb at three privacy levels.

All our privacy method applications set $\delta=0$ and the bin parameter $\binparam=2/3$. For MV Histogram and Hybrid methods the privacy parameter is $\epsilon=1$. We implement the \longdpmb method at three privacy levels $\epsilon\in \{1,2,9\}$. 
 With privacy budget $\epsilon=1$, the privacy allocation to the covariate data is $\epsilon_{\bcov}=0.5$ and the privacy-preserving dataset is denoted \dpmb{1}. The two synthetic datasets made with the \longdpmb algorithm with $\epsilon=2$ and $\epsilon=9$ are denoted \dpmb{2} and \dpmb{9} respectively. In both cases, the MV Histogram step to sanitize the covariate data has the same privacy budget allocated as the Hybrid and MV Histogram algorithms ($\epsilon_{\bcov}=1$). For all three applications of the \longdpmb algorithm, after the multivariate histogram step, the remaining budget is split evenly over the eight response variables ($\epsilon_{\bfy}=\frac{{\epsilon}-\epsilon_{\bcov}}{8}$). Within a response, each treatment coefficient and the standard error are allocated 10\% of the response-level budget ($\epsilon_{var,\sigma^2}=\epsilon_{range,sigma^2}=\epsilon_{\sigma^1}=\frac{\epsilon-\epsilon_{\bcov}}{240}$ and $\epsilon_{range,\treff,k}=\epsilon_{\treff,k}=\frac{\epsilon-\epsilon_{\bcov}}{160}$ for $k=1,2,3$). The remaining budget is split evenly among the covariate coefficients ($\epsilon_{range,\covcoef,\ell}=\epsilon_{\covcoef,\ell}=\frac{3(\epsilon-\epsilon_{\bcov})}{80*(p+1)}$ for $\ell=1,\ldots,p$).  
We use 5,000 iterations to generate proxy response variables. The standard deviation bounds for sanitizing the residual variance for each model are $(2^{-15},2^{15})$. All models use $\alpha=0.05$ and $\bdmean=50$ for the parameters to sanitize the range of each coefficient in the model (Algorithm \ref{algo:dprange}) These parameter choices correspond to those in the simulated experiments in Section \ref{sec:evaluating}. The allocation of the privacy budget and the specification of other parameters are summarized in Appendix Table \ref{tab:algoparams}.

We report the privacy-preserving treatment effect statistics for each of the sanitizing algorithms, covariate sets, and model response and the original data results in the Appendix (Tables \ref{tab:apdx-itt-orig} to \ref{tab:apdx-itt-dpmb2}).

\subsubsection{Selected Response Variable Results}
In this section we will focus on only four response variables: \texttt{fam\_asb\_lt, stealnb\_ltav, carryweapon\_ltav,} and \texttt{asbhostilstd\_ltav}. The anti-social behavior summary index is included because it is a main focus of the analysis done by BJS. The aggressive behavior index (\texttt{asbhostilstd\_ltav}) is chosen since it satisfies the regression model assumptions with the original data. Then we select two responses which could be better modeled as discrete count data and a binary variable respectively: the count of stealing activities (\texttt{stealnb\_ltav}) and the indicator of carrying a weapon (\texttt{carryweapon\_ltav}).  BJS treat these response variables as continuous, which is one reason why they do not satisfy the regression model assumptions. 

From Section \ref{sec:evaluating}, we expect Hybrid method to produce the highest utility. With 60 predictors, the \longdpmb is not likely to do well with the full baseline models. For the three responses where model assumptions are violated the utility of the privacy-preserving datasets from the Hybrid and \longdpmb model-informed algorithm's utility will suffer.   

\input{tables-main/liberia_itt_pval_reducedsubset}

 For the reduced subset of covariates and the \texttt{fam\_asb\_lt} response, the Hybrid synthetic data produce a privacy-preserving treatment effect close to the the original data treatment effect estimate, and both are significant at $\alpha=0.05$ level ($\widehat{\treff_{both}}=-0.25,~\widetilde{\treff}_{both,Hybrid}=-0.43$ from Table \ref{tab:itt-pval-reducedsubset}). The MV Histogram synthetic data results suggest that the only cash treatment increases anti-social behavior ($\widetilde{\treff}_{cash,MV}=0.34$ for \texttt{fam\_asb\_lt} and it is significant at $\alpha=0.05$). The Hybrid method also indicated the cash only treatment increases anti-social behavior ($\widetilde{\treff}_{cash,Hybrid}=0.24$ for \texttt{fam\_asb\_lt} and it is significant at $\alpha=0.05$). This contradicts the original data results.
  Almost all of the privacy-preserving treatment effects from \longdpmb synthetic data are significant with the reduced subset of covariates. This is likely due to the concerns about the estimated standard errors that we discussed in Section \ref{sec:evaluating}. For \dpmb{1} and \texttt{asbhostilstd\_ltav}, the privacy-preserving treatment effect for combined therapy and cash has the opposite sign of the original estimated treatment effects ( $\widehat{\treff_{both,\dpmb{1}}}=1.48$). The privacy budgets for this may be too low to handle eight responses and 12 predictors. For the models with \texttt{stealnb\_ltav} and \texttt{carryweapon\_ltav} as responses, the \longdpmb algorithm produces some extreme privacy-preserving treatment effects (e.g. $\widetilde{\treff}_{therapy,\dpmb{1}}=-45.28$ and $\widetilde{\treff}_{cash,\dpmb{9}}=-46.43$ for \texttt{carryweapon\_ltav}). This is likely due to the poor fit of these models to the original data.
\begin{figure}[!htb]
    \centering
    $\phantom{abfek}$\includegraphics[width=0.8\linewidth]{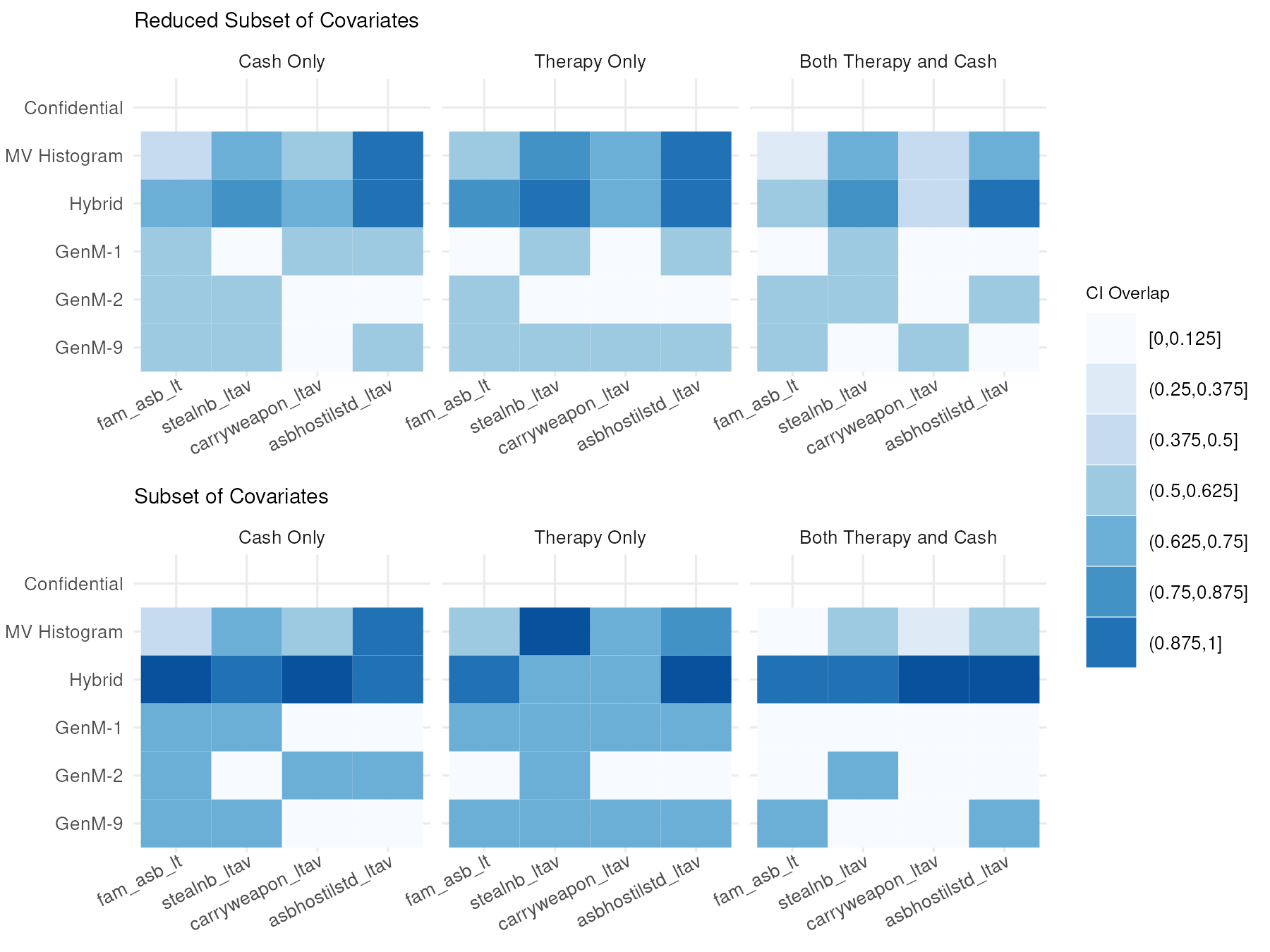}\\
    \includegraphics[width=0.7\linewidth]{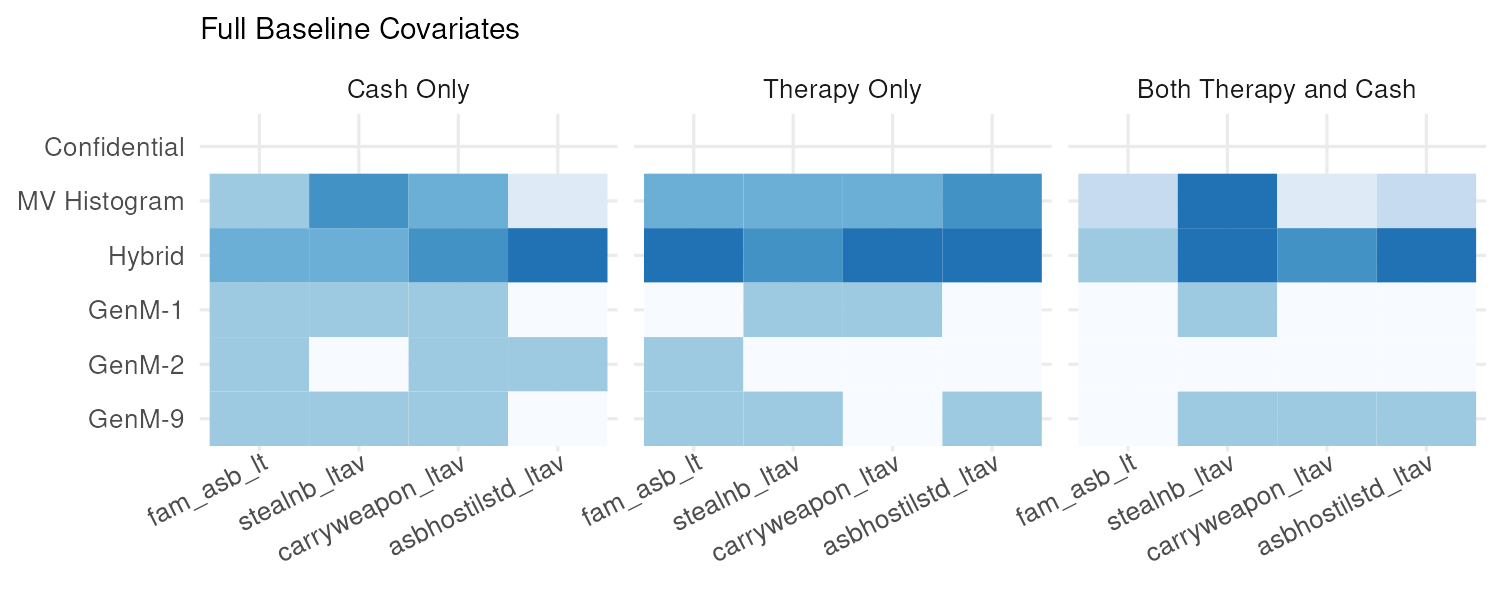}$\phantom{avre}$
    \caption{Confidence interval overlap for all response variables across all five synthetic datasets and all three covariate sets.}
    \label{fig:liberia-ciover-subred}
\end{figure}

\begin{figure}[!htb]
    \centering
    $\phantom{abfek}$\includegraphics[width=0.8\linewidth]{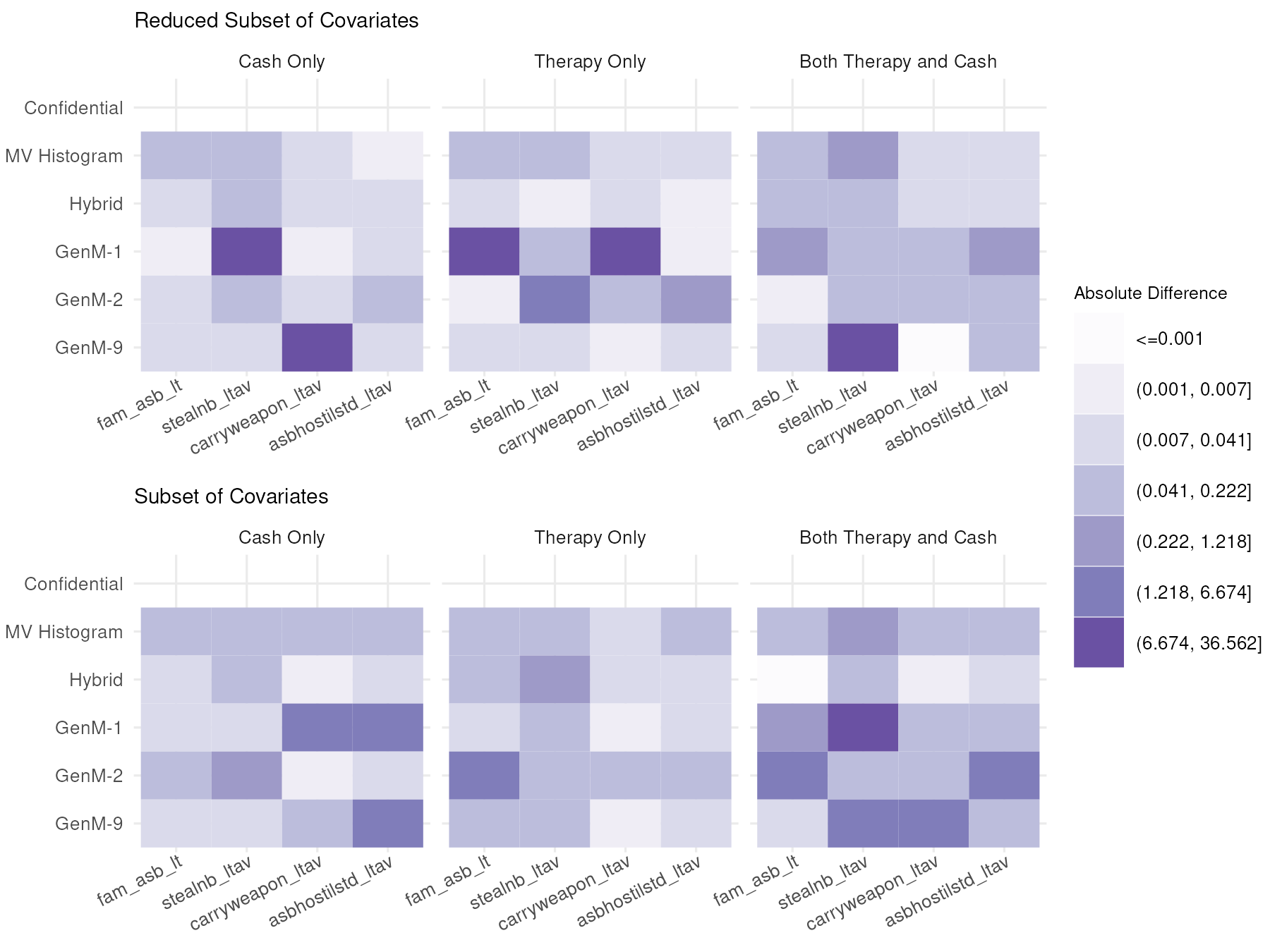}\\
    \includegraphics[width=0.7\linewidth]{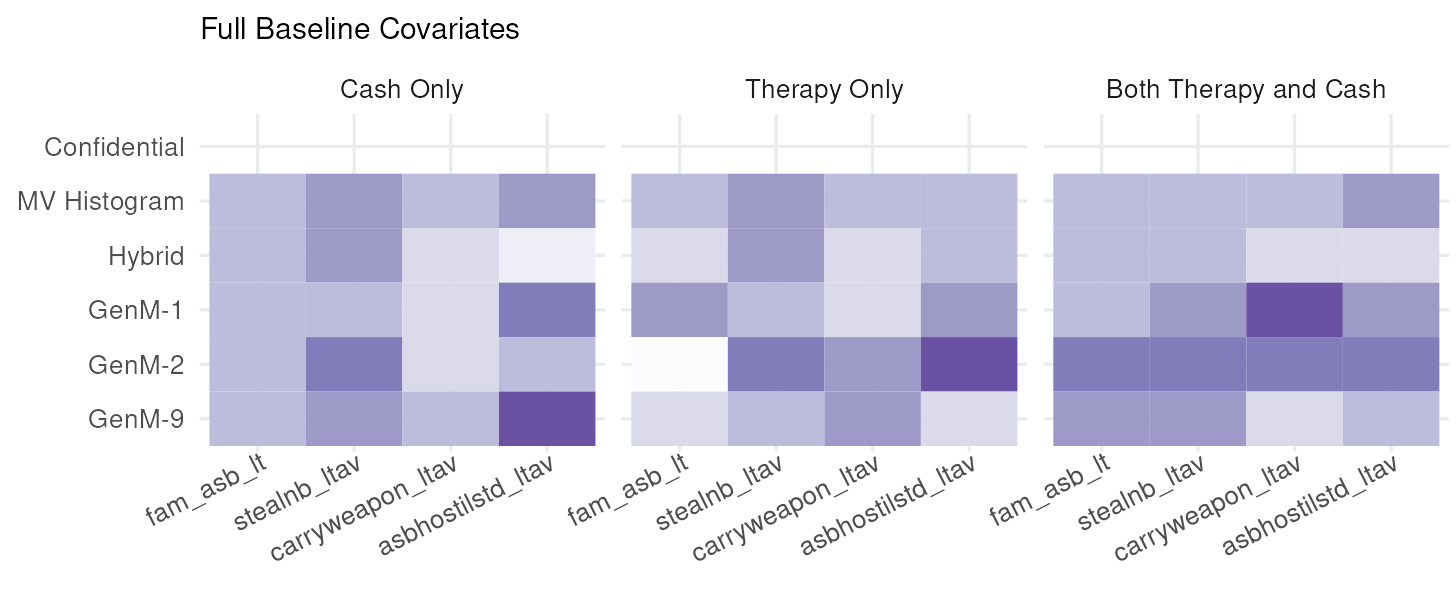}$\phantom{avre}$
    \caption{Absolute difference between privacy-preserving treatment effect from a synthetic dataset and estimated treatment effect from the original dataset is shaded in accordance to a log-transformed scale for all covariate sets.}
    \label{fig:liberia-absdiff-subred}
\end{figure}

Much like our experimental results in Section \ref{sec:evaluating}, the Hybrid synthetic dataset almost always has the highest CI overlap with the original estimated treatment effects (Figure \ref{fig:liberia-ciover-subred}). The MV Histogram generally performs better than the \longdpmb methods. The \longdpmb datasets have confidence interval overlaps that are $0$ or near $0.5$. The \dpmb{9} has similar or better results then the \longdpmb privacy-preserving datasets with lower privacy budgets. The Hybrid method once again low  absolute differences between the privacy-preserving treatment effect and the original treatment effect (Figure \ref{fig:liberia-absdiff-subred}). With a few exceptions, \dpmb{9} has similar or better absolute difference than the MV Histogram synthetic data. 
The absolute difference values tend to increase as the number of covariates increase for all of the methods because the MV histogram will discretize more variables which adds more noise.

\begin{figure}[!htb]
    \centering
    \includegraphics[width=0.45\linewidth]{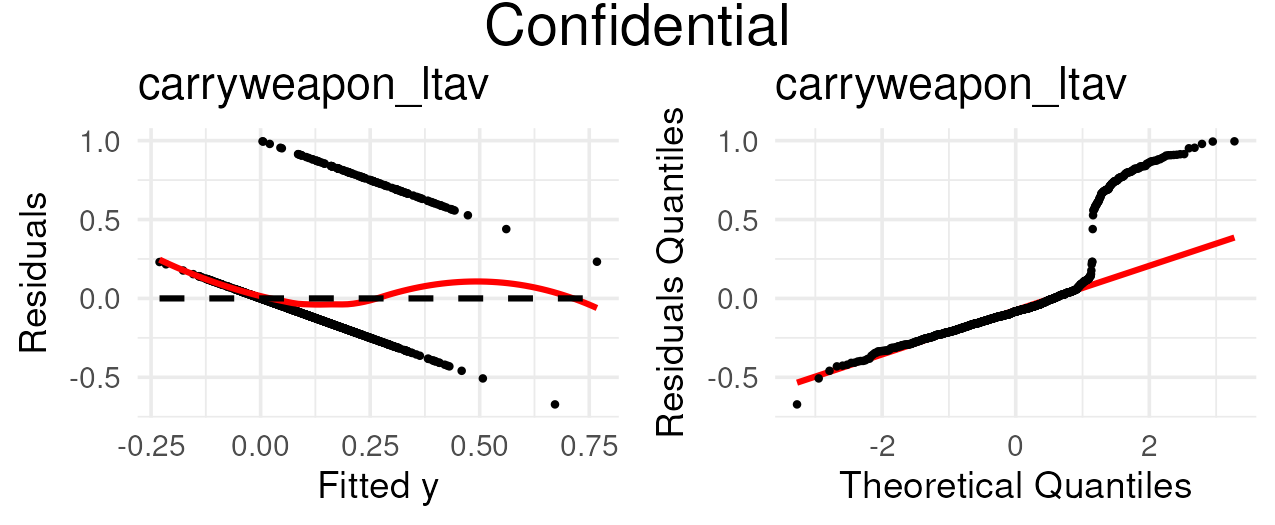}\includegraphics[width=0.45\linewidth]{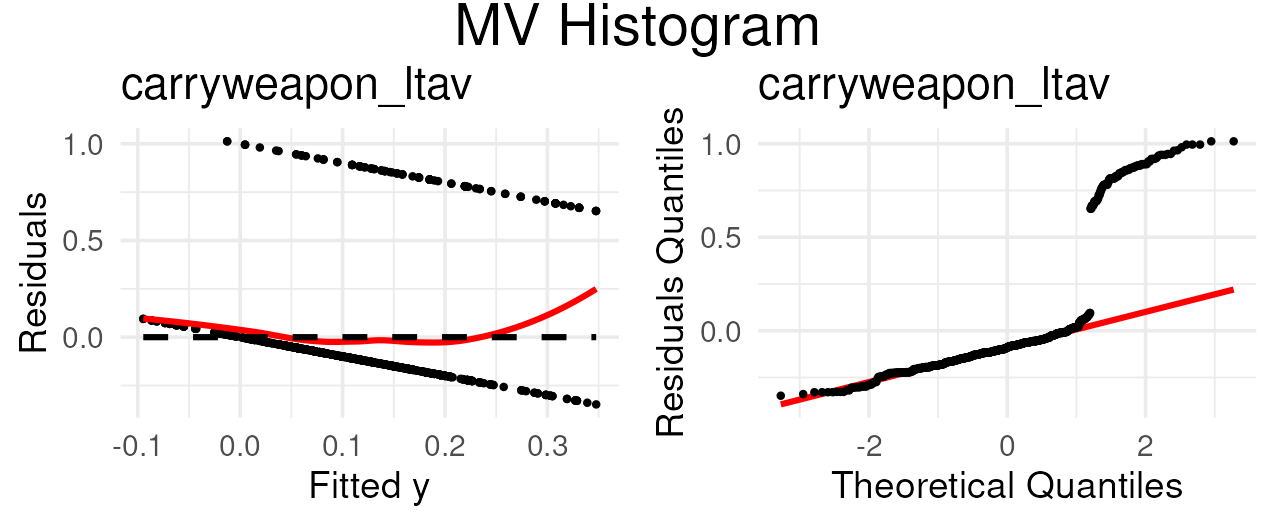}
    \includegraphics[width=0.45\linewidth]{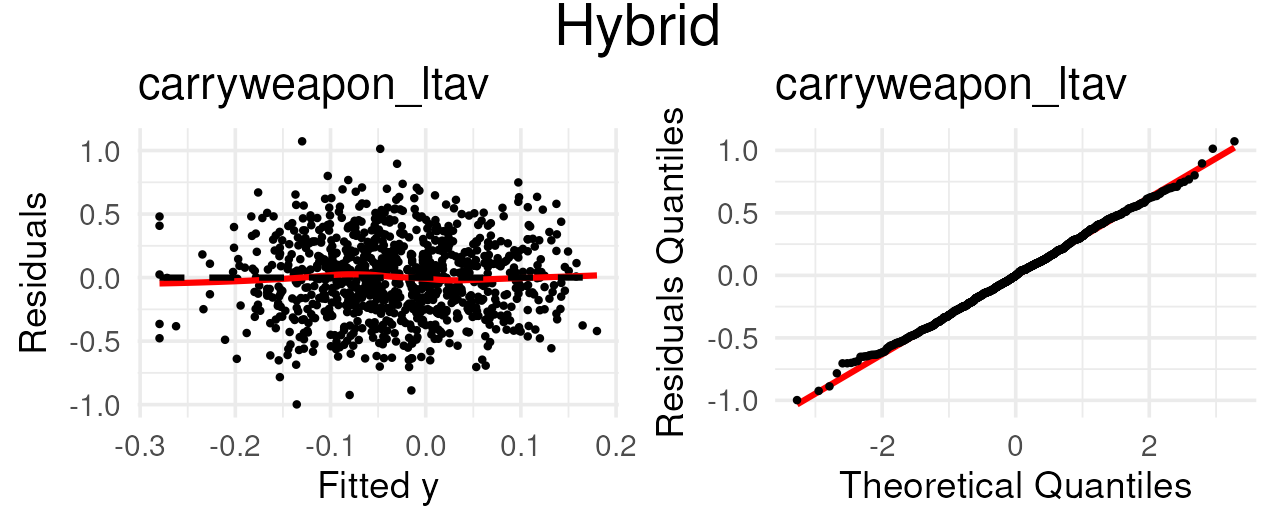}\includegraphics[width=0.45\linewidth]{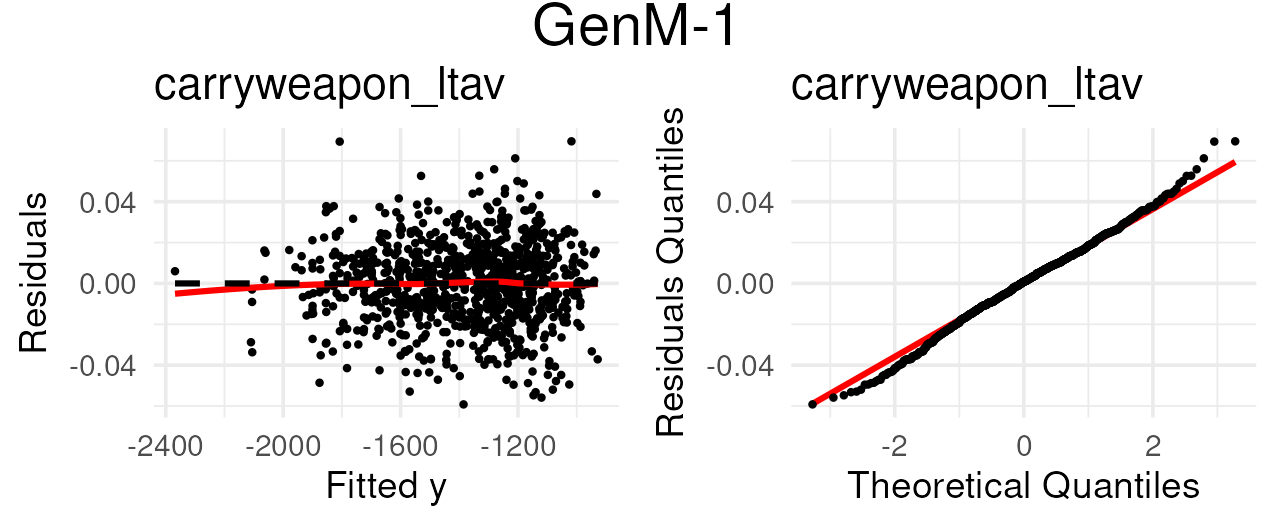}
    \caption{Diagnostic plots for the model with \texttt{carryweapon\_ltav} response variable and the reduced subset of covariates across some of the synthetic datasets. }
    \label{fig:liberia-some-diag-plt}
\end{figure}

Another concern with model-informed sanitization algorithms is how they influence diagnostic checks of the released data. For example, if BJS only released the synthetic data generated with the Hybrid or \longdpmb algorithms, a researcher replicating the study would be unaware of the issues with the model fit in the underlying original data (bottom plots in Figure \ref{fig:liberia-some-diag-plt}). When a model is used to generate synthetic data and is fitted to the resulting synthetic data, the underlying assumptions of regression model will always meet by the synthetic data ones whether or not they were in the original data.  The model-agnostic procedures such as the MV Histogram are more robust; they are likely to retain the data characteristics that result in the poor fit of the regression model. This can be seen in the diagnostic plots for the model of the indicator of carrying a weapon with the reduced covariate subset fitted to the original and three synthetic datasets in Figure \ref{fig:liberia-some-diag-plt}. The MV Histogram synthetic data captures the same issues about the model fit as our analysis of the original data, while the Hybrid and \dpmb{1} synthetic datasets suggest that the model fits well to the data.


\subsection{Computational Feasibility}

To address Goal 3 outlined in Section \ref{sec:definitions-problemsetup}, we discuss the computational feasibility of our algorithms that are summarized in Table \ref{tab:liberia-comptimes-summary}. The simulations were run on personal laptop with a quadcore Intel Core i7-1165G7 and 16 GB of RAM. The Hybrid method has approximately consistent computation time across the covariate sets. With the removal of three covariates from the subset of covariates, generating the \longdpmb privacy-preserving data for the reduced subset of covariates is on average 2\% faster. The MV Histogram takes twice as long to generate a synthetic dataset with the full covariate set with 60 predictors than the subset with 15 predictors. Comparing the full and subset times for the \longdpmb methods, each additional covariate takes on average $9.8$ additional seconds. 

\input{tables-main/liberia_summary_time}

We can break down the privacy-preserving data generation times into smaller sub-steps for \longdpmb. As a model-informed algorithm, \longdpmb must be repeated to sanitize each response variable. However, the privacy-preserving covariates and treatments can be shared across these iterations. Table \ref{tab:liberia-comptimes-byresponse} breaks down the computation times for sub-steps of the process of generating the privacy-preserving responses.  In the \longdpmb algorithm, generating the proxy responses and then estimating the model parameters is the most time consuming sub-step. Sanitizing the model parameters is the second most time consuming step. Once the model parameters are protected, generating the privacy-preserving responses takes a couple seconds.

\include{tables-main/liberia_byresponse_short}

Overall, all three methods are computationally tractable on commodity computer hardware. The \longdpmb method is the most computationally taxing. This computational time will also likely increase with the number of iterations of proxies generated. Both the Hybrid and the MV Histogram algorithms are very quick to apply to the original dataset by BJS.

%% file: tables-main/table-response-variables.tex
\begin{table}[!htb]
\begin{tabular}{>{\raggedright\arraybackslash}p{0.21\textwidth} >{\raggedright\arraybackslash}p{0.70\textwidth}}
\toprule
\multicolumn{2}{c}{Response Variables: Long Term Average (12-13 month)}\\
Variable Name & Description\\
\midrule
fam\_asb\_lt & Standardized antisocial behavior summary index \\
drugssellever\_ltav & Indicator for usually sells drugs \\
stealnb\_ltav & Count of stealing activities in past 2 weeks \\
disputes\_all\_z\_ltav & Standardized index of all disputes in the past 2 weeks \\
carryweapon\_ltav$\phantom{l}^*$ & Indicator of carrying a weapon  \\
arrested\_ltav & Indicator of arrested by the police or in jail \\
asbhostilstd\_ltav$\phantom{l}^*$ & Standardized aggressive behavior index \\
domabuse\_z\_ltav$\phantom{l}^{**}$ & Standardized Index of domestic verbal abuse \\
\bottomrule
\end{tabular}
\caption{BJS analyzed eight outcome variables. Each is the long term average across 12-13 months. There are four ($\phantom{.}^*$) and 217 ($\phantom{.}^{**}$) missing observations from some variables.}\label{tab:response-vars}
\end{table}

%% file: tables-main/liberia_itt_pval_reducedsubset.tex
\begin{table}[!htb]
\centering

\begin{tabular}{>{\raggedleft\arraybackslash}p{0.005\textwidth}>{\raggedleft\arraybackslash}p{0.25\textwidth}>{\raggedleft\arraybackslash}p{0.15\textwidth}>{\raggedleft\arraybackslash}p{0.15\textwidth}>{\raggedleft\arraybackslash}p{0.15\textwidth}}
\toprule
Response & Method & Therapy Only & Cash Only & Both\\
\midrule
\addlinespace\multirow{6}{*}{\rotatebox{90}{fam\_asb\_lt}} & Confidential & \textcolor{gray}{-0.07} & \textcolor{gray}{0.12} & \textcolor{black}{-0.25}\\
 & MV Histogram & \textcolor{gray}{0.07} & \textcolor{black}{0.34} & \textcolor{gray}{-0.04}\\
 & Hybrid & \textcolor{gray}{0} & \textcolor{black}{0.24} & \textcolor{black}{-0.43}\\
 & \dpmb{1} & \textcolor{black}{-36.93} & \textcolor{black}{0.1} & \textcolor{black}{-1.51}\\
 & \dpmb{2} & \textcolor{black}{-0.09} & \textcolor{black}{0.17} & \textcolor{black}{-0.28}\\
 & \dpmb{9} & \textcolor{gray}{0} & \textcolor{black}{0.08} & \textcolor{black}{-0.13}\\
\addlinespace\multirow{6}{*}{\rotatebox{90}{stealnb\_ltav}} & Confidential & \textcolor{gray}{0.06} & \textcolor{gray}{0.28} & \textcolor{gray}{-0.84}\\
 & MV Histogram & \textcolor{gray}{-0.24} & \textcolor{gray}{0.72} & \textcolor{gray}{-0.25}\\
 & Hybrid & \textcolor{gray}{0.08} & \textcolor{gray}{0} & \textcolor{gray}{-0.7}\\
 & \dpmb{1} & \textcolor{black}{-0.37} & \textcolor{black}{-22.15} & \textcolor{black}{-0.97}\\
 & \dpmb{2} & \textcolor{black}{3.06} & \textcolor{black}{-0.22} & \textcolor{black}{-1.22}\\
 & \dpmb{9} & \textcolor{black}{0.18} & \textcolor{black}{0.21} & \textcolor{black}{-14.38}\\
\addlinespace\multirow{6}{*}{\rotatebox{90}{carryweapon\_ltav}} & Confidential & \textcolor{gray}{-0.05} & \textcolor{gray}{0.03} & \textcolor{gray}{-0.07}\\
 & MV Histogram & \textcolor{gray}{-0.02} & \textcolor{gray}{0.08} & \textcolor{gray}{0}\\
 & Hybrid & \textcolor{black}{-0.1} & \textcolor{gray}{-0.01} & \textcolor{gray}{0.02}\\
 & \dpmb{1} & \textcolor{black}{-45.28} & \textcolor{black}{0.04} & \textcolor{black}{-0.59}\\
 & \dpmb{2} & \textcolor{black}{-0.4} & \textcolor{black}{0.12} & \textcolor{black}{0.28}\\
 & \dpmb{9} & \textcolor{black}{-0.03} & \textcolor{black}{-46.43} & \textcolor{black}{-0.07}\\
\addlinespace\multirow{6}{*}{\rotatebox{90}{asbhostilstd\_ltav}} & Confidential & \textcolor{gray}{-0.15} & \textcolor{gray}{-0.07} & \textcolor{black}{-0.31}\\
 & MV Histogram & \textcolor{gray}{-0.11} & \textcolor{gray}{-0.06} & \textcolor{gray}{-0.2}\\
 & Hybrid & \textcolor{gray}{-0.17} & \textcolor{gray}{-0.12} & \textcolor{gray}{-0.25}\\
 & \dpmb{1} & \textcolor{black}{-0.18} & \textcolor{black}{-0.11} & \textcolor{black}{1.48}\\
 & \dpmb{2} & \textcolor{black}{1.33} & \textcolor{black}{-0.53} & \textcolor{black}{-0.16}\\
 & \dpmb{9} & \textcolor{black}{-0.05} & \textcolor{black}{0.05} & \textcolor{black}{-0.57}\\
\bottomrule
\end{tabular}
\caption{The estimated treatment effect for each treatment is shown across the various datasets for each response and the Reduced Subset covariates. Values in \textcolor{gray}{ gray } are not significant at the $\alpha= 0.05 $ level.}
\label{tab:itt-pval-reducedsubset}
\end{table}

%% file: tables-main/liberia_summary_time.tex
\begin{table}[!htb]
\centering

\begin{tabular}{>{\raggedright\arraybackslash}p{0.16\textwidth}>{\centering\arraybackslash}p{0.11\textwidth}>{\centering\arraybackslash}p{0.11\textwidth}>{\centering\arraybackslash}p{0.1\textwidth}>{\centering\arraybackslash}p{0.1\textwidth}>{\centering\arraybackslash}p{0.1\textwidth}>{\centering\arraybackslash}p{0.1\textwidth}}
\toprule
Covariate Set & Number of Predictors  & MV Histogram & Hybrid & \dpmb{1} & \dpmb{2} & \dpmb{9} \\
\midrule
Full Baseline & 60  & 0.16 & 9.29 & 628.76 & 652.89 & 578.14 \\
Subset & 15 & 0.08 & 10.11 & 172.46 & 183.75 & 184.90 \\
Reduced Subset & 12  & 0.08 & 9.10 & 171.43 & 205.40 & 159.79 \\
\bottomrule
\end{tabular}
\caption{Summary of computation times per algorithm and covariate set. The simulations were run on personal laptop with a quadcore Intel Core i7-1165G7 and 16 GB of RAM.}
\label{tab:liberia-comptimes-summary}
\end{table}

%% file: tables-main/liberia_byresponse_short.tex
\begin{table}[!htb]
\centering
\small
\begin{tabular}{>{\raggedright\arraybackslash}p{0.27\textwidth} >{\centering\arraybackslash}p{0.06\textwidth}>{\centering\arraybackslash}p{0.06\textwidth}>{\centering\arraybackslash}p{0.06\textwidth}>{\centering\arraybackslash}p{0.06\textwidth}>{\centering\arraybackslash}p{0.06\textwidth}>{\centering\arraybackslash}p{0.06\textwidth}>{\centering\arraybackslash}p{0.06\textwidth}>{\centering\arraybackslash}p{0.06\textwidth}}
\toprule
Step & \rotatebox{83}{fam\_asb\_lt} & \rotatebox{83}{drugssellever\_ltav} & \rotatebox{83}{stealnb\_ltav} & \rotatebox{83}{disputes\_all\_z\_ltav} & \rotatebox{83}{carryweapon\_ltav} & \rotatebox{83}{arrested\_ltav} & \rotatebox{83}{asbhostilstd\_ltav} & \rotatebox{83}{domabuse\_z\_ltav}\\
\midrule
\addlinespace[0.3em]
\hline
\multicolumn{9}{l}{\textbf{Reduced Subset}}\\
\hline
\addlinespace[0.3em]
\multicolumn{9}{l}{\textit{\dpmb{1}}}\\
\hspace{1em}Total & 41.416 & 48.978 & 28.766 & 40.329 & 43.852 & 46.963 & 30.674 & 35.053\\
\hspace{1em}Generate and fit proxy & 33.165 & 41.653 & 22.547 & 37.838 & 39.309 & 28.038 & 24.104 & 23.145\\
\hspace{1em}Sanitize model parameters & 6.079 & 5.104 & 4.040 & 0.338 & 2.644 & 16.633 & 4.539 & 9.984\\
\hspace{1em}Generate $\sanby$ & 2.137 & 2.186 & 2.155 & 2.130 & 1.874 & 2.122 & 2.009 & 1.908\\
\addlinespace[0.3em]
\multicolumn{9}{l}{\textit{\dpmb{2}}}\\
\hspace{1em}Total & 33.010 & 45.512 & 28.970 & 59.728 & 34.462 & 72.582 & 28.557 & 27.469\\
\hspace{1em}Generate and fit proxy & 27.967 & 41.311 & 24.406 & 36.748 & 29.153 & 38.280 & 25.232 & 23.644\\
\hspace{1em}Sanitize model parameters & 3.023 & 2.188 & 2.517 & 20.881 & 3.255 & 32.409 & 1.096 & 1.896\\
\hspace{1em}Generate $\sanby$ & 1.986 & 1.978 & 2.024 & 2.076 & 2.030 & 1.871 & 2.207 & 1.911\\
\addlinespace[0.3em]
\multicolumn{9}{l}{\textit{\dpmb{9}}}\\
\hspace{1em}Total & 30.956 & 60.853 & 25.478 & 28.481 & 45.448 & 41.336 & 48.481 & 29.008\\
\hspace{1em}Generate and fit proxy & 28.174 & 41.456 & 23.411 & 25.292 & 38.205 & 36.591 & 22.910 & 25.355\\
\hspace{1em}Sanitize model parameters & 0.767 & 17.304 & 0.082 & 1.185 & 5.026 & 2.556 & 23.536 & 1.745\\
\hspace{1em}\hspace{1em}Generate $\sanby$ & 1.980 & 2.060 & 1.962 & 1.981 & 2.193 & 2.166 & 2.012 & 1.890\\
\addlinespace[0.3em]
\hline
\multicolumn{9}{l}{\textbf{Full}}\\
\hline
\addlinespace[0.3em]
\multicolumn{9}{l}{\textit{\dpmb{1}}}\\
\hspace{1em}Total & 134.983 & 188.491 & 140.402 & 118.147 & 121.335 & 209.954 & 115.378 & 111.915\\
\hspace{1em}Generate and fit proxy & 68.040 & 67.214 & 79.116 & 68.464 & 75.997 & 69.083 & 64.321 & 64.428\\
\hspace{1em}Sanitize model parameters & 64.930 & 119.153 & 59.317 & 47.506 & 43.363 & 138.800 & 49.075 & 45.485\\
\hspace{1em}Generate $\sanby$ & 1.926 & 2.038 & 1.914 & 1.987 & 1.921 & 2.014 & 1.928 & 1.959\\
\addlinespace[0.3em]
\multicolumn{9}{l}{\textit{\dpmb{2}}}\\
\hspace{1em}Total & 103.026 & 169.649 & 113.208 & 187.061 & 161.744 & 182.704 & 147.964 & 113.241\\
\hspace{1em}Generate and fit proxy & 67.963 & 81.237 & 64.929 & 76.278 & 79.403 & 78.373 & 77.381 & 63.401\\
\hspace{1em}Sanitize model parameters & 32.957 & 86.399 & 46.194 & 108.684 & 80.129 & 102.396 & 68.532 & 47.850\\
\hspace{1em}Generate $\sanby$ & 2.019 & 1.926 & 2.032 & 2.046 & 2.158 & 1.881 & 1.998 & 1.947\\
\addlinespace[0.3em]
\multicolumn{9}{l}{\textit{\dpmb{9}}}\\
\hspace{1em}Total & 95.475 & 132.798 & 86.823 & 117.154 & 135.931 & 218.693 & 123.830 & 109.342\\
\hspace{1em}Generate and fit proxy & 67.758 & 80.959 & 65.413 & 76.283 & 79.126 & 82.739 & 66.407 & 66.147\\
\hspace{1em}Sanitize model parameters & 25.425 & 49.680 & 19.156 & 38.859 & 54.597 & 133.893 & 55.357 & 41.225\\
\hspace{1em}Generate $\sanby$ & 2.209 & 2.077 & 2.200 & 1.958 & 2.074 & 2.006 & 2.012 & 1.928\\
\bottomrule
\end{tabular}
\caption{Computation time per response variable for each steps of the model-based algorithms.}
\label{tab:liberia-comptimes-byresponse}
\end{table}

%% file: section-discussion.tex



Much of the literature (in economics) publishes replication packages with either weakly protected (de-identified) data (contrary to our stated Goal 2), or withholds the data out of privacy concerns (impeding the ability for others to investigate inference, related to what we called Goal 1). The trigger for the analysis conducted here was the need for privacy protection in the presence of a requirement to publish data underlying an \ac{RCT}, while maintaining reasonably broad usability of the data. We start with a specific focus on economists and social scientists interested in replication of RCTs. We explore the use of one of the simplest DP mechanisms (histogram count perturbation with Laplace noise) for generation of the protected sensitive covariate data. We show that with very careful consideration of the type of data and the problem at hand, the original algorithm needs detailed tweaking, but that our proposed model-agnostic mechanism (MV Histogram) can produce a protected dataset (Goal 2) and that analyses from such data would sufficiently maintain precise inference (Goal 1). Even for low values of the privacy-loss budget (i.e., stronger privacy), we can obtain comparable estimates in our regression models of interest. 

The \longdpmb method, one of the model-informed mechanisms, infuses noise to the estimated model parameters to provide additional confidentiality protections and has comparable utility to the multivariate histogram for larger privacy budgets and when the number of covariates is small for well-specified models like in Section \ref{sec:evaluating}. This method uses privacy-preserving covariates from the MV Histogram method and then generates proxy fitted model coefficients in order to sanitize each model parameter. However, in Section \ref{sec:real-world}, we see when the model does not fit the data well or the model assumptions are not met, the \longdpmb preforms poorly. Additionally, the standard errors of the privacy-preserving treatment effect under the \longdpmb are so small that the $p$-values do not accurately reflect those from the original analysis. Further work should be done to adjust these for the infused noise. 

The other model-informed mechanism, the Hybrid mechanism offers strong privacy protection at conventional levels of $\epsilon$ for covariate data, but does not infuse noise into the estimated model parameters. This yields much better utility than the \longdpmb mechanism. In the Hybrid mechanism, outcomes are imputed based on the privacy-preserving covariates and confidential parameters.
In the real-world experiment, there are between 14 and 62 estimated parameters (point estimates and associated standard errors).
The Hybrid mechanism works quite well to allow for reasonable inferences in both simulations and real-world examples, in general close to the original (unprotected) inferences on population (intent-to-treat) parameters. This is achieved by leveraging the targeted structure, focusing on a small number of parameters of interest. 
The mechanism allows for the release of the protected data and the perturbed parameters as part of a privacy-preserving replication package. 

Publication of replication package is a requirement in an increasing number of social science journals. While we replicated the analysis from BJS, which includes a large number of covariates, there are mixed recommendations among the literature about whether covariates should be included, how they should be incorporated, and how many should be included, especially in clinical trial research  \citep[e.g.,][]{covariate_nonlinear,covariate_increasepower,covariate_interactions,EU_covariate_guidelines}. Our algorithms work best with a small number of covariates which aligns with \cite{EU_covariate_guidelines} guidelines on the analysis of clinical trial data. However, the flexibility of our model allows for a variety of model structures. If interaction terms between treatments and covariates are included in the model as recommended by \cite{covariate_blocking} and \cite{covariate_interactions}, only a small adjustment to the model-informed algorithms is needed. In this case, the treatments should be sanitized with the covariates in the MV Histogram stage, instead of randomly assigned after the fact. 

Some caveats apply. Firstly, both model-informed methods assume the model used to sanitize the data accurately portrays the underlying relationships between the variables. When the model fits poorly or the model assumptions are not met, this can lead to less utility in the privacy-preserving data. The \longdpmb is especially sensitive to violations of the normality assumption of the residuals since the theory behind the sanitization of the model parameters relies heavily on this assumption. When fitting the model used to produce the privacy-preserving data on the privacy-preserving output of the the Hybrid or \longdpmb mechanisms, the model assumptions will always be met by the fitted privacy-preserving regression. Thus those using the privacy-preserving data from a replication package must trust that model fit to the original data met the model assumptions. They will not be able to replicate checking those model assumptions.  Additionally, the Hybrid mechanism can be extended to additional types of regression such as logistic regression and Poisson regression since it does not rely on the normality assumptions that the \longdpmb mechanism does. 

Additionally, some major assumptions have been made in our implementation of the MV Histogram (Algorithm \ref{algo:perturbMVhist}). Since the  discretization used in Algorithm \ref{algo:perturbMVhist} uses the observed range of the continuous variables, the range is not strictly protected under DP. We have treated the range of the observed values as not sensitive. When $\delta=0$, what combinations of categorical and discretized continuous values appear in the observed data is also treated as not sensitive. This is because Algorithm \ref{algo:perturbMVhist} only adds noise to the proportions of unique values rows observed in the original dataset, rather than all possible combinations of the categorical and discretized continuous values. Thus the MV Histogram and by extension the \longdpmb method do not protect against all possible attacks. 
For an algorithm, $M$, to satisfy approximate differential privacy, $P(M(\data_1)=\dpout)\leq e^{\epsilon}P(M(\data_2)=\dpout)$ (Equation \ref{eqn:DPdefn}) for \emph{all} possible outputs, $\dpout$, and for \emph{all} pairs of datasets, $\data_1$,$\data_2$ that differ on on record (in this case a row of participant information). Consider the case where there are 12 possible combinations of covariates and $\data_1$ only has 10 of those unique combinations of covariates in its rows. If the output of the MV Histogram has 11 unique covariate combinations, an adversary knows it did not come from $\data_1$. The MV Histogram algorithm satisfies Equation \ref{eqn:DPdefn} for \emph{some} outputs $\dpout$ and \emph{some} pairs of datasets $\data_1$ and $\data_2$ given the following three conditions are met.
First, we assume the bounds for the discretization of continuous variables as not sensitive. Second, $\delta$ must be greater than the ratio of two over the number of unique possible combinations of column values. This informs the choice of the binning parameter, $\zeta$. The third condition requires at least one privacy-preserving proportion to be above the initial threshold of $2\frac{\log(2/\delta)}{n\epsilon}+\frac{1}{n}$. This is  dependent on parameters we set for the algorithm ($\epsilon$, $\delta$, $\zeta$), on the properties inherent to the study design and participant pool (e.g., $n$ and the distribution of participants across the covariate combinations), and the specific random draw when implementing the algorithm. With these three conditions, the MV Histogram and the model-informed algorithms satisfy the inequality that characterizes approximate DP, $P(M(D_1)=\dpout)\leq e^\epsilon P(M(D_2)=\dpout)+\delta$. 
However, since it only holds for specific pairs of neighboring datasets and depends on the randomly drawn noise of the mechanism, these methods do not strictly satisfy approximate DP. 

Finally, how useful is having a privacy-protected replication package to address broader questions posed by other researchers? Does the privacy-protected replication package allow for robustness tests, for instance, through other specifications? The currently proposed mechanisms are tightly bound to the original authors' proposed specifications, and may not perform well when other researchers apply non-congenial specifications. In such cases, access to the original data and additional privacy budget may be necessary in order to create additional protected datasets that allow for such investigation. Transparency of sanitization, however, is increased, as the privacy-preserving mechanism would be released as part of the replication package. This could in turn better support adjustments of inference due to additional randomness present, and allowing for replication in other contexts, including informing new data collections.  


We demonstrate that reasonably simple methods with strong privacy guarantees could become   “out-of-the-box” methods. 
The ideas and results reported here are the first step towards better understanding of feasible privacy-preservation of \acp{RCT}-based data, ensuring that confidentiality of data contributors to \acp{RCT}, often from \ac{LMIC}, will be more strongly protected while maintaining the ability of researchers to draw meaningful inferences. While policy-oriented stakeholders are primarily interested in the latter, citizens that contribute their data to \acp{RCT} and companies, such as fin-tech providers, that provide key data to researchers are also heavily invested in protecting privacy. Consumer and citizen protection agencies, ethic review boards, and other regulators, should be interested in knowing of the existence of privacy-enhancing methods, possibly facilitating approval of studies in the presence of strong privacy guarantees.

%% file: acknowledgements.tex
\section{Acknowledgement and Disclosure of funding}
This work was supported through Digital Credit Observatory (CEGA, University of California, Berkeley/Bill and Melinda Gates
Foundation (MP)), and in-part by NSF awards SES-1853209 and CNS-1702760 to Penn State University. We would like to thank Luqi Emanuele and Shuhang Lou for their contributions towards this work and its extensions. We declare no known conflicts of interest.

%% file: section-appendix-syntheticdata.tex
Two algorithms proposed by \citet{karwa2017finite} are used for our DP Model-based Method (Algorithm \ref{algo:fullmodelbased}). 

In text, we refer to the first as the DP Variance Algorithm\footnote{The only adjustment from the original algorithm by \citet{karwa2017finite} that we make is we square the privacy-preserving output to get the variance rather than the standard deviation.} which uses $\epsilon_{var}$ and $\delta_{var}$ privacy parameters as well as $\sigma_{min}$ and $\sigma_{max}$ bound parameters to produce a privacy-preserving estimate of the standard deviation of some input vector. In our experiments and real data example, $\sigma_{min}$ and $\sigma_{max}$ are conservatively set as $2^{-15}$ and $2^{15}$.

For a vector of observations from a Gaussian distributed with the variance bounded by $(\sigma_{min}^2,\sigma_{max}^2)$ where $\sigma_{max}>\sigma_{min}>0$, \cite{karwa2017finite} propose an Algorithm that satifies $(\epsilon,\delta)$-DP to get a privacy-preserving estimate of the variance. This algorithm shown in Algorithm \ref{algo:dpvar} uses a perturbed histogram algorithm in lines 2 to 10. This is the same structure as Algorithm \ref{algo:perturbMVhist}, but the bins are specified by $B_j$. 

\begin{algorithm}
    \caption{\textit{(DP Variance Algorithm)} Differentially Private Estimate of Variance Algorithm is proposed by \cite{karwa2017finite} (Algorithm 2 in paper).}\label{algo:dpvar}
    \KwData{$\bx=(x_1 \dots x_n)^T$ such that $x_i\overset{iid}{\sim}N(\mu, \sigma^2)$ and $\sigma\in (\sigma_{min},\sigma_{\max})$}
    \KwIn{$\epsilon>0,\delta>0,\sigma_{min}>0,\sigma_{max}>\sigma_{min}$}
    $y_i=x_{2i}-x_{2i-1}$ for $i=1,\ldots, \lfloor n/2 \rfloor$\;
    $B_j\gets \left(2^j,2^{j+1}\right]$ for $j=j_{min},\ldots, j_{max}$ where $j_{min}=\lfloor \log_2(\sigma_{min})-2\rfloor$ and $j_{max}=\lceil \log_2(\sigma_{max})+1\rceil$\;
    $\eta=\lfloor n/2\rfloor$\;
    \If{$\delta>0$ and $(j_{max}-j_{min}+1)>2/\delta$}{
    $c=\frac{2\log(2/\delta)+\epsilon}{\eta\epsilon}$
    }\Else{
    $c=0$
    }
    \ForEach{$j=j_{min},\ldots,j_{max}$}{
    $p_j=\sum_{i=1}^{\eta}\indic\{|x_{2i}-x_{2i-1}|\in B_j\}/\eta\lfloor$\;
    $\widetilde{p_j}\gets \left(p_j+Z_j\right)\indic\{p_j>\max(0,c-Z_j)\}$ where $Z_j\sim $Laplace($0,2/(\eta\epsilon)$)\;
    }
    $\hat{\ell}=\argmax_{j=j_{min},\ldots, j_{max}}\widetilde{p_j}$\;
    \KwResult{$\widetilde{\sigma^2}\gets 2^{2\hat{\ell}+2}$}
\end{algorithm}

The next algorithm used is referred to in text as the DP Range Algorithm which satisfies $\epsilon,\delta$-DP and uses a parameter, $\bdmean$, to bound the mean of a input vector
For a vector of observations from a Gaussian distributed with the mean, $\mu$, bounded by $\bdmean>|\mu|$, \cite{karwa2017finite} propose an algorithm that satisfies $(\epsilon,\delta)$-DP to get a privacy-preserving range which contains all the observations with confidence $1-\alpha$. This algorithm is show in Algorithm \ref{algo:dpvar} and uses a perturbed histogram algorithm in lines 2 to 10. 
A privacy-preserving estimate of $\sigma^2$ can be used instead of $\sigma$ to get a privacy-preserving approximate range. If Algorithm \ref{algo:dpvar} satisfies $(\epsilon_{var},\delta_{var})$-DP and Algorithm \ref{algo:dprange} satisfies $(\epsilon_{range},\delta_{range})$-DP, then combining the two algorithms satisfies $(\epsilon_{var}+\epsilon_{range},\delta_{var}+\delta_{range})$-DP.

\begin{algorithm}
    \caption{Differentially Private estimate of range with known variance is proposed by \cite{karwa2017finite} (Algorithm 1 in paper). A privacy-preserving estimate of $\sigma$ can also be used in the algorithm. }\label{algo:dprange}
    \KwData{$\bx=(x_1 \dots x_n)^T$ such that $x_i\overset{iid}{\sim}N(\mu, \sigma^2)$ and $|\mu|<\bdmean$}
    \KwIn{$\epsilon>0,\delta>0,\alpha\in (0,1), \sigma>0,\bdmean>0$}
    $r=\lceil \bdmean/\sigma\rceil$\;
    \If{$\delta>0$ and $(2r+1)>2/\delta$}{
    $c=\frac{2\log(2/\delta)+\epsilon}{n\epsilon}$
    }\Else{
    $c=0$
    }
    $B_j\gets \left(\sigma(j-\frac{1}{2}),\sigma(j+\frac{1}{2})\right]$ for $j=-r,\ldots,r$\;
    \ForEach{$j=-r,\ldots,r$}{
    $p_j=\sum_{i=1}^n\indic\{x_{i}\in B_j\}/n$\;
    $\widetilde{p_j}\gets \left(p_j+Z_j\right)\indic\{p_j>\max(0,c-Z_j)\}$ where $Z_j\sim $Laplace($0,2/(n\epsilon)$)\;
    }
    $\hat{\ell}=\argmax_{j=-r,\ldots,t}\widetilde{p_j}$\;
    $\kappa=4\sigma\sqrt{\log(n/\alpha)}$\;
    \KwResult{$(x_{min},x_{max})$ where $x_{min}=\sigma\hat{\ell}-\kappa$ and $x_{max}=\sigma\hat{\ell}+\kappa$}
\end{algorithm}

%% file: section-appendix-real-world.tex
\subsection{Variables}
The treatment indicators are described in the main text and are denoted as \texttt{cashassonly}, \texttt{tpassonly}, \texttt{tpcashass}, \texttt{control}. Similarly, the blocking variables are \texttt{cg\_strata} with 20 levels, and \texttt{tp\_strata\_alt} with 55 levels.  The 55 additional covariates are described in Tables \ref{tab:bincovariates} and \ref{tab:covariatevars-nonbinary}.

\input{tables-and-figures-appendix/table-covariates}

\subsection{Replicating Original Analysis}

The original study by BJS is performed using Stata, and the replication files are publicly available \citep{ReducecrimePkg}.  Before we test our proposed algorithms, we replicate the results of the 12-13 month section of Table 2 by \citet{Reducecrime}. Our replication achieve the same intent-to-treat (ITT) and standard error values as BJS (Table \ref{tab:replicate2b}). We use a Bonferroni correction instead of a Westfall-Young correction to adjust the p-values. 

\input{tables-and-figures-appendix/liberia_replicate2b}

The model used to replicate the results of BJS treats all the baseline covariates in Tables \ref{tab:bincovariates} and \ref{tab:covariatevars-nonbinary} as numeric. The blocking variables, \texttt{cg\_strata} and \texttt{tp\_strata\_alt}, are treated as categorical variables. The eight response variables in Table \ref{tab:response-vars} are also all treated as continuous. Let $y_i$ be the response value for observation $i$. The model for observation $i=1,\ldots,n$ is of the form:
\begin{align*}
    y_{i}=&\gamma_0+\tau_1\indic\{\texttt{tpassonly}_i==1\}+\tau_2\indic\{\texttt{cashassonly}_i==1\}+\tau_3\indic\{\texttt{tpcashass}_i==1\}\\
    &+\gamma_1\texttt{drugsellever\_b}_i+\cdots [\text{Table \ref{tab:bincovariates} variables}]\cdots+\gamma_{15}\texttt{harddrugsdailyuser\_b}_i\\
    &+\gamma_{16}\texttt{age\_b}_i+\gamma_{17}\texttt{ashostil\_b}_i+\cdots [\text{Table \ref{tab:covariatevars-nonbinary} variables}]\cdots+\gamma_{45}\texttt{risk\_game\_resc\_b}_i\\
    &+\gamma_{46}\indic\{\texttt{cg\_strata}_i==2\}+\gamma_{47}\indic\{\texttt{cg\_strata}_i==3\}+\cdots +\gamma_{64}\indic\{\texttt{cg\_strata}_i=20\}\\
    &+\gamma_{65}\indic\{\texttt{tp\_strata\_alt}_i=2\}+\cdots +\gamma_{118}\indic\{\texttt{tp\_strata\_alt}_i=55\}\\
    &+e_i, \hspace{5em}\text{ where }e_i\overset{iid}{\sim}N(0,\sigma^2).
\end{align*}
The intent to treat (ITT) estimates in Table \ref{tab:replicate2b} are the estimated coefficients, $\hat{\tau_1},\hat{\tau_2},$ and $\hat{\tau_3}$. We refer to these as the estimated treatment effects in our discussions. The control mean reported in Table \ref{tab:replicate2b} is the mean of the response variable for the control group. This value is not model dependent. The standard errors presented are estimated using type "HC1"  Heteroskedasticity-Consistent Covariance Matrix Estimator proposed by \citet{MACKINNON1985305}. This is the default robust standard error in Stata \citep{scottervin2000}.

\begin{figure}[!htb]
    \centering
    \includegraphics[width=0.99\linewidth]{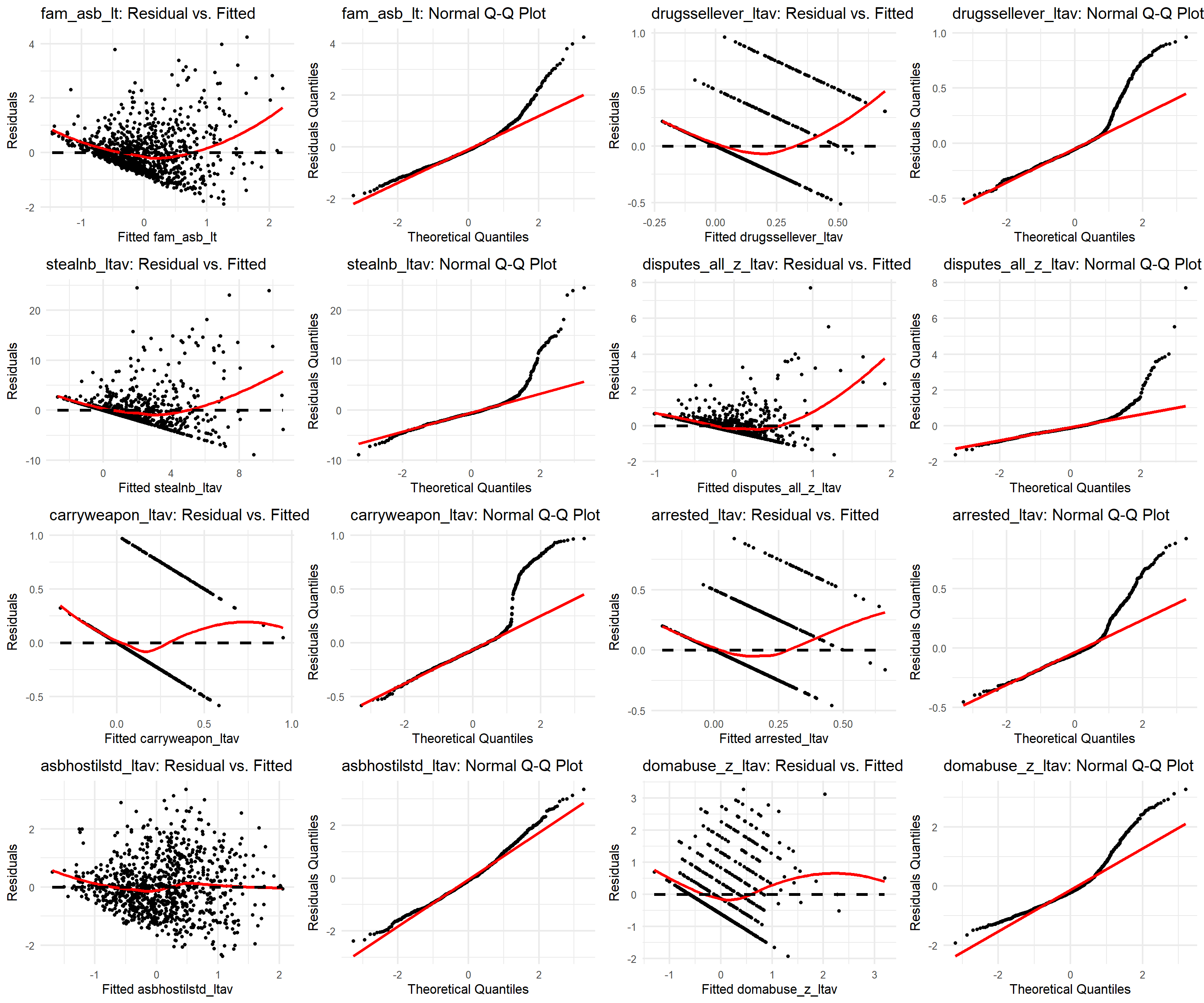}
    \caption{For each of the eight response variables, we check the assumptions on the residuals. For the residuals v.s. fitted plots, the red line is the smoothed fit of the residual points. This should closely follow the dashed line at $0$ to make the assumption that residuals have mean $0$ reasonable. Additionally, the spread of the points should not change drastically over the fitted values to make the assumption of constant error variance reasonable.  For the Normal Q-Q plots the residual points should closely follow the red line to make the assumption that the residuals are normally distributed reasonable. Only \texttt{asbhostilstd} in the bottom left two plots appears to meet these assumptions.}
    \label{fig:2bdiagnostics}
\end{figure}

In order for model assumptions to met in a multiple linear regression model. The residuals should be approximately normal distributed with mean 0 and constant variance. We use plots of the fitted values against the residual values and a quantile-quantile plot for normality of the residuals to check these assumptions (Figure \ref{fig:2bdiagnostics}).We find that the assumptions for a multiple linear regression model do not seem reasonable for seven out of eight response variables. Only the standardized aggressive behavior index (\texttt{asbhostilstd}) appears to have normally distributed residuals with mean $0$ and constant variance. Many other response variables have residuals which deviate from the normal distribution at the upper tail. The variance of the residuals appears to increase with the fitted value for several of the models as well.

The fit of these models may be improved if the baseline covariate variable about how often the respondent sees their family (\texttt{famseeoften\_b}) is treated as a categorical or ordinal variable. Additionally, the average indicator variable responses (\texttt{drugsellever\_ltav}, \texttt{carryweapon\_ltav}, and \texttt{arrested\_ltav}) could benefit from being fit as a logistic model. The average count of stealing activities in the last 2 weeks (\texttt{stealnb\_ltav}) might need to be used in a Poisson regression model. Predictor variable selection techniques, the addition of interaction terms, and variable transformations may also need to be included to improve the model fits. Since we are interested in sanitization algorithms that can get similar analysis results to the analysis preformed on the confidential data, we do not attempt to improve the model fits in this paper. This could be an area of further research.

\subsection{Sanitization of Original Data and Replication of Analysis}\label{sec:appendix-sanitize}

Although we are able to replicate the analysis of BJS, when we compare our sanitizing methods to original data we remove one observation from the original dataset. This observation reported a noninteger value of number of women the respondent is financially supporting. Table \ref{tab:apdx-itt-orig} reports the treatment effect (ITT), Standard Error, and Adjusted P-values for the models of the full baseline covariates, the subset of covariates, and the reduced subset of covariates with the 947 observations of the original data. For each algorithm and privacy budget considered, we have produced a similar table which now include confidence interval overlap metrics (see MV histogram in Table \ref{tab:apdx-itt-mvhistogram}; Hybrid in Table \ref{tab:apdx-itt-hybrid}; and the DP Model-based Algorithm with $\epsilon=1$ in Table \ref{tab:apdx-itt-dpmb1}, $\epsilon=2$ in Table \ref{tab:apdx-itt-dpmb2}, and $\epsilon=9$ in Table \ref{tab:apdx-itt-dpmb2}).





We believe improvement to the sanitization of the standard error of the residuals could make this algorithm feasible. Further work can be done in this area. However, due to this scale of error between the original estimates and the Fully DP Model-based estimates, we do not report the results for this model for all the covariate sets.

\begin{landscape}
    \input{tables-and-figures-appendix/liberia_allcovsets_confidential}
\end{landscape}

\begin{landscape}
    \input{tables-and-figures-appendix/liberia_allcovsets_sanitized_mvhistogram}
\end{landscape}

\begin{landscape}
    \input{tables-and-figures-appendix/liberia_allcovsets_sanitized_hybrid}
\end{landscape}

\begin{landscape}
    \input{tables-and-figures-appendix/liberia_allcovsets_sanitized_dpmb1}
\end{landscape}

\begin{landscape}
    \input{tables-and-figures-appendix/liberia_allcovsets_sanitized_dpmb2}
\end{landscape}


\begin{landscape}
\include{tables-main/liberia_itt_fam_asb_lt.tex}
\end{landscape}

\begin{landscape}
  \include{tables-main/liberia_itt_asbhostilstd_ltav.tex}  
\end{landscape}

\begin{figure}[!htb]
    \centering
    \includegraphics[width=0.9\linewidth]{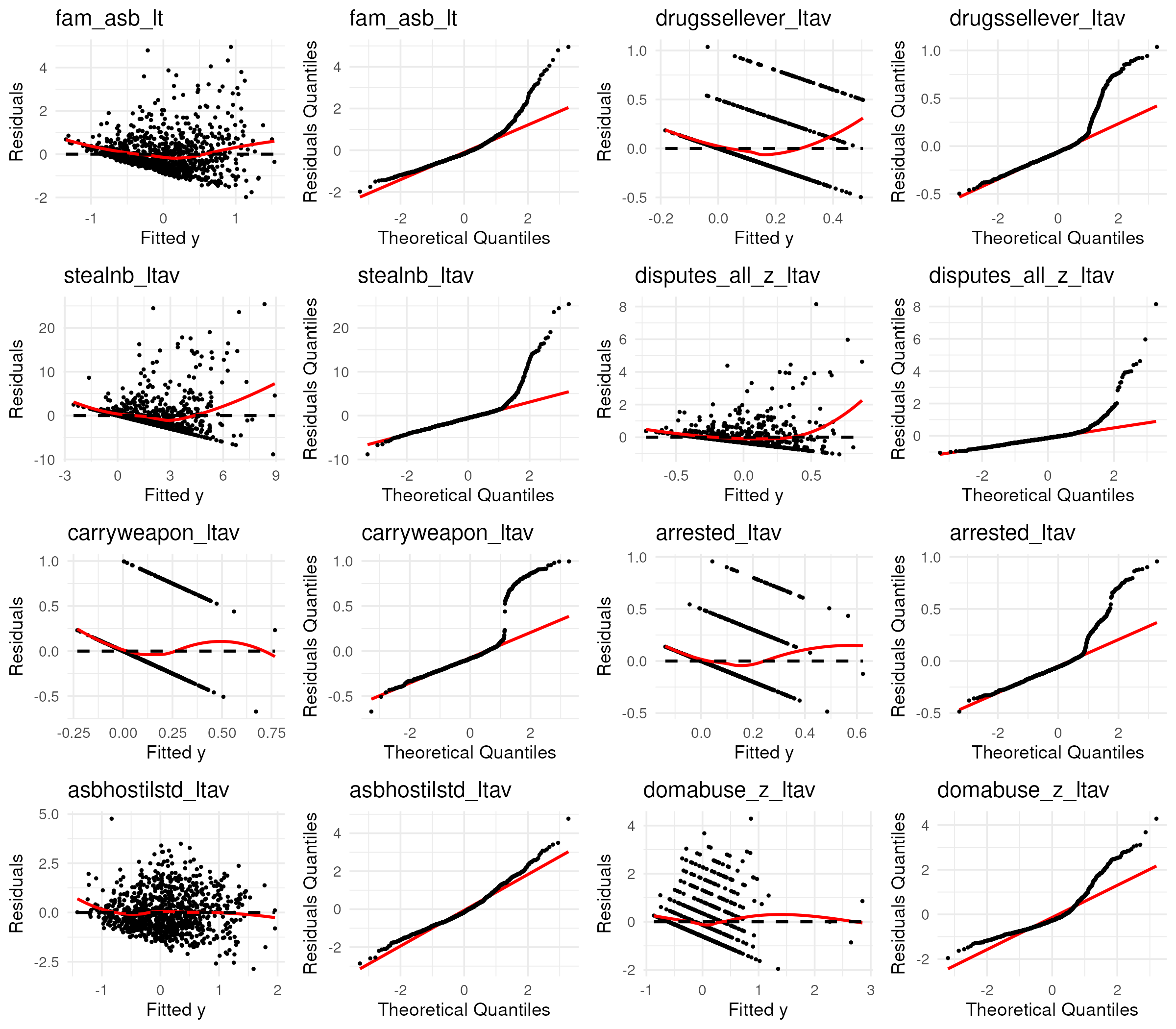}
    \caption{With the reduced subset of covariates, the original data has similar diagnostic plots to those produced using the full baseline covariates}
    \label{fig:apdx-diag-red-conf}
\end{figure}

\begin{figure}[!htb]
    \centering
\includegraphics[width=0.9\linewidth]{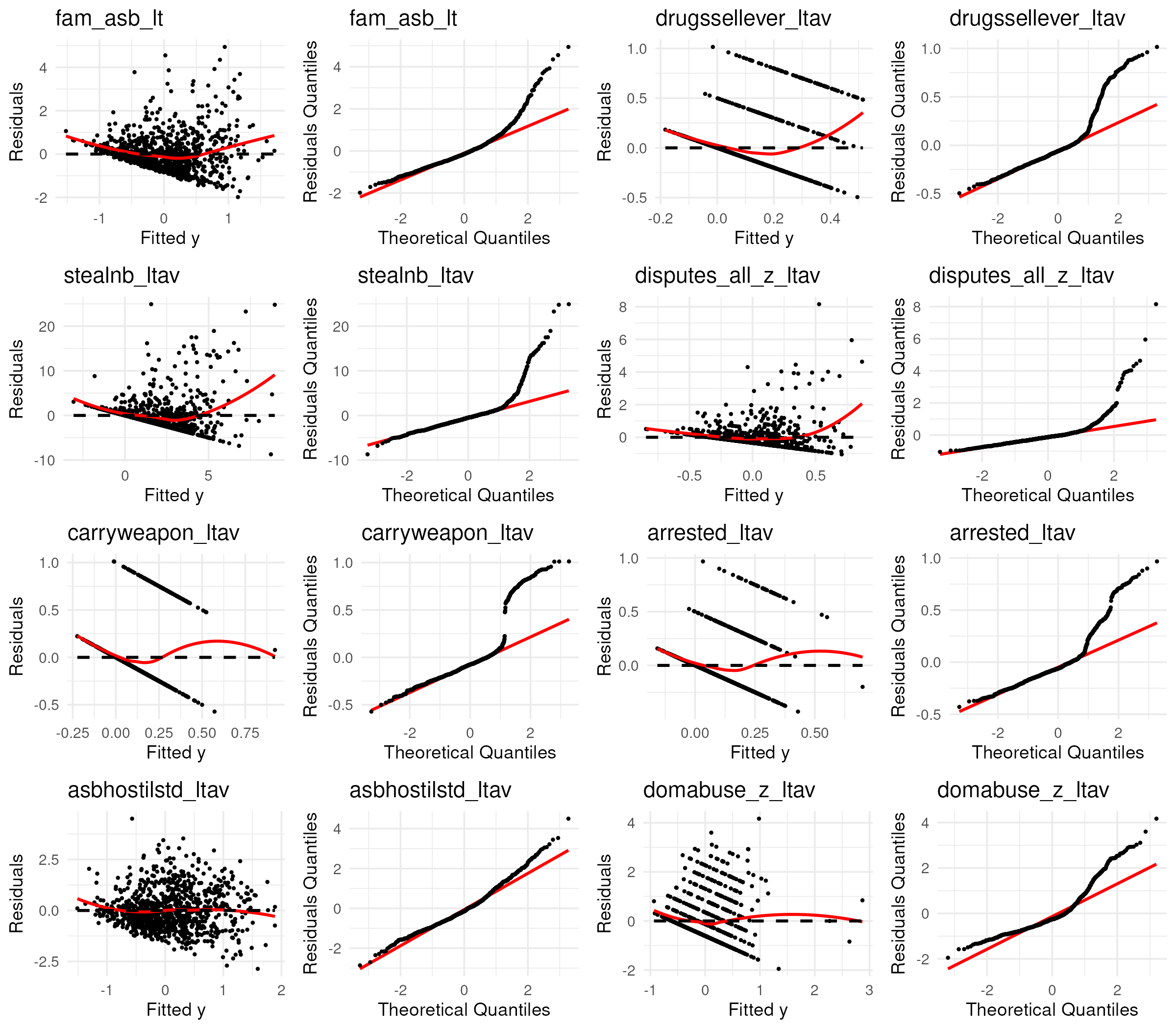}
    \caption{With the subset of covariates, the original data has similar diagnostic plots to those produced using the full baseline covariates and the reduced subset of covariates}
    \label{fig:apdx-diag-sub-conf}
\end{figure}

\begin{figure}[!htb]
    \centering
\includegraphics[width=0.9\linewidth]{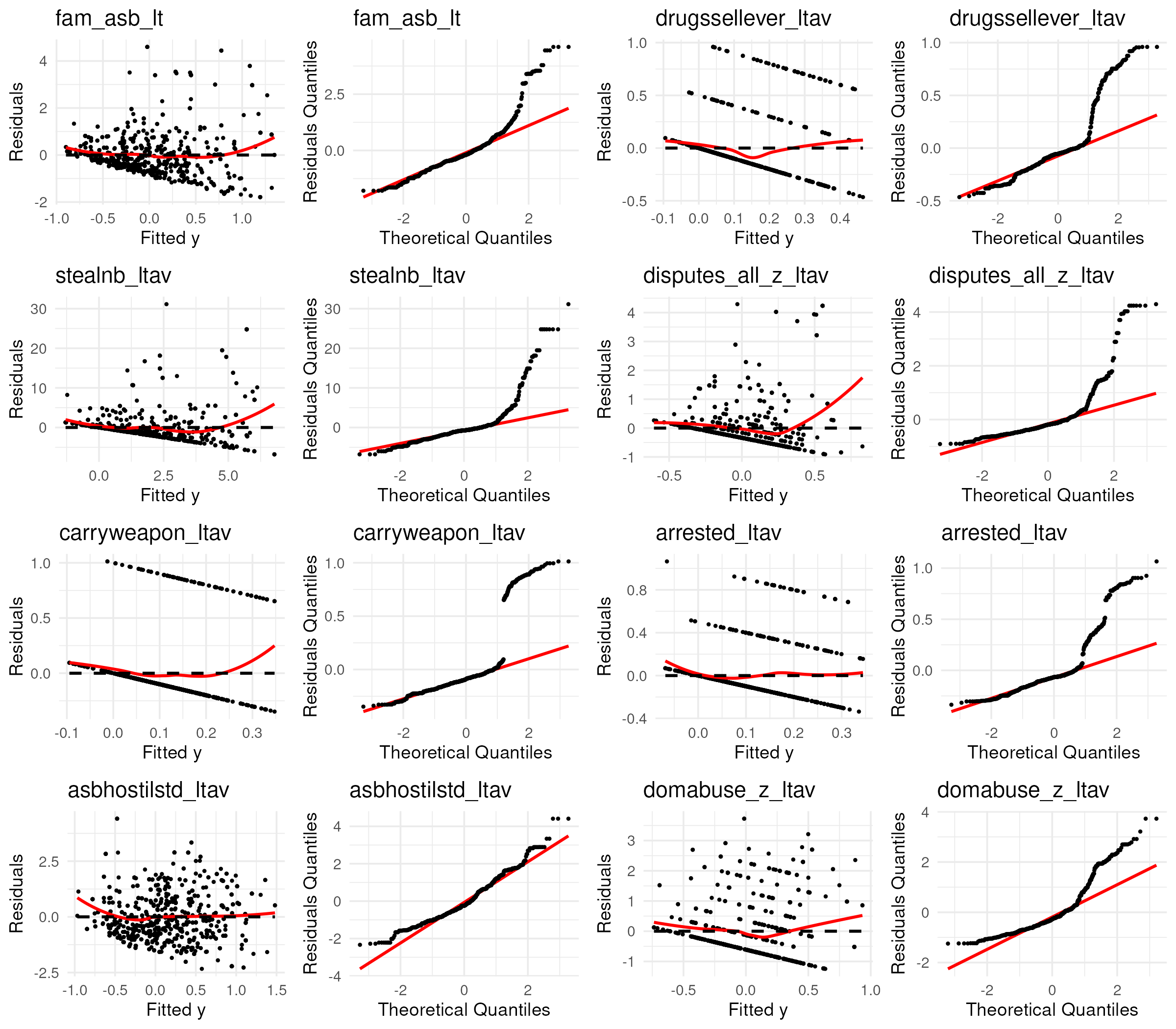}
    \caption{With the reduced subset of covariates, the synthetic data from the MV Histogram has similar diagnostic plots to those produced using the original data. On the residual v. fitted value plots, the range of the fitted values mirrors that of the models from the original data.}
    \label{fig:apdx-diag-ref-mvh}
\end{figure}

\begin{figure}[!htb]
    \centering
    \includegraphics[width=0.9\linewidth]{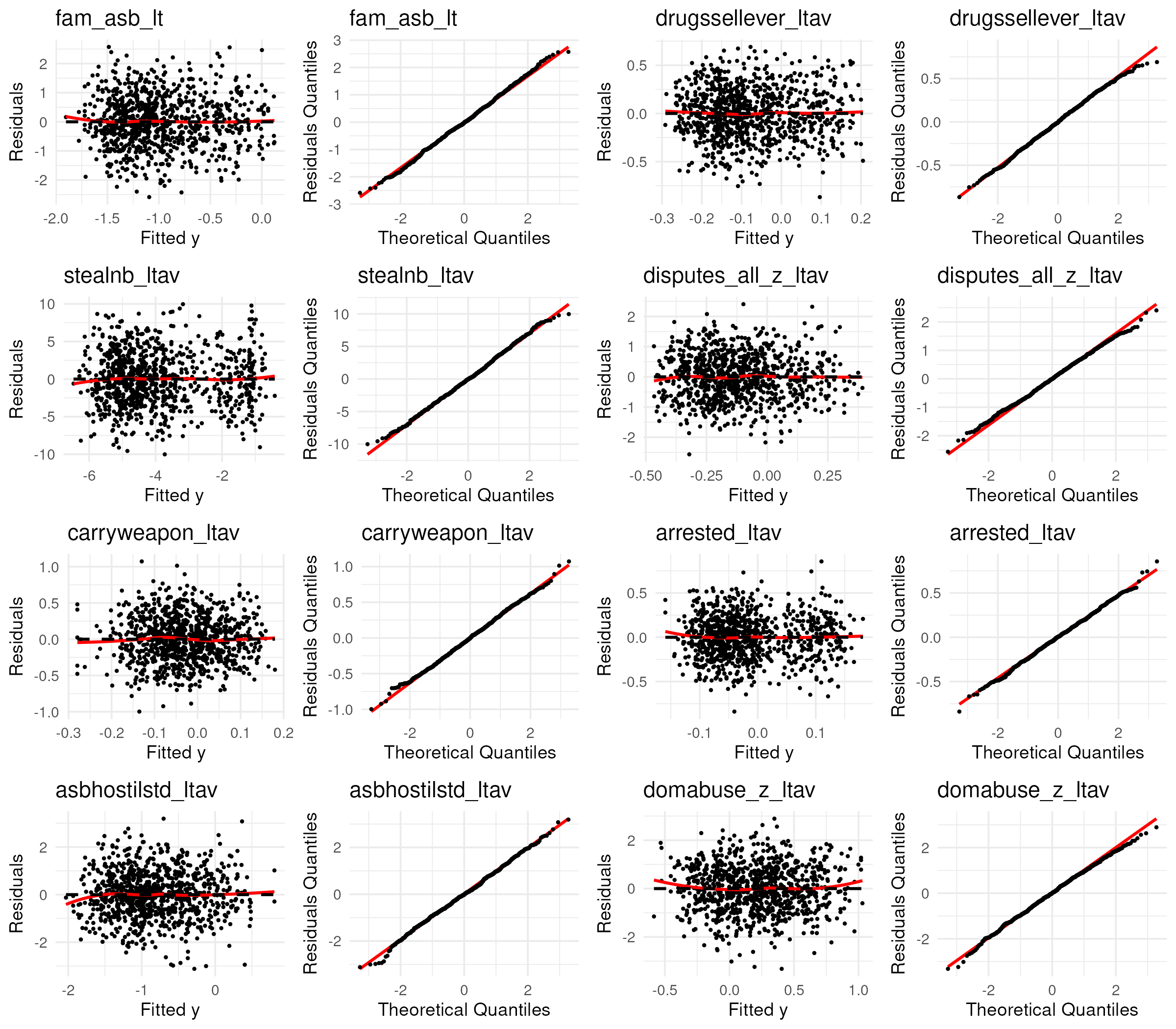}
    \caption{With the reduced subset of covariates, the synthetic data from the Hybrid method does not have similar diagnostic plots to those produced using the original data. In fact, these plots suggest that the regression assumptions are reasonable.}
    \label{fig:apdx-diag-ref-hybrid}
\end{figure}

\begin{figure}[!htb]
    \centering
    \includegraphics[width=0.9\linewidth]{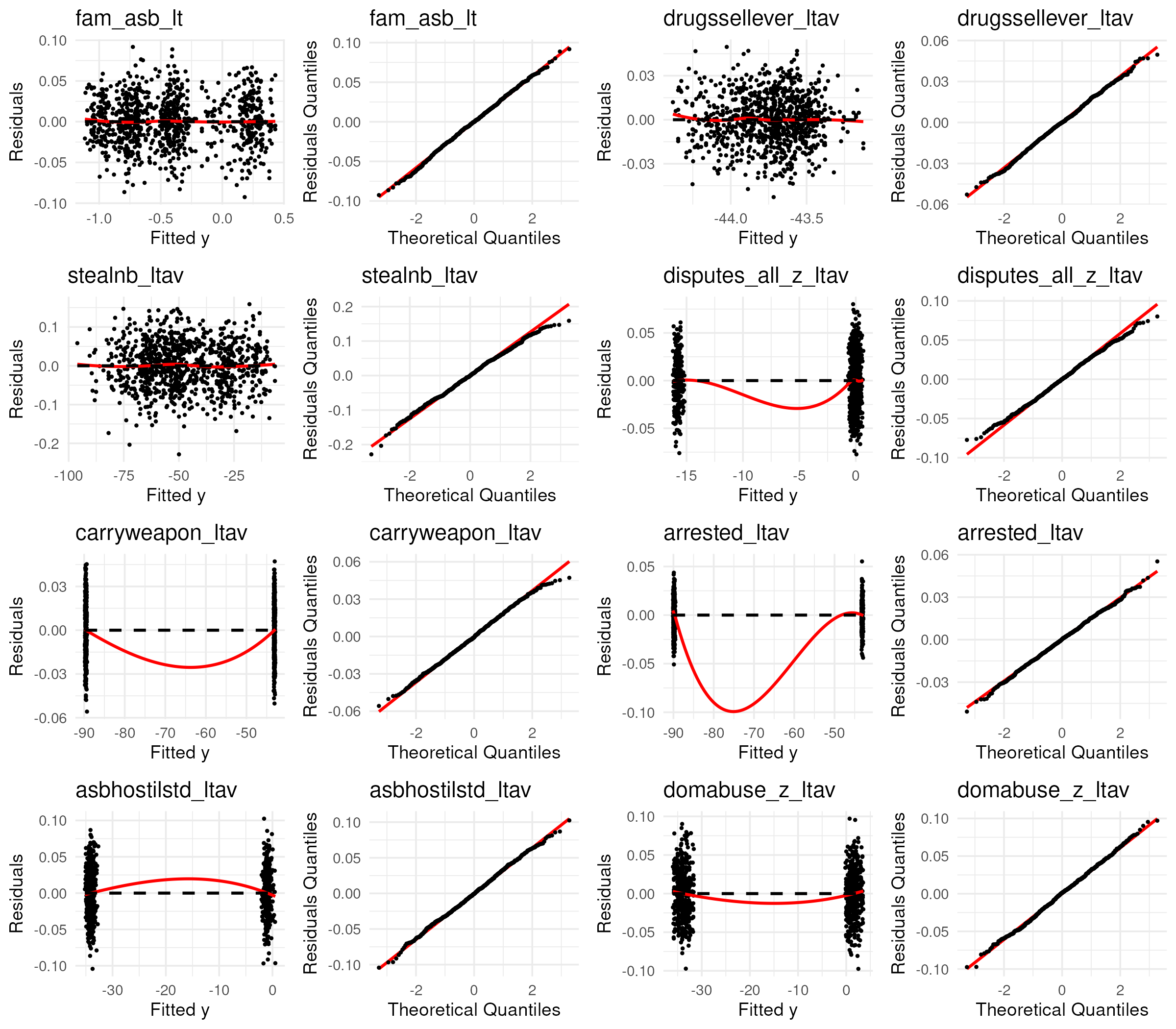}
    \caption{With the reduced subset of covariates, the synthetic data from the DP-Mb-(9) method does not have similar diagnostic plots to those produced using the original data. In fact, these plots suggest that the regression assumptions are reasonable.}
    \label{fig:apdx-diag-ref-dpmb}
\end{figure}

\begin{figure}[!htb]
    \centering
    \includegraphics[width=0.9\linewidth]{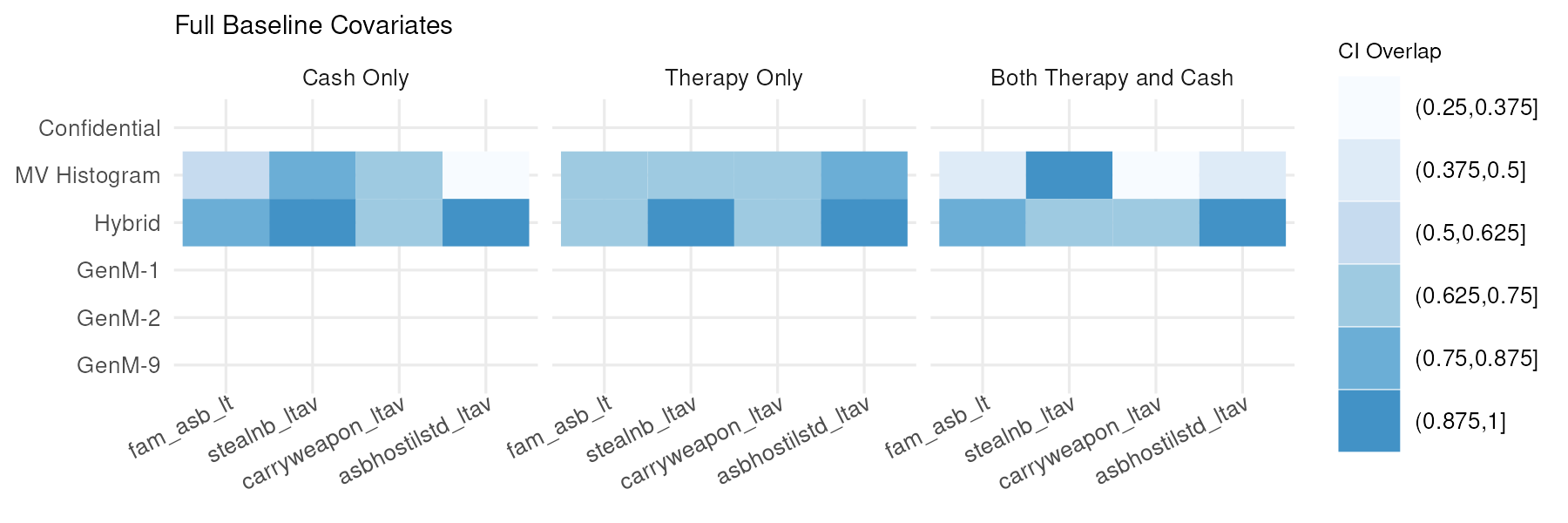}
    \caption{Confidence interval overlap for subset and reduced subset.}
    \label{fig:liberia-ciover-full}
\end{figure}

\begin{figure}[!htb]
    \centering
    \includegraphics[width=0.9\linewidth]{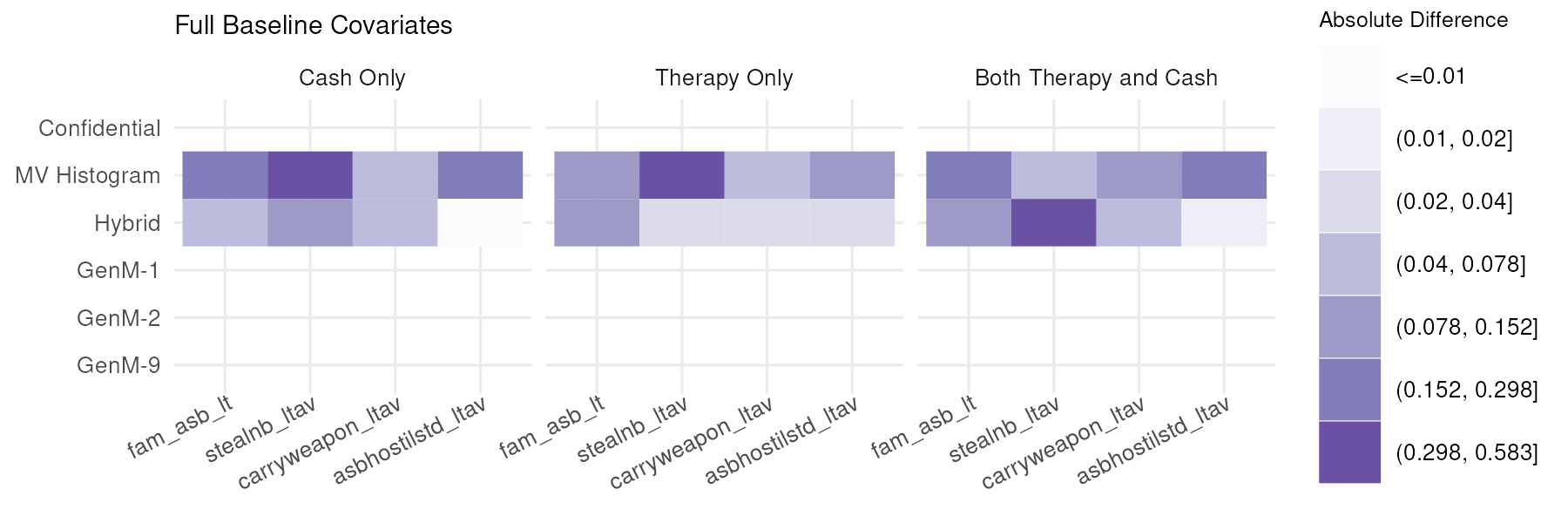}
    \caption{Absolute difference between privacy-preserving and confidential treatment effect for subset and reduced subset.}
    \label{fig:liberia-absdiff-full}
\end{figure}

%% file: tables-and-figures-appendix/table-covariates.tex
\begin{table}[!htb]
\centering
\begin{tabular}{>{\raggedright\arraybackslash}p{0.23\textwidth} >{\raggedright\arraybackslash}p{0.43\textwidth}}
\toprule
\multicolumn{2}{c}{Binary Covariates}\\
Variable Name & Indicator for...\\
\midrule
\addlinespace[0.3em]
\cellcolor{redsubsetcol}{\textcolor{black}{drugssellever\_b}} & Usually sells drugs\\
\cellcolor{redsubsetcol}{\textcolor{black}{drinkboozeself\_b}} & Drinks liquor, beer or palm wine\\
\cellcolor{redsubsetcol}{\textcolor{black}{druggrassself\_b}} & If you can smoke opium/grass\\
\cellcolor{redsubsetcol}{\textcolor{black}{harddrugsever\_b}} & Usually takes hard drugs\\
\cellcolor{redsubsetcol}{\textcolor{black}{steals\_b}} & Any stealing in past 2 weeks\\ 
livepartner\_b & Married or living with a partner\\
muslim\_b & Muslim\\
schoolbasin\_b & Currently in school\\
disabled\_b & Any serious disabilities\\
faction\_b & Part of army or faction since DOE\\ 
homeless\_b & Sleeping on the streets now\\
loan50\_b & Could get \$50 USD within a month\\
loan300\_b & Could get \$300 USD loan within a month\\
grassdailyuser\_b & Smokes grass daily\\
harddrugsdailyuser\_b & Take hard drugs daily\\
\bottomrule
\end{tabular}
\caption{There are 15 binary covariate variables in the full set of baselines variables used by \citet{Reducecrime} in their regression models. Five of these (in blue) are in our subset and reduced subset of the covariates covariates.}\label{tab:bincovariates}
\end{table}

\begin{table}[!htb]
\begin{tabular}{>{\raggedright\arraybackslash}p{0.25\textwidth} >{\raggedright\arraybackslash}p{0.72\textwidth}}
\toprule
Variable Name & Description\\
\midrule
\addlinespace[0.3em]
\cellcolor{redsubsetcol}{\textcolor{black}{age\_b}} & Age\\
\cellcolor{redsubsetcol}{\textcolor{black}{asbhostil\_b}} & Standardized antisocial behavior index\\
\cellcolor{subsetcol}{\textcolor{black}{school\_b}} & Years of schooling\\
\cellcolor{subsetcol}{\textcolor{black}{wealth\_indexstd\_b}} & Standardized index of wealth - housing quality and assets\\
\cellcolor{subsetcol}{\textcolor{black}{cognitive\_score\_b}} & Standardized index of 4 cognitive function measures\\ 
\cellcolor{white}{\textcolor{black}{mpartners\_b}} & Count of women you are supporting\\
\cellcolor{white}{\textcolor{black}{hhunder15\_b}} & Count of household children under age of 15\\
\cellcolor{white}{\textcolor{black}{illicit7da\_zero\_b}} & Average weekly illicit work hours\\
\cellcolor{white}{\textcolor{black}{agricul7da\_zero\_b}} & Average weekly agricultural work hours\\
\cellcolor{white}{\textcolor{black}{nonagwage7da\_zero\_b}} & Average weekly hours, non-agricultural low-skill wage labor\\ 
\cellcolor{white}{\textcolor{black}{allbiz7da\_zero\_b}} & All weekly hours, business for self\\
\cellcolor{white}{\textcolor{black}{nonaghigh7da\_zero\_b}} & Average weekly hours, non-agricultural high-skill work\\
\cellcolor{white}{\textcolor{black}{agriculeveramt\_b}} & Years of experience in agricultural\\
\cellcolor{white}{\textcolor{black}{nonagbizeveramt\_b}} & Years of experience in non-agricultural low-skill business\\
\cellcolor{white}{\textcolor{black}{nonaghigheveramt\_b}} & Years of experience in non-agricultural high-skill work\\ 
\cellcolor{white}{\textcolor{black}{stealnb\_nonviol\_b}} & Count of non-violent stealing activities in past 4 weeks\\
\cellcolor{white}{\textcolor{black}{stealnb\_felony\_b}} & Count of felony stealing activities in 4 weeks\\
\cellcolor{white}{\textcolor{black}{disputes\_all\_b}} & Index of all disputes and fights in the past 2 weeks\footnote{Disputes in the last 2 week are on a scale 0-9. However, this variable takes values between 0 and 30 with non-integer values.}\\
\cellcolor{white}{\textcolor{black}{timedecl\_b}} & Declared patience\\
\cellcolor{white}{\textcolor{black}{riskdecl\_b}} & Declared risk appetite\\ 
\cellcolor{white}{\textcolor{black}{ef\_score\_b}} & Standardized index of executive function\\
\cellcolor{white}{\textcolor{black}{famseeoften\_b}} & Categorical: How often do you see members of your family? (4 levels)\\
\cellcolor{white}{\textcolor{black}{literacy\_b}} & Literacy index (0-2)\\
\cellcolor{white}{\textcolor{black}{mathscore\_b}} & Math index (0-6)\\
\cellcolor{white}{\textcolor{black}{health\_resc\_b}} & Index of health (difficulty with daily life activities) (0-6)\\ 
\cellcolor{white}{\textcolor{black}{depression\_b}} & Depression index (0-17)\\
\cellcolor{white}{\textcolor{black}{distress\_b}} & Local Distress index (0-21)\\
\cellcolor{white}{\textcolor{black}{rel\_commanders\_b}} & Index of close relationships with commanders (0-4)\\
\cellcolor{white}{\textcolor{black}{warexper\_b}} & Count of various war experiences (out of 12)\\
\cellcolor{white}{\textcolor{black}{slphungry7dx\_b}} & Count of days you sleep hungry (in last 7 days)?\\ 
\cellcolor{white}{\textcolor{black}{conscientious\_b}} & Conscientiousness index (0-23)\\
\cellcolor{white}{\textcolor{black}{neurotic\_b}} & Neurotism index (0-22)\\
\cellcolor{white}{\textcolor{black}{grit\_b}} & Perseverance index (0-7)\\
\cellcolor{white}{\textcolor{black}{rewardresp\_b}} & Reward responsiveness (0-23)\\
\cellcolor{white}{\textcolor{black}{locuscontr\_b}} & Locus of control index (0-24)\\ 
\cellcolor{white}{\textcolor{black}{impulsive\_b}} & Impulsiveness index (0-20)\\
\cellcolor{white}{\textcolor{black}{selfesteem\_b}} & Self-esteem index (0-24)\\
\cellcolor{white}{\textcolor{black}{patient\_game\_real\_b}} & Patience total index (IBM) (0-6)\\
\cellcolor{white}{\textcolor{black}{inconsistent\_game\_resc\_b}} & Time inconsistency index (IBM) (0-6)\\
\cellcolor{white}{\textcolor{black}{risk\_game\_resc\_b}} & Risk total index (0-3)\\
\bottomrule
\end{tabular}
\caption{There are 40 covariates that are not binary in the full set of baseline variables used by \citet{Reducecrime}. Our subset of covariates only uses five of the covariates (light and dark blue). The reduced subset only includes two of them (dark blue).}\label{tab:covariatevars-nonbinary}
\end{table}

%% file: tables-and-figures-appendix/liberia_replicate2b.tex
\begin{landscape}
\begin{table}
    \centering
\begin{tabular}{ p{0.45\textwidth}>{\raggedleft\arraybackslash}p{0.07\textwidth} >{\raggedleft\arraybackslash}p{0.07\textwidth} >{\raggedleft\arraybackslash}p{0.075\textwidth} >{\raggedleft\arraybackslash}p{0.07\textwidth}>{\raggedleft\arraybackslash}p{0.07\textwidth} >{\raggedleft\arraybackslash}p{0.075\textwidth} >{\raggedleft\arraybackslash}p{0.07\textwidth}>{\raggedleft\arraybackslash}p{0.07\textwidth} >{\raggedleft\arraybackslash}p{0.075\textwidth} >{\raggedleft\arraybackslash}p{0.07\textwidth}}
\toprule
\multicolumn{2}{c}{ } & \multicolumn{3}{c}{Therapy Only} & \multicolumn{3}{c}{Cash Only} & \multicolumn{3}{c}{Both} \\
\cmidrule(l{3pt}r{3pt}){3-5} \cmidrule(l{3pt}r{3pt}){6-8} \cmidrule(l{3pt}r{3pt}){9-11}
Outcome & Control mean & ITT & Std. Err & Adj.p-value & ITT & Std. Err & Adj.p-value & ITT & Std. Err & Adj.p-value\\
\midrule
Anti-social Behaviors, z-score & 0.032 & -0.083 & {}[0.093] & 1.000 & 0.132 & {}[0.097]* & 0.784 & -0.247 & {}[0.088]*** & 0.023\\
\hspace{1em}Usually Sells Drugs & 0.135 & -0.034 & {}[0.029] & 1.000 & 0.035 & {}[0.030] & 1.000 & -0.059 & {}[0.029]** & 0.428\\
\hspace{1em}\# of thefts/robberies in past 2 weeks & 1.839 & 0.073 & {}[0.395] & 1.000 & 0.352 & {}[0.388] & 1.000 & -0.728 & {}[0.363]** & 0.471\\
Disputes and fights in past 2 weeks, z-score & -0.060 & -0.026 & {}[0.091] & 1.000 & 0.100 & {}[0.090] & 1.000 & -0.100 & {}[0.077]* & 1.000\\
\hspace{1em}Carries a weapon on body & 0.148 & -0.059 & {}[0.031]** & 0.632 & 0.043 & {}[0.035] & 1.000 & -0.066 & {}[0.033]** & 0.506\\
\addlinespace
\hspace{1em}Arrested in past 2 weeks & 0.118 & -0.006 & {}[0.024] & 1.000 & 0.007 & {}[0.025] & 1.000 & -0.033 & {}[0.024]* & 1.000\\
\hspace{1em}Aggressive behaviors, z-score & 0.188 & -0.153 & {}[0.110]* & 1.000 & -0.043 & {}[0.107] & 1.000 & -0.339 & {}[0.109]*** & 0.020\\
\hspace{1em}Verbal/physical abuse of partner, z-score & -0.071 & 0.142 & {}[0.100]* & 1.000 & 0.233 & {}[0.113]** & 0.416 & 0.059 & {}[0.104] & 1.000\\
\bottomrule
\end{tabular}
\caption{We replicated the analysis of BJS to recreate the 12 month portion of Table 2 \citep{Reducecrime}. However, we use a Bonferroni correction instead of a Westfall-Young correction for the p-values. The anti-social behavior index is adjusted for 9 comparisons while the other outcomes are adjusted for 21 comparisons.}\label{tab:replicate2b}
\end{table}
\end{landscape}

%% file: tables-and-figures-appendix/liberia_allcovsets_confidential.tex
\begin{table}
\centering
\small

\begin{tabular}{>{\raggedleft\arraybackslash}p{0.23\textwidth}>{\raggedleft\arraybackslash}p{0.07\textwidth}>{\raggedleft\arraybackslash}p{0.07\textwidth}>{\raggedleft\arraybackslash}p{0.07\textwidth}>{\raggedleft\arraybackslash}p{0.07\textwidth}>{\raggedleft\arraybackslash}p{0.07\textwidth}>{\raggedleft\arraybackslash}p{0.07\textwidth}>{\raggedleft\arraybackslash}p{0.07\textwidth}>{\raggedleft\arraybackslash}p{0.07\textwidth}>{\raggedleft\arraybackslash}p{0.07\textwidth}}
\toprule
\multicolumn{1}{c}{ } & \multicolumn{3}{c}{Therapy Only} & \multicolumn{3}{c}{Cash Only} & \multicolumn{3}{c}{Both} \\
\cmidrule(l{3pt}r{3pt}){2-4} \cmidrule(l{3pt}r{3pt}){5-7} \cmidrule(l{3pt}r{3pt}){8-10}
Response & ITT & Std. Error & Adj. p-value & ITT & Std. Error & Adj. p-value & ITT & Std. Error & Adj. p-value\\
\midrule
\addlinespace[0.3em]
\multicolumn{10}{l}{\textbf{Reduced Subset}}\\
\hspace{1em}fam\_asb\_lt & -0.07 & 0.09 & 0.65 & 0.12 & 0.10 & 0.34 & -0.25 & 0.08 & 0.00\\
\hspace{1em}drugssellever\_ltav & -0.04 & 0.03 & 1.00 & 0.04 & 0.03 & 1.00 & -0.06 & 0.03 & 0.16\\
\hspace{1em}stealnb\_ltav & 0.06 & 0.39 & 1.00 & 0.28 & 0.38 & 1.00 & -0.84 & 0.34 & 0.15\\
\hspace{1em}disputes\_all\_z\_ltav & 0.00 & 0.09 & 1.00 & 0.11 & 0.10 & 1.00 & -0.10 & 0.08 & 1.00\\
\hspace{1em}carryweapon\_ltav & -0.05 & 0.03 & 1.00 & 0.03 & 0.03 & 1.00 & -0.07 & 0.03 & 0.48\\
\hspace{1em}arrested\_ltav & 0.00 & 0.02 & 1.00 & 0.01 & 0.03 & 1.00 & -0.03 & 0.02 & \vphantom{1} 1.00\\
\hspace{1em}asbhostilstd\_ltav & -0.15 & 0.11 & 1.00 & -0.07 & 0.11 & 1.00 & -0.31 & 0.11 & 0.04\\
\hspace{1em}domabuse\_z\_ltav & 0.14 & 0.10 & 1.00 & 0.23 & 0.11 & 0.38 & 0.02 & 0.10 & 1.00\\
\addlinespace[0.3em]
\multicolumn{10}{l}{\textbf{Subset}}\\
\hspace{1em}fam\_asb\_lt & -0.08 & 0.09 & 0.61 & 0.13 & 0.10 & 0.27 & -0.26 & 0.08 & 0.00\\
\hspace{1em}drugssellever\_ltav & -0.04 & 0.03 & 1.00 & 0.04 & 0.03 & 1.00 & -0.07 & 0.03 & 0.14\\
\hspace{1em}stealnb\_ltav & 0.03 & 0.38 & 1.00 & 0.29 & 0.37 & 1.00 & -0.87 & 0.34 & 0.12\\
\hspace{1em}disputes\_all\_z\_ltav & -0.01 & 0.09 & 1.00 & 0.11 & 0.10 & 1.00 & -0.11 & 0.07 & 1.00\\
\hspace{1em}carryweapon\_ltav & -0.05 & 0.03 & 1.00 & 0.03 & 0.03 & 1.00 & -0.07 & 0.03 & 0.38\\
\hspace{1em}arrested\_ltav & 0.00 & 0.02 & 1.00 & 0.01 & 0.03 & 1.00 & -0.03 & 0.02 & 1.00\\
\hspace{1em}asbhostilstd\_ltav & -0.15 & 0.11 & 1.00 & -0.06 & 0.10 & 1.00 & -0.32 & 0.11 & 0.03\\
\hspace{1em}domabuse\_z\_ltav & 0.14 & 0.10 & 1.00 & 0.24 & 0.11 & 0.29 & 0.01 & 0.10 & 1.00\\
\addlinespace[0.3em]
\multicolumn{10}{l}{\textbf{Full Baseline}}\\
\hspace{1em}fam\_asb\_lt & -0.07 & 0.09 & 0.70 & 0.15 & 0.10 & 0.18 & -0.24 & 0.09 & 0.01\\
\hspace{1em}drugssellever\_ltav & -0.03 & 0.03 & 1.00 & 0.04 & 0.03 & 1.00 & -0.06 & 0.03 & 0.42\\
\hspace{1em}stealnb\_ltav & 0.11 & 0.40 & 1.00 & 0.40 & 0.39 & 1.00 & -0.71 & 0.37 & 0.53\\
\hspace{1em}disputes\_all\_z\_ltav & -0.01 & 0.09 & 1.00 & 0.12 & 0.09 & 1.00 & -0.09 & 0.08 & 1.00\\
\hspace{1em}carryweapon\_ltav & -0.06 & 0.03 & 0.73 & 0.05 & 0.03 & 1.00 & -0.07 & 0.03 & 0.51\\
\hspace{1em}arrested\_ltav & -0.01 & 0.02 & 1.00 & 0.01 & 0.03 & 1.00 & -0.03 & 0.02 & 1.00\\
\hspace{1em}asbhostilstd\_ltav & -0.14 & 0.11 & 1.00 & -0.03 & 0.11 & 1.00 & -0.33 & 0.11 & 0.03\\
\hspace{1em}domabuse\_z\_ltav & 0.17 & 0.10 & 1.00 & 0.26 & 0.11 & 0.24 & 0.05 & 0.11 & 1.00\\
\bottomrule
\end{tabular}
\caption{Original data results across the three covariate sets.}\label{tab:apdx-itt-orig}
\end{table}

%% file: tables-and-figures-appendix/liberia_allcovsets_sanitized_mvhistogram.tex
\begin{table}
\centering
\small

\begin{tabular}{>{\raggedleft\arraybackslash}p{0.21\textwidth}>{\raggedleft\arraybackslash}p{0.072\textwidth}>{\raggedleft\arraybackslash}p{0.055\textwidth}>{\raggedleft\arraybackslash}p{0.055\textwidth}>{\raggedleft\arraybackslash}p{0.068\textwidth}>{\raggedleft\arraybackslash}p{0.072\textwidth}>{\raggedleft\arraybackslash}p{0.055\textwidth}>{\raggedleft\arraybackslash}p{0.055\textwidth}>{\raggedleft\arraybackslash}p{0.068\textwidth}>{\raggedleft\arraybackslash}p{0.072\textwidth}>{\raggedleft\arraybackslash}p{0.055\textwidth}>{\raggedleft\arraybackslash}p{0.055\textwidth}>{\raggedleft\arraybackslash}p{0.068\textwidth}}
\toprule
\multicolumn{1}{c}{ } & \multicolumn{4}{c}{Therapy Only} & \multicolumn{4}{c}{Cash Only} & \multicolumn{4}{c}{Both} \\
\cmidrule(l{3pt}r{3pt}){2-5} \cmidrule(l{3pt}r{3pt}){6-9} \cmidrule(l{3pt}r{3pt}){10-13}
Response & ITT & Std. Error & Adj. p-value & CI overlap & ITT & Std. Error & Adj. p-value & CI overlap & ITT & Std. Error & Adj. p-value & CI overlap\\
\midrule
\addlinespace[0.3em]
\multicolumn{13}{l}{\textbf{Reduced Subset}}\\
\hspace{1em}fam\_asb\_lt & 0.07 & 0.09 & 0.64 & 0.60 & 0.34 & 0.11 & 0.00 & 0.46 & -0.04 & 0.08 & 0.93 & 0.35\\
\hspace{1em}drugssellever\_ltav & 0.00 & 0.03 & 1.00 & 0.62 & 0.09 & 0.03 & 0.04 & 0.60 & -0.01 & 0.03 & 1.00 & 0.44\\
\hspace{1em}stealnb\_ltav & -0.24 & 0.41 & 1.00 & 0.81 & 0.72 & 0.44 & 1.00 & 0.73 & -0.25 & 0.45 & 1.00 & 0.63\\
\hspace{1em}disputes\_all\_z\_ltav & 0.19 & 0.07 & 0.07 & 0.40 & 0.35 & 0.09 & 0.00 & 0.31 & -0.02 & 0.05 & 1.00 & 0.68\\
\hspace{1em}carryweapon\_ltav & -0.02 & 0.03 & 1.00 & 0.70 & 0.08 & 0.03 & 0.21 & 0.62 & 0.00 & 0.03 & 1.00 & 0.46\\
\hspace{1em}arrested\_ltav & 0.01 & 0.02 & 1.00 & 0.87 & 0.04 & 0.03 & 1.00 & 0.73 & -0.02 & 0.02 & 1.00 & 0.87\\
\hspace{1em}asbhostilstd\_ltav & -0.11 & 0.11 & 1.00 & 0.89 & -0.06 & 0.10 & 1.00 & 0.97 & -0.20 & 0.10 & 0.58 & 0.72\\
\hspace{1em}domabuse\_z\_ltav & 0.28 & 0.09 & 0.03 & 0.62 & 0.37 & 0.10 & 0.00 & 0.66 & 0.08 & 0.10 & 1.00 & 0.85\\
\addlinespace[0.3em]
\multicolumn{13}{l}{\textbf{Subset}}\\
\hspace{1em}fam\_asb\_lt & 0.11 & 0.09 & 0.32 & 0.47 & 0.42 & 0.10 & 0.00 & 0.27 & 0.03 & 0.08 & 1.00 & 0.10\\
\hspace{1em}drugssellever\_ltav & 0.01 & 0.03 & 1.00 & 0.57 & 0.10 & 0.03 & 0.01 & 0.52 & 0.00 & 0.03 & 1.00 & 0.32\\
\hspace{1em}stealnb\_ltav & -0.14 & 0.41 & 1.00 & 0.89 & 0.93 & 0.43 & 0.33 & 0.60 & -0.05 & 0.46 & 1.00 & 0.49\\
\hspace{1em}disputes\_all\_z\_ltav & 0.19 & 0.07 & 0.06 & 0.36 & 0.37 & 0.09 & 0.00 & 0.26 & -0.01 & 0.05 & 1.00 & 0.62\\
\hspace{1em}carryweapon\_ltav & 0.00 & 0.03 & 1.00 & 0.59 & 0.10 & 0.03 & 0.03 & 0.48 & 0.02 & 0.03 & 1.00 & 0.25\\
\hspace{1em}arrested\_ltav & 0.02 & 0.02 & 1.00 & 0.76 & 0.06 & 0.03 & 0.36 & 0.59 & 0.00 & 0.02 & 1.00 & 0.72\\
\hspace{1em}asbhostilstd\_ltav & -0.05 & 0.11 & 1.00 & 0.75 & 0.04 & 0.10 & 1.00 & 0.76 & -0.08 & 0.10 & 1.00 & 0.42\\
\hspace{1em}domabuse\_z\_ltav & 0.28 & 0.10 & 0.04 & 0.61 & 0.40 & 0.10 & 0.00 & 0.63 & 0.10 & 0.10 & 1.00 & 0.78\\
\addlinespace[0.3em]
\multicolumn{13}{l}{\textbf{Full Baseline}}\\
\hspace{1em}fam\_asb\_lt & 0.02 & 0.09 & 1.00 & 0.74 & 0.34 & 0.10 & 0.00 & 0.52 & -0.06 & 0.09 & 0.68 & 0.48\\
\hspace{1em}drugssellever\_ltav & -0.02 & 0.03 & 1.00 & 0.87 & 0.07 & 0.03 & 0.33 & 0.76 & -0.03 & 0.03 & 1.00 & 0.70\\
\hspace{1em}stealnb\_ltav & -0.47 & 0.46 & 1.00 & 0.66 & 0.78 & 0.45 & 0.88 & 0.78 & -0.76 & 0.41 & 0.69 & 0.94\\
\hspace{1em}disputes\_all\_z\_ltav & 0.16 & 0.06 & 0.06 & 0.45 & 0.25 & 0.07 & 0.00 & 0.61 & 0.04 & 0.06 & 1.00 & 0.52\\
\hspace{1em}carryweapon\_ltav & -0.01 & 0.03 & 1.00 & 0.65 & 0.09 & 0.03 & 0.05 & 0.66 & 0.02 & 0.03 & 1.00 & 0.34\\
\hspace{1em}arrested\_ltav & -0.02 & 0.02 & 1.00 & 0.88 & 0.00 & 0.03 & 1.00 & 0.94 & -0.03 & 0.02 & 1.00 & 0.93\\
\hspace{1em}asbhostilstd\_ltav & -0.03 & 0.11 & 1.00 & 0.75 & 0.26 & 0.11 & 0.19 & 0.33 & -0.08 & 0.10 & 1.00 & 0.39\\
\hspace{1em}domabuse\_z\_ltav & 0.33 & 0.11 & 0.03 & 0.61 & 0.40 & 0.14 & 0.04 & 0.72 & 0.04 & 0.11 & 1.00 & 0.96\\
\bottomrule
\end{tabular}
\caption{Synthetic data from the MV Histogram method results across the three covariate sets.} \label{tab:apdx-itt-mvhistogram}\end{table}

%% file: tables-and-figures-appendix/liberia_allcovsets_sanitized_hybrid.tex
\begin{table}
\centering
\small

\begin{tabular}{>{\raggedleft\arraybackslash}p{0.21\textwidth}>{\raggedleft\arraybackslash}p{0.072\textwidth}>{\raggedleft\arraybackslash}p{0.055\textwidth}>{\raggedleft\arraybackslash}p{0.055\textwidth}>{\raggedleft\arraybackslash}p{0.068\textwidth}>{\raggedleft\arraybackslash}p{0.072\textwidth}>{\raggedleft\arraybackslash}p{0.055\textwidth}>{\raggedleft\arraybackslash}p{0.055\textwidth}>{\raggedleft\arraybackslash}p{0.068\textwidth}>{\raggedleft\arraybackslash}p{0.072\textwidth}>{\raggedleft\arraybackslash}p{0.055\textwidth}>{\raggedleft\arraybackslash}p{0.055\textwidth}>{\raggedleft\arraybackslash}p{0.068\textwidth}}
\toprule
\multicolumn{1}{c}{ } & \multicolumn{4}{c}{Therapy Only} & \multicolumn{4}{c}{Cash Only} & \multicolumn{4}{c}{Both} \\
\cmidrule(l{3pt}r{3pt}){2-5} \cmidrule(l{3pt}r{3pt}){6-9} \cmidrule(l{3pt}r{3pt}){10-13}
Response & ITT & Std. Error & Adj. p-value & CI overlap & ITT & Std. Error & Adj. p-value & CI overlap & ITT & Std. Error & Adj. p-value & CI overlap\\
\midrule
\addlinespace[0.3em]
\multicolumn{13}{l}{\textbf{Reduced Subset}}\\
\hspace{1em}fam\_asb\_lt & 0.00 & 0.08 & 1.00 & 0.80 & 0.24 & 0.08 & 0.01 & 0.66 & -0.43 & 0.11 & 0.00 & 0.57\\
\hspace{1em}drugssellever\_ltav & -0.05 & 0.02 & 0.39 & 0.86 & 0.01 & 0.03 & 1.00 & 0.71 & -0.05 & 0.03 & 1.00 & 0.89\\
\hspace{1em}stealnb\_ltav & 0.08 & 0.33 & 1.00 & 0.93 & 0.00 & 0.33 & 1.00 & 0.80 & -0.70 & 0.46 & 1.00 & 0.87\\
\hspace{1em}disputes\_all\_z\_ltav & 0.03 & 0.07 & 1.00 & 0.89 & 0.09 & 0.07 & 1.00 & 0.87 & -0.10 & 0.10 & 1.00 & 0.88\\
\hspace{1em}carryweapon\_ltav & -0.10 & 0.03 & 0.01 & 0.63 & -0.01 & 0.03 & 1.00 & 0.69 & 0.02 & 0.04 & 1.00 & 0.41\\
\hspace{1em}arrested\_ltav & -0.02 & 0.02 & 1.00 & 0.85 & 0.02 & 0.02 & 1.00 & 0.93 & -0.01 & 0.03 & 1.00 & 0.86\\
\hspace{1em}asbhostilstd\_ltav & -0.17 & 0.09 & 0.57 & 0.91 & -0.12 & 0.09 & 1.00 & 0.88 & -0.25 & 0.13 & 0.51 & 0.88\\
\hspace{1em}domabuse\_z\_ltav & 0.13 & 0.09 & 1.00 & 0.97 & 0.31 & 0.09 & 0.01 & 0.82 & 0.13 & 0.13 & 1.00 & 0.76\\
\addlinespace[0.3em]
\multicolumn{13}{l}{\textbf{Subset}}\\
\hspace{1em}fam\_asb\_lt & 0.00 & 0.08 & 1.00 & 0.77 & 0.09 & 0.08 & 0.40 & 0.89 & -0.26 & 0.11 & 0.03 & 0.87\\
\hspace{1em}drugssellever\_ltav & -0.09 & 0.02 & 0.00 & 0.44 & 0.03 & 0.02 & 1.00 & 0.88 & -0.06 & 0.03 & 0.89 & 0.90\\
\hspace{1em}stealnb\_ltav & 0.74 & 0.33 & 0.28 & 0.50 & 0.50 & 0.33 & 1.00 & 0.86 & -1.21 & 0.45 & 0.08 & 0.79\\
\hspace{1em}disputes\_all\_z\_ltav & -0.15 & 0.07 & 0.30 & 0.55 & -0.01 & 0.07 & 1.00 & 0.66 & 0.05 & 0.09 & 1.00 & 0.54\\
\hspace{1em}carryweapon\_ltav & 0.00 & 0.03 & 1.00 & 0.59 & 0.04 & 0.03 & 1.00 & 0.92 & -0.07 & 0.04 & 0.77 & 0.92\\
\hspace{1em}arrested\_ltav & 0.00 & 0.02 & 1.00 & 0.96 & 0.04 & 0.02 & 0.63 & 0.73 & -0.04 & 0.03 & 1.00 & 0.88\\
\hspace{1em}asbhostilstd\_ltav & -0.20 & 0.09 & 0.33 & 0.89 & -0.12 & 0.09 & 1.00 & 0.86 & -0.29 & 0.13 & 0.28 & 0.92\\
\hspace{1em}domabuse\_z\_ltav & 0.09 & 0.09 & 1.00 & 0.86 & 0.33 & 0.08 & 0.00 & 0.79 & 0.02 & 0.12 & 1.00 & 0.93\\
\addlinespace[0.3em]
\multicolumn{13}{l}{\textbf{Full Baseline}}\\
\hspace{1em}fam\_asb\_lt & -0.03 & 0.07 & 0.96 & 0.90 & 0.25 & 0.07 & 0.00 & 0.73 & -0.40 & 0.10 & 0.00 & 0.57\\
\hspace{1em}drugssellever\_ltav & -0.04 & 0.02 & 1.00 & 0.90 & -0.01 & 0.02 & 1.00 & 0.54 & -0.02 & 0.03 & 1.00 & 0.66\\
\hspace{1em}stealnb\_ltav & -0.13 & 0.32 & 1.00 & 0.84 & -0.03 & 0.34 & 1.00 & 0.70 & -0.56 & 0.46 & 1.00 & 0.89\\
\hspace{1em}disputes\_all\_z\_ltav & -0.02 & 0.06 & 1.00 & 0.85 & 0.09 & 0.06 & 1.00 & 0.84 & 0.00 & 0.09 & 1.00 & 0.72\\
\hspace{1em}carryweapon\_ltav & -0.05 & 0.03 & 0.86 & 0.93 & 0.07 & 0.03 & 0.07 & 0.79 & -0.10 & 0.04 & 0.14 & 0.79\\
\hspace{1em}arrested\_ltav & -0.05 & 0.02 & 0.17 & 0.52 & -0.03 & 0.02 & 1.00 & 0.60 & 0.03 & 0.03 & 1.00 & 0.43\\
\hspace{1em}asbhostilstd\_ltav & -0.18 & 0.09 & 0.39 & 0.89 & -0.02 & 0.09 & 1.00 & 0.92 & -0.29 & 0.12 & 0.20 & 0.92\\
\hspace{1em}domabuse\_z\_ltav & 0.21 & 0.09 & 0.21 & 0.88 & 0.29 & 0.09 & 0.02 & 0.90 & 0.10 & 0.13 & 1.00 & 0.91\\
\bottomrule
\end{tabular}
\caption{Synthetic data from the Hybrid method results across the three covariate sets.} \label{tab:apdx-itt-hybrid}\end{table}

%% file: tables-and-figures-appendix/liberia_allcovsets_sanitized_dpmb1.tex
\begin{table}
\centering
\small

\begin{tabular}{>{\raggedleft\arraybackslash}p{0.21\textwidth}>{\raggedleft\arraybackslash}p{0.072\textwidth}>{\raggedleft\arraybackslash}p{0.055\textwidth}>{\raggedleft\arraybackslash}p{0.055\textwidth}>{\raggedleft\arraybackslash}p{0.068\textwidth}>{\raggedleft\arraybackslash}p{0.072\textwidth}>{\raggedleft\arraybackslash}p{0.055\textwidth}>{\raggedleft\arraybackslash}p{0.055\textwidth}>{\raggedleft\arraybackslash}p{0.068\textwidth}>{\raggedleft\arraybackslash}p{0.072\textwidth}>{\raggedleft\arraybackslash}p{0.055\textwidth}>{\raggedleft\arraybackslash}p{0.055\textwidth}>{\raggedleft\arraybackslash}p{0.068\textwidth}}
\toprule
\multicolumn{1}{c}{ } & \multicolumn{4}{c}{Therapy Only} & \multicolumn{4}{c}{Cash Only} & \multicolumn{4}{c}{Both} \\
\cmidrule(l{3pt}r{3pt}){2-5} \cmidrule(l{3pt}r{3pt}){6-9} \cmidrule(l{3pt}r{3pt}){10-13}
Response & ITT & Std. Error & Adj. p-value & CI overlap & ITT & Std. Error & Adj. p-value & CI overlap & ITT & Std. Error & Adj. p-value & CI overlap\\
\midrule
\addlinespace[0.3em]
\multicolumn{13}{l}{\textbf{Reduced Subset}}\\
\hspace{1em}fam\_asb\_lt & -36.93 & 0.00 & 0 & 0.00 & 0.10 & 0.00 & 0.00 & 0.52 & -1.51 & 0.00 & 0 & 0.00\\
\hspace{1em}drugssellever\_ltav & -0.04 & 0.00 & 0 & 0.54 & -42.41 & 0.00 & 0.00 & 0.00 & 0.49 & 0.00 & 0 & 0.00\\
\hspace{1em}stealnb\_ltav & -0.37 & 0.00 & 0 & 0.50 & -22.15 & 0.00 & 0.00 & 0.00 & -0.97 & 0.00 & 0 & 0.50\\
\hspace{1em}disputes\_all\_z\_ltav & -0.17 & 0.00 & 0 & 0.49 & 0.14 & 0.00 & 0.00 & 0.51 & -0.48 & 0.00 & 0 & 0.00\\
\hspace{1em}carryweapon\_ltav & -45.28 & 0.00 & 0 & 0.00 & 0.04 & 0.00 & 0.00 & 0.53 & -0.59 & 0.00 & 0 & 0.00\\
\hspace{1em}arrested\_ltav & 0.16 & 0.00 & 0 & 0.00 & 0.11 & 0.00 & 0.00 & 0.00 & 0.08 & 0.00 & 0 & 0.00\\
\hspace{1em}asbhostilstd\_ltav & -0.18 & 0.00 & 0 & 0.51 & -0.11 & 0.00 & 0.00 & 0.51 & 1.48 & 0.00 & 0 & 0.00\\
\hspace{1em}domabuse\_z\_ltav & -0.49 & 0.00 & 0 & 0.00 & -1.96 & 0.00 & 0.00 & 0.00 & 0.11 & 0.00 & 0 & 0.50\\
\addlinespace[0.3em]
\multicolumn{13}{l}{\textbf{Subset}}\\
\hspace{1em}fam\_asb\_lt & -0.04 & 0.00 & 0 & 0.52 & 0.19 & 0.00 & 0.00 & 0.52 & -5.86 & 0.00 & 0 & 0.00\\
\hspace{1em}drugssellever\_ltav & -0.09 & 0.00 & 0 & 0.53 & -0.23 & 0.00 & 0.00 & 0.00 & -0.15 & 0.00 & 0 & 0.00\\
\hspace{1em}stealnb\_ltav & 0.49 & 0.01 & 0 & 0.51 & 0.28 & 0.01 & 0.00 & 0.51 & -675.49 & 0.01 & 0 & 0.00\\
\hspace{1em}disputes\_all\_z\_ltav & -0.03 & 0.00 & 0 & 0.52 & 0.10 & 0.00 & 0.00 & 0.52 & -0.11 & 0.00 & 0 & 0.53\\
\hspace{1em}carryweapon\_ltav & -0.05 & 0.00 & 0 & 0.53 & -31.33 & 0.00 & 0.00 & 0.00 & 0.00 & 0.00 & 1 & 0.00\\
\hspace{1em}arrested\_ltav & -0.13 & 0.00 & 0 & 0.00 & 0.12 & 0.00 & 0.00 & 0.00 & -0.08 & 0.00 & 0 & 0.00\\
\hspace{1em}asbhostilstd\_ltav & -0.13 & 0.00 & 0 & 0.50 & -20.99 & 0.00 & 0.00 & 0.00 & -0.60 & 0.00 & 0 & 0.00\\
\hspace{1em}domabuse\_z\_ltav & -34.58 & 0.00 & 0 & 0.00 & 0.25 & 0.00 & 0.00 & 0.50 & -202.64 & 0.00 & 0 & 0.00\\
fam\_asb\_lt & 0.46 & 0.00 & 0 & 0.00 & 0.29 & 0.00 & 0.00 & 0.51 & -0.06 & 0.00 & 0 & 0.10\\
drugssellever\_ltav & -47.11 & 0.00 & 0 & 0.00 & -0.31 & 0.00 & 0.00 & 0.00 & -0.50 & 0.00 & 0 & 0.00\\
stealnb\_ltav & 0.27 & 0.01 & 0 & 0.51 & 0.59 & 0.01 & 0.00 & 0.51 & -0.07 & 0.01 & 0 & 0.51\\
disputes\_all\_z\_ltav & -0.29 & 0.00 & 0 & 0.00 & -0.06 & 0.00 & 0.00 & 0.00 & 86.60 & 0.00 & 0 & 0.00\\
carryweapon\_ltav & -0.08 & 0.00 & 0 & 0.52 & 0.01 & 0.00 & 0.00 & 0.52 & -36.62 & 0.00 & 0 & 0.00\\
arrested\_ltav & 0.09 & 0.00 & 0 & 0.00 & 0.00 & 0.00 & 0.09 & 0.53 & -129.27 & 0.00 & 0 & 0.00\\
asbhostilstd\_ltav & -0.54 & 0.00 & 0 & 0.00 & 1.71 & 0.00 & 0.00 & 0.00 & -0.05 & 0.00 & 0 & 0.00\\
domabuse\_z\_ltav & 0.16 & 0.00 & 0 & 0.52 & -26.76 & 0.00 & 0.00 & 0.00 & 0.65 & 0.01 & 0 & 0.00\\
\bottomrule
\end{tabular}
\caption{Synthetic data from the \dpmb{1} method results across the three covariate sets.} \label{tab:apdx-itt-dpmb1}\end{table}

%% file: tables-and-figures-appendix/liberia_allcovsets_sanitized_dpmb2.tex
\begin{table}
\centering
\small

\begin{tabular}{>{\raggedleft\arraybackslash}p{0.21\textwidth}>{\raggedleft\arraybackslash}p{0.072\textwidth}>{\raggedleft\arraybackslash}p{0.055\textwidth}>{\raggedleft\arraybackslash}p{0.055\textwidth}>{\raggedleft\arraybackslash}p{0.068\textwidth}>{\raggedleft\arraybackslash}p{0.072\textwidth}>{\raggedleft\arraybackslash}p{0.055\textwidth}>{\raggedleft\arraybackslash}p{0.055\textwidth}>{\raggedleft\arraybackslash}p{0.068\textwidth}>{\raggedleft\arraybackslash}p{0.072\textwidth}>{\raggedleft\arraybackslash}p{0.055\textwidth}>{\raggedleft\arraybackslash}p{0.055\textwidth}>{\raggedleft\arraybackslash}p{0.068\textwidth}}
\toprule
\multicolumn{1}{c}{ } & \multicolumn{4}{c}{Therapy Only} & \multicolumn{4}{c}{Cash Only} & \multicolumn{4}{c}{Both} \\
\cmidrule(l{3pt}r{3pt}){2-5} \cmidrule(l{3pt}r{3pt}){6-9} \cmidrule(l{3pt}r{3pt}){10-13}
Response & ITT & Std. Error & Adj. p-value & CI overlap & ITT & Std. Error & Adj. p-value & CI overlap & ITT & Std. Error & Adj. p-value & CI overlap\\
\midrule
\addlinespace[0.3em]
\multicolumn{13}{l}{\textbf{Reduced Subset}}\\
\hspace{1em}fam\_asb\_lt & -0.09 & 0.00 & 0 & 0.52 & 0.17 & 0.00 & 0 & 0.52 & -0.28 & 0.00 & 0 & 0.52\\
\hspace{1em}drugssellever\_ltav & 0.15 & 0.00 & 0 & 0.00 & 0.01 & 0.00 & 0 & 0.53 & -0.10 & 0.00 & 0 & 0.54\\
\hspace{1em}stealnb\_ltav & 3.06 & 0.01 & 0 & 0.00 & -0.22 & 0.01 & 0 & 0.51 & -1.22 & 0.01 & 0 & 0.51\\
\hspace{1em}disputes\_all\_z\_ltav & 0.15 & 0.00 & 0 & 0.51 & -1.48 & 0.00 & 0 & 0.00 & -83.74 & 0.00 & 0 & 0.00\\
\hspace{1em}carryweapon\_ltav & -0.40 & 0.00 & 0 & 0.00 & 0.12 & 0.00 & 0 & 0.00 & 0.28 & 0.00 & 0 & 0.00\\
\hspace{1em}arrested\_ltav & 0.38 & 0.00 & 0 & 0.00 & -0.06 & 0.00 & 0 & 0.00 & -0.09 & 0.00 & 0 & 0.00\\
\hspace{1em}asbhostilstd\_ltav & 1.33 & 0.00 & 0 & 0.00 & -0.53 & 0.00 & 0 & 0.00 & -0.16 & 0.00 & 0 & 0.52\\
\hspace{1em}domabuse\_z\_ltav & -0.66 & 0.00 & 0 & 0.00 & 0.46 & 0.00 & 0 & 0.00 & 7.09 & 0.00 & 0 & 0.00\\
\addlinespace[0.3em]
\multicolumn{13}{l}{\textbf{Subset}}\\
\hspace{1em}fam\_asb\_lt & -20.00 & 0.00 & 0 & 0.00 & 0.31 & 0.00 & 0 & 0.51 & -20.38 & 0.00 & 0 & 0.00\\
\hspace{1em}drugssellever\_ltav & -0.02 & 0.00 & 0 & 0.53 & 0.05 & 0.00 & 0 & 0.53 & 0.41 & 0.00 & 0 & 0.00\\
\hspace{1em}stealnb\_ltav & -0.29 & 0.01 & 0 & 0.51 & -5.10 & 0.01 & 0 & 0.00 & -0.26 & 0.01 & 0 & 0.51\\
\hspace{1em}disputes\_all\_z\_ltav & -36.21 & 0.00 & 0 & 0.00 & 0.03 & 0.00 & 0 & 0.51 & 4.81 & 0.00 & 0 & 0.00\\
\hspace{1em}carryweapon\_ltav & 0.09 & 0.00 & 0 & 0.00 & 0.03 & 0.00 & 0 & 0.52 & 0.01 & 0.00 & 0 & 0.00\\
\hspace{1em}arrested\_ltav & -46.29 & 0.00 & 0 & 0.00 & 0.04 & 0.00 & 0 & 0.53 & -25.62 & 0.00 & 0 & 0.00\\
\hspace{1em}asbhostilstd\_ltav & -0.74 & 0.00 & 0 & 0.00 & -0.05 & 0.00 & 0 & 0.51 & -7.91 & 0.00 & 0 & 0.00\\
\hspace{1em}domabuse\_z\_ltav & -2.15 & 0.00 & 0 & 0.00 & 0.34 & 0.00 & 0 & 0.51 & -5.07 & 0.00 & 0 & 0.00\\
fam\_asb\_lt & -0.07 & 0.00 & 0 & 0.50 & 0.04 & 0.00 & 0 & 0.50 & -4.78 & 0.00 & 0 & 0.00\\
drugssellever\_ltav & 0.03 & 0.00 & 0 & 0.00 & 0.13 & 0.00 & 0 & 0.00 & -0.45 & 0.00 & 0 & 0.00\\
stealnb\_ltav & 1.91 & 0.01 & 0 & 0.00 & 2.12 & 0.01 & 0 & 0.00 & -6.90 & 0.01 & 0 & 0.00\\
disputes\_all\_z\_ltav & 0.01 & 0.00 & 0 & 0.51 & -22.46 & 0.00 & 0 & 0.00 & -22.43 & 0.00 & 0 & 0.00\\
carryweapon\_ltav & -0.64 & 0.00 & 0 & 0.00 & 0.06 & 0.00 & 0 & 0.53 & -4.48 & 0.00 & 0 & 0.00\\
arrested\_ltav & -0.23 & 0.00 & 0 & 0.00 & -0.34 & 0.00 & 0 & 0.00 & -4.74 & 0.00 & 0 & 0.00\\
asbhostilstd\_ltav & -24.57 & 0.00 & 0 & 0.00 & 0.15 & 0.00 & 0 & 0.51 & -3.44 & 0.00 & 0 & 0.00\\
domabuse\_z\_ltav & -35.53 & 0.00 & 0 & 0.00 & 0.28 & 0.00 & 0 & 0.51 & -29.88 & 0.00 & 0 & 0.00\\
\bottomrule
\end{tabular}
\caption{Synthetic data from the \dpmb{2} method results across the three covariate sets.} \label{tab:apdx-itt-dpmb2}\end{table}

%% file: tables-main/liberia_itt_fam_asb_lt.tex
\begin{table}
\centering
\small

\begin{tabular}{>{\raggedright\arraybackslash}p{0.17\textwidth}|>{\raggedleft\arraybackslash}p{0.061\textwidth}>{\raggedleft\arraybackslash}p{0.055\textwidth}>{\raggedleft\arraybackslash}p{0.055\textwidth}>{\raggedleft\arraybackslash}p{0.055\textwidth}|>{\raggedleft\arraybackslash}p{0.061\textwidth}>{\raggedleft\arraybackslash}p{0.055\textwidth}>{\raggedleft\arraybackslash}p{0.055\textwidth}>{\raggedleft\arraybackslash}p{0.055\textwidth}|>{\raggedleft\arraybackslash}p{0.061\textwidth}>{\raggedleft\arraybackslash}p{0.055\textwidth}>{\raggedleft\arraybackslash}p{0.055\textwidth}>{\raggedleft\arraybackslash}p{0.055\textwidth}}
\toprule
\multicolumn{1}{c}{ } & \multicolumn{4}{c}{Therapy Only} & \multicolumn{4}{c}{Cash Only} & \multicolumn{4}{c}{Both} \\
\cmidrule(l{3pt}r{3pt}){2-5} \cmidrule(l{3pt}r{3pt}){6-9} \cmidrule(l{3pt}r{3pt}){10-13}
Method & ITT & Std. Error & CI overlap & Abs. Diff. & ITT & Std. Error & CI overlap & Abs. Diff. & ITT & Std. Error & CI overlap & Abs. Diff.\\
\midrule
\addlinespace[0.3em]
\multicolumn{13}{l}{\textbf{Reduced Subset}}\\
\hspace{1em}MV Histogram & 0.07 & 0.09 & 0.60 & 0.14 & 0.34 & 0.11 & 0.46 & 0.22 & -0.04 & 0.08 & 0.35 & 0.21\\
\hspace{1em}Hybrid & 0.00 & 0.08 & 0.80 & 0.07 & 0.24 & 0.08 & 0.66 & 0.12 & -0.43 & 0.11 & 0.57 & 0.17\\
\hspace{1em}\dpmb{1} & -36.93 & 0.00 & 0.00 & 36.86 & 0.10 & 0.00 & 0.52 & 0.02 & -1.51 & 0.00 & 0.00 & 1.26\\
\hspace{1em}\dpmb{2} & -0.09 & 0.00 & 0.52 & 0.02 & 0.17 & 0.00 & 0.52 & 0.05 & -0.28 & 0.00 & 0.52 & 0.03\\
\hspace{1em}\dpmb{9} & 0.00 & 0.00 & 0.51 & 0.07 & 0.08 & 0.00 & 0.51 & 0.04 & -0.13 & 0.00 & 0.52 & 0.12\\
\addlinespace[0.3em]
\multicolumn{13}{l}{\textbf{Subset}}\\
\hspace{1em}MV Histogram & 0.11 & 0.09 & 0.47 & 0.19 & 0.42 & 0.10 & 0.27 & 0.29 & 0.03 & 0.08 & 0.10 & 0.29\\
\hspace{1em}Hybrid & 0.00 & 0.08 & 0.77 & 0.08 & 0.09 & 0.08 & 0.89 & 0.04 & -0.26 & 0.11 & 0.87 & 0.00\\
\hspace{1em}\dpmb{1} & -0.04 & 0.00 & 0.52 & 0.04 & 0.19 & 0.00 & 0.52 & 0.06 & -5.86 & 0.00 & 0.00 & 5.59\\
\hspace{1em}\dpmb{2} & -20.00 & 0.00 & 0.00 & 19.92 & 0.31 & 0.00 & 0.51 & 0.18 & -20.38 & 0.00 & 0.00 & 20.12\\
\hspace{1em}\dpmb{9} & 0.00 & 0.00 & 0.52 & 0.08 & 0.08 & 0.00 & 0.51 & 0.05 & -0.22 & 0.00 & 0.52 & 0.04\\
\addlinespace[0.3em]
\multicolumn{13}{l}{\textbf{Full Baseline}}\\
\hspace{1em}MV Histogram & 0.02 & 0.09 & 0.74 & 0.09 & 0.34 & 0.10 & 0.52 & 0.18 & -0.06 & 0.09 & 0.48 & 0.18\\
\hspace{1em}Hybrid & -0.03 & 0.07 & 0.90 & 0.03 & 0.25 & 0.07 & 0.73 & 0.09 & -0.40 & 0.10 & 0.57 & 0.16\\
\dpmb{1} & 0.46 & 0.00 & 0.00 & 0.52 & 0.29 & 0.00 & 0.51 & 0.14 & -0.06 & 0.00 & 0.10 & 0.18\\
\dpmb{2} & -0.07 & 0.00 & 0.50 & 0.00 & 0.04 & 0.00 & 0.50 & 0.11 & -4.78 & 0.00 & 0.00 & 4.54\\
\dpmb{9} & -0.09 & 0.00 & 0.52 & 0.02 & 0.21 & 0.00 & 0.51 & 0.06 & 0.05 & 0.00 & 0.00 & 0.29\\
\bottomrule
\end{tabular}
\caption{Utility measure for synthetic datasets from the various methods across the three covariate sets for the \texttt{fam\_asb\_lt} response variable.} \label{tab:apdx-itt-famasblt}
\end{table}

%% file: tables-main/liberia_itt_asbhostilstd_ltav.tex
\begin{table}
\centering
\small

\begin{tabular}{>{\raggedright\arraybackslash}p{0.17\textwidth}|>{\raggedleft\arraybackslash}p{0.061\textwidth}>{\raggedleft\arraybackslash}p{0.055\textwidth}>{\raggedleft\arraybackslash}p{0.055\textwidth}>{\raggedleft\arraybackslash}p{0.055\textwidth}|>{\raggedleft\arraybackslash}p{0.061\textwidth}>{\raggedleft\arraybackslash}p{0.055\textwidth}>{\raggedleft\arraybackslash}p{0.055\textwidth}>{\raggedleft\arraybackslash}p{0.055\textwidth}|>{\raggedleft\arraybackslash}p{0.061\textwidth}>{\raggedleft\arraybackslash}p{0.055\textwidth}>{\raggedleft\arraybackslash}p{0.055\textwidth}>{\raggedleft\arraybackslash}p{0.055\textwidth}}
\toprule
\multicolumn{1}{c}{ } & \multicolumn{4}{c}{Therapy Only} & \multicolumn{4}{c}{Cash Only} & \multicolumn{4}{c}{Both} \\
\cmidrule(l{3pt}r{3pt}){2-5} \cmidrule(l{3pt}r{3pt}){6-9} \cmidrule(l{3pt}r{3pt}){10-13}
Method & ITT & Std. Error & CI overlap & Abs. Diff. & ITT & Std. Error & CI overlap & Abs. Diff. & ITT & Std. Error & CI overlap & Abs. Diff.\\
\midrule
\addlinespace[0.3em]
\multicolumn{13}{l}{\textbf{Reduced Subset}}\\
\hspace{1em}MV Histogram & -0.11 & 0.11 & 0.89 & 0.05 & -0.06 & 0.10 & 0.97 & 0.01 & -0.20 & 0.10 & 0.72 & 0.12\\
\hspace{1em}Hybrid & -0.17 & 0.09 & 0.91 & 0.02 & -0.12 & 0.09 & 0.88 & 0.05 & -0.25 & 0.13 & 0.88 & 0.06\\
\hspace{1em}\dpmb{1} & -0.18 & 0.00 & 0.51 & 0.02 & -0.11 & 0.00 & 0.51 & 0.04 & 1.48 & 0.00 & 0.00 & 1.79\\
\hspace{1em}\dpmb{2} & 1.33 & 0.00 & 0.00 & 1.48 & -0.53 & 0.00 & 0.00 & 0.46 & -0.16 & 0.00 & 0.52 & 0.15\\
\hspace{1em}\dpmb{9} & -0.05 & 0.00 & 0.51 & 0.10 & 0.05 & 0.00 & 0.51 & 0.12 & -0.57 & 0.00 & 0.00 & 0.26\\
\addlinespace[0.3em]
\multicolumn{13}{l}{\textbf{Subset}}\\
\hspace{1em}MV Histogram & -0.05 & 0.11 & 0.75 & 0.11 & 0.04 & 0.10 & 0.76 & 0.10 & -0.08 & 0.10 & 0.42 & 0.24\\
\hspace{1em}Hybrid & -0.20 & 0.09 & 0.89 & 0.04 & -0.12 & 0.09 & 0.86 & 0.06 & -0.29 & 0.13 & 0.92 & 0.04\\
\hspace{1em}\dpmb{1} & -0.13 & 0.00 & 0.50 & 0.02 & -20.99 & 0.00 & 0.00 & 20.93 & -0.60 & 0.00 & 0.00 & 0.27\\
\hspace{1em}\dpmb{2} & -0.74 & 0.00 & 0.00 & 0.58 & -0.05 & 0.00 & 0.51 & 0.02 & -7.91 & 0.00 & 0.00 & 7.59\\
\hspace{1em}\dpmb{9} & -0.17 & 0.00 & 0.51 & 0.02 & -35.50 & 0.00 & 0.00 & 35.44 & -0.43 & 0.00 & 0.52 & 0.11\\
\addlinespace[0.3em]
\multicolumn{13}{l}{\textbf{Full Baseline}}\\
\hspace{1em}MV Histogram & -0.03 & 0.11 & 0.75 & 0.11 & 0.26 & 0.11 & 0.33 & 0.28 & -0.08 & 0.10 & 0.39 & 0.25\\
\hspace{1em}Hybrid & -0.18 & 0.09 & 0.89 & 0.04 & -0.02 & 0.09 & 0.92 & 0.01 & -0.29 & 0.12 & 0.92 & 0.04\\
\dpmb{1} & -0.54 & 0.00 & 0.00 & 0.40 & 1.71 & 0.00 & 0.00 & 1.74 & -0.05 & 0.00 & 0.00 & 0.27\\
\dpmb{2} & -24.57 & 0.00 & 0.00 & 24.43 & 0.15 & 0.00 & 0.51 & 0.17 & -3.44 & 0.00 & 0.00 & 3.11\\
\dpmb{9} & -0.13 & 0.00 & 0.51 & 0.01 & -35.99 & 0.00 & 0.00 & 35.97 & -0.24 & 0.00 & 0.52 & 0.09\\
\bottomrule
\end{tabular}
\caption{Utility measure for synthetic datasets from the various methods across the three covariate sets for the \texttt{asbhostilstd\_ltav} response variable.} \label{tab:apdx-itt-asbhostilstdltav}
\end{table}

%% file: section-appendix-plots-and-tables.tex
In Sections \ref{sec:evaluating} and \ref{sec:real-world}, we consider several parametrization of the \longdpmb method as summarized in \ref{tab:algoparams}. Experiment 1 in Section \ref{sec:evaluating} refers to the simulations which vary the privacy budget (Section \ref{sec:sim-privbudget-compare}. Experiment 2 in Section \ref{sec:evaluating} refers to the simulations which vary the model and consider two privacy budgets for \longdpmb (Section \ref{sec:sim-model-compare}).

\input{tables-and-figures-appendix/method_parameters}

\begin{figure}[!htb]
    \centering
    \includegraphics[width=0.9\linewidth]{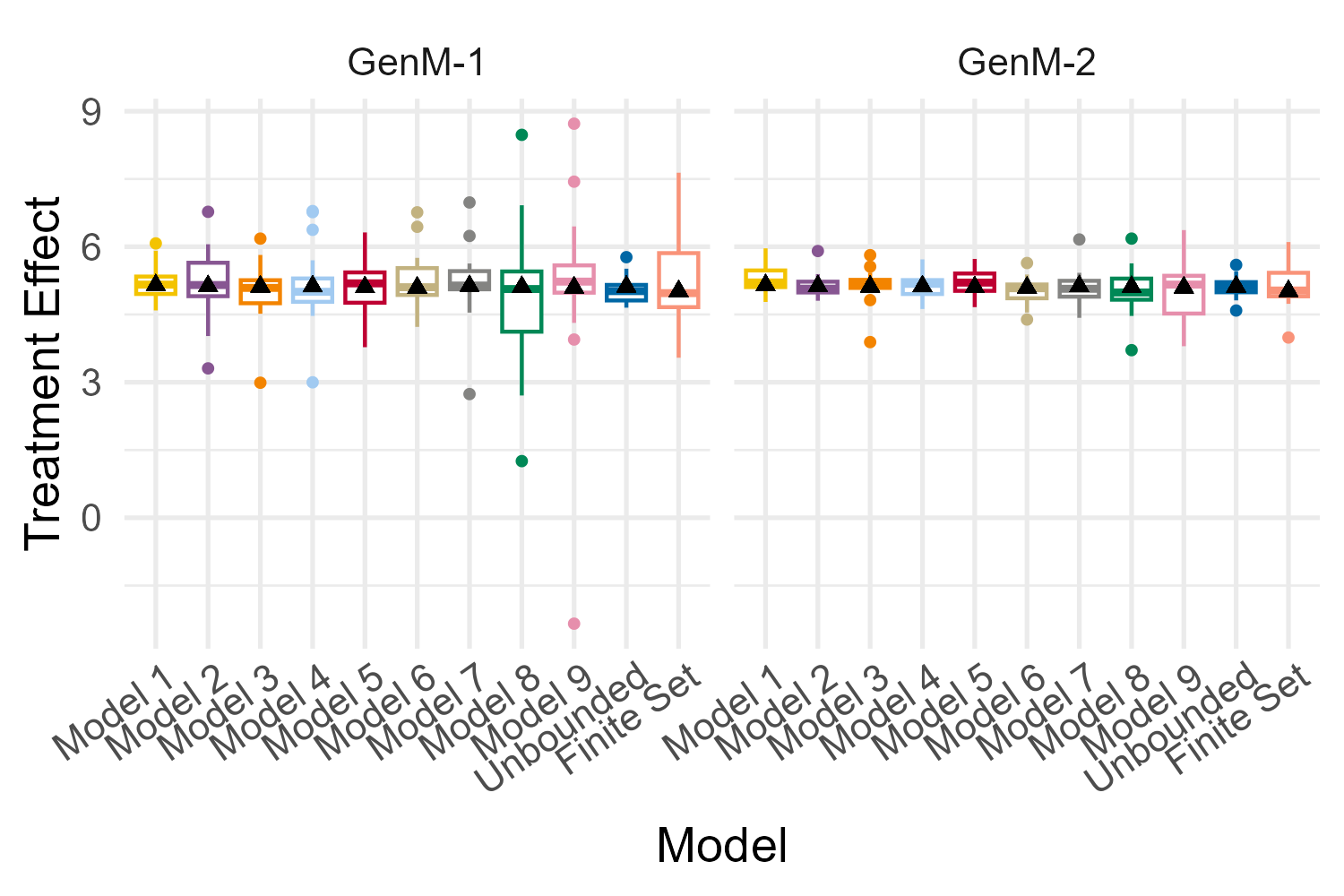}
    \caption{There are several outliers in the estimated treatment effects which were not included in Figure \ref{fig:sim-privbudget-boxplots} in Section \ref{sec:sim-privbudget-compare}.}
    \label{fig:apdx-sim-privbudget}
\end{figure}

In Figure \ref{fig:sim-privbudget-boxplots}, we remove outliers from the \longdpmb plots to improve the ease of interpretation. For full disclosure, we include the plots with outliers in Figure \ref{fig:apdx-sim-privbudget}.

\subsection{Additional Plots for Experiment 2 in Section \ref{sec:sim-model-compare}}

\begin{figure}[!htb]
    \centering
    \includegraphics[width=0.98\linewidth]{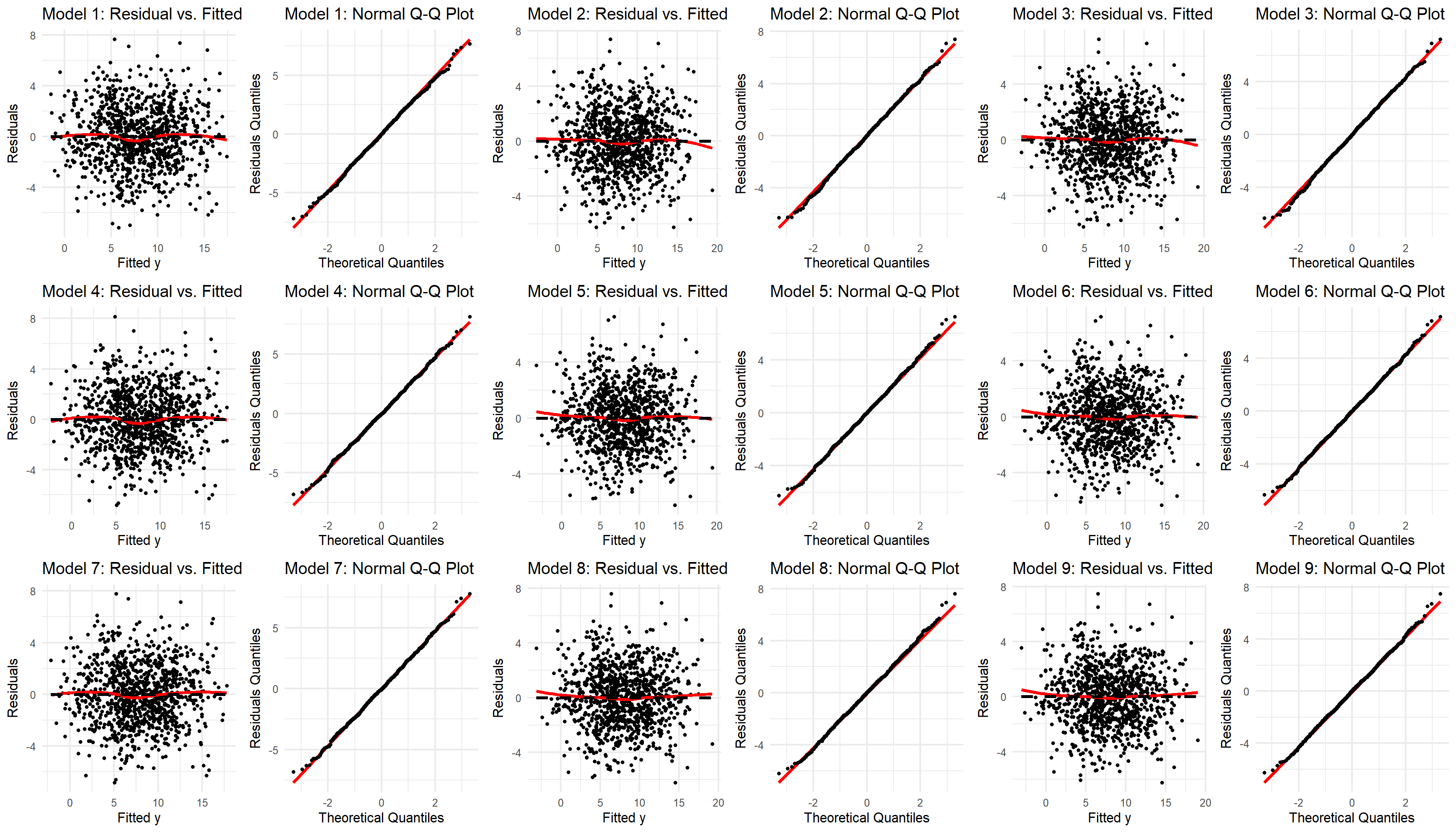}
    
    \includegraphics[width=0.65\linewidth]{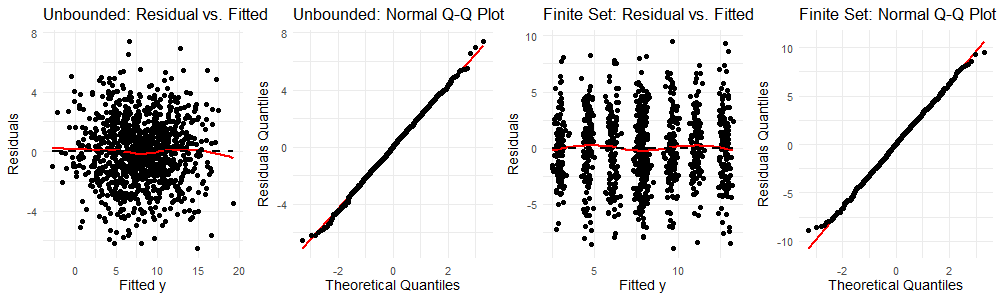}
    \caption{We consider the diagnostic plots of the residuals for each of the 11 regression models (Models 1 through 9 and the Unbounded and Finite Set models). The assumption that the residuals are normally distributed with mean zero and equal variance appears to be reasonable for all of the models. Despite several models missing covariate variables used in the generation of the response variable, the models do not suffer from omitted variable bias.}
    \label{fig:sim-model-fits}
\end{figure}

When we generate the "confidential" dataset in Section \ref{sec:sim-model-compare} we use the variables found in Model 9. However, when we then fit other models to the dataset we risk omitted variable bias. The diagnostic plots in Figure \ref{fig:sim-model-fits} suggest that the residuals are normally distributed with mean zero and constant variance. The models with less covariates do not appear to suffer from omitted variable bias.

%% file: tables-and-figures-appendix/method_parameters.tex
\begin{table}[!htb]
\begin{tabular}{cl}
    \toprule
     \multicolumn{2}{c}{\longdpmb in Section \ref{sec:evaluating}}\\
     \midrule
     $\epsilon_{\bX}=\epsilon_{\bfy}=\epsilon/2$, $\delta_{\bX}=\delta_{\bfy}=0$, & for experiment 1 in Section \ref{sec:sim-privbudget-compare}\\
     $\epsilon=1$, $\delta=0$ $\epsilon_{\bX}=\epsilon_{\bfy}=1/2$ & for \dpmb{1} in Section \ref{sec:sim-model-compare} \\ 
     $\epsilon=2$, $\delta=0$ $\epsilon_{\bX}=\epsilon_{\bfy}=1$ & for \dpmb{2} in Section \ref{sec:sim-model-compare} \\ \midrule
     \multicolumn{2}{c}{\longdpmb in Section \ref{sec:real-world}}\\ \midrule
     $\epsilon=1$, $\delta=0$ $\epsilon_{\bX}=1/2$, $\epsilon_{\bfy}=1/16$ & for \dpmb{1} \\
     $\epsilon=2$, $\delta=0$, $\epsilon_{\bX}=1$,$\epsilon_{\bfy}=1/8$& for \dpmb{2} \\
     $\epsilon=9$, $\delta=0$, $\epsilon_{\bX}=1$, $\epsilon_{\bfy}=1$ & for \dpmb{9}\\ \midrule
     \multicolumn{2}{c}{\longdpmb in Sections \ref{sec:evaluating} and \ref{sec:real-world}}\\ \midrule
        $\binparam=2/3$, $B=5000$ & for MV Histogram and proxies\\
         $(\sigma_{min},\sigma_{max})=(2^{-15},2^{15})$, $\epsilon_{\sigma,var}=\epsilon_{\bfy}/(3*10)$, $\delta_{\sigma}=0$ & for DP Variance \\ $\epsilon_{range,\sigma}=\epsilon_{\bfy}/(3*10)$, $\delta_{range, \sigma}=0$ & for DP Range \\
         $\epsilon_{\sigma}==\epsilon_{\bfy}/(3*10)$ for Laplace noise\\
$\alpha_{range,\sigma^2}=\alpha_{range,\beta}=0.05$, $\bdmean_{\sigma^2}=\bdmean_{\beta}=50$ & for DP Range \\
$\epsilon_{range,\treff,k}=\frac{3\epsilon_{\bfy}}{20t}$, $\delta_{range,\treff,k}=0$, $\epsilon_{\treff,k}=\frac{3\epsilon_{\bfy}}{20t}$ for $k=1,\dots,t$ treatments & for each $\treff$ \\
$\epsilon_{range,\covcoef,\ell}=\frac{3\epsilon_{\bfy}}{10p}$, $\delta_{range,\covcoef,\ell}=0$,  $\epsilon_{\covcoef,\ell}=\frac{3\epsilon_{\bfy}}{10p}$ for $\ell=1,\dots,p$ covariates & for each $\covcoef$ \\ \bottomrule
    \end{tabular}
    \caption{Parameters are set for our applications of \longdpmb (Algorithm \ref{algo:fullmodelbased}) in Sections \ref{sec:evaluating} and \ref{sec:real-world}. In the first experiment we vary $\epsilon$ with $t=1$ and $p=8$. In the second experiment we vary $p$ with $t=1$. In our application to Section \ref{sec:real-world}, $t=3$ and $p$ varies. Also, there are eight response variables being generated, so $\epsilon=\epsilon_{\bX}+8\epsilon_{\bfy}$.}
    \label{tab:algoparams}
\end{table}